\documentclass[amssymb,10pt,amsmath,bibnotes,twocolumn,aps,pra]{revtex4-2}
\usepackage{graphicx}
\usepackage{dcolumn}
\usepackage{bm}
\usepackage{xcolor} 
\usepackage{physics}
\usepackage{relsize}
\usepackage{hyperref}
\hypersetup{
  colorlinks   = true, 
  urlcolor     = black, 
  linkcolor    = blue, 
  citecolor    = red 
}

\newcommand{\uu}[1]{\underline{\underline{#1}}}

\begin{document}

\title{Tracking Adiabaticity in Non-Equilibrium Many-Body Systems: The Hard Case of the X-ray Photoemission in Metals}
\author{Gustavo Diniz¹}
\author{Felipe D. Picoli¹}
\author{Luiz N. Oliveira¹}
\author{Irene D'Amico²}
\altaffiliation[ ]{}
\affiliation{¹Instituto de Física de São Carlos, Universidade de São Paulo, São Carlos, São Paulo, Brazil \\
²School of Physics, Engineering and Technology, University of York, York, United Kingdom}%
\begin{abstract}
The level of adiabaticity determines many properties of time-dependent quantum systems. However, a reliable and easy-to-apply criterion to check and track it remains an open question, especially for complex many-body systems. Here we test techniques based on metrics which have been recently proposed to quantitatively characterize and track adiabaticity.
We investigate the time evolution of x-ray photoemission in metals, which displays a strongly out-of-equilibrium character, continuum energy spectrum, and experiences the Anderson orthogonality catastrophe: a nightmarish scenario for this type of test.
Our results show that the metrics-based methods remains valid. In particular, we demonstrate that the natural local density distance is able not only to track adiabaticity, but also to provide information not captured by the corresponding Bures' or trace distances about the system's dynamics.  
In the process, we establish an explicit upper limit for this local density distance in terms of the trace distance, and derive a simple analytical solution that accurately describes the time evolution of a Fermi gas with a localized scattering potential for a large range of parameters. We also demonstrate that, for x-ray photoemission, the quantum adiabatic criterion, as commonly used, fails to predict and track adiabaticity. The local particle density is typically much simpler to compute than the corresponding quantum state and it is experimentally measurable: this makes the method tested extremely appealing.


\end{abstract}
\maketitle

\section{Introduction}
Understanding adiabaticity is important in nearly all areas of physics. Quantum adiabatic processes have garnered significant interest due to their relevance in quantum computing \cite{farhi2002quantum, RevModPhys.90.015002, PhysRevA.107.032422,RevModPhys.91.045001, TORRONTEGUI2013117, Grace_2010}, quantum thermodynamics \cite{Su_2018,PhysRevE.79.041129,PhysRevE.87.022143,Skelt_2019,PhysRevLett.118.100602,PhysRevA.93.032118} and non-equilibrium condensed matter physics \cite{PhysRevLett.101.175701, Eisert2015, PhysRevB.45.9413}.

The pioneers on this topic were Born and Fock in 1928 \cite{Born}. In their paper, they articulated the Adiabatic Theorem. Under the assumption of a discrete and non-degenerate spectrum (or at most a linear crossing between the key eigenvalue and the rest of the spectrum), they state that if a quantum system is initially in an eigenstate and the external parameters of the system change slowly enough, the system will remain in an instantaneous eigenstate throughout the process. The theorem has been further refined in a set of seminal papers \footnote{See e.g.  J. E. Avron and A. Elgart, Communications in Mathematical Physics 203, 445 (1999) and references therein.}, including the restriction on degeneracy being lifted and part of the spectrum allowed to be continuous \cite{JPSJ.5.435}, and the possibility of having energy bands \cite{JPhysAMath.13.L15,CommunMathPhys.110.3349,CommunMathPhys.3.649650}. In particular, Avron and Elgart \cite{CommMathPhys.203.445463} reformulated the Adiabatic Theorem without the need for a gap condition. 

At zero temperature, the adiabatic theorem is frequently applied in the form of the Quantum Adiabatic Criterion (QAC) 
\begin{eqnarray}\label{QAC}
 \frac{\hbar|\bra{m(t)} \Dot{H}(t) \ket{m'(t)}|}{(E_{m'}(t) - E_m(t))^2} \ll 1,
\end{eqnarray}
with an extension to finite-temperatures provided in Ref.~\cite{AdvQuantumTechnol.3.1900139}.
In this context, $\Dot H$ represents the time derivative of the Hamiltonian, while $\ket{m}$ and $\ket{m'}$ denote the instantaneous eigenstates of $H$ with corresponding eigenenergies $E_m$ and $E_{m'}$. The QAC is commonly approximated by including the ground and first excited states only.
Using Eq.~\eqref{QAC} in many-body systems presents unique challenges, due to its perturbative nature and the complexity of the many-body energy spectrum. Also, the continuous spectrum limit cannot be studied using Eq. \eqref{QAC}.

Recently, the validity and sufficiency of QAC for certain systems have been questioned \cite{PhysRevLett.101.060403, PhysRevLett.95.110407,PhysRevA.80.012106,PhysRevLett.93.160408,PhysRevA.86.032121,PhysRevLett.102.220401,JMathPhys.48.102111}. In addition, new approaches have emerged to characterize adiabaticity \cite{PhysRevB.98.024307,PhysRevA.104.L030202,AdvQuantumTechnol.3.1900139,PhysRevA.98.012104,TheorMathPhys.211.545}, in which the focus is on quantities more directly related to the evolution of the state of the system than to the spectrum of eigenenergies. This helps with  characterizing adiabaticity for systems with complex or continuum energy spectra and with accounting for memory effects, which are important when the dynamics encompasses both periods of adiabaticity and of non-adiabaticity.  

In their work \cite{PhysRevA.98.012104, AdvQuantumTechnol.3.1900139,BrazJPhys.48.467471}, the authors use the metric-space approach to quantum mechanics\cite{PhysRevLett.106.050401,PhysRevB.89.115137} to characterize the level of adiabaticity of dynamical quantum many-body systems operating at zero and finite-temperatures. The metric-space approach is based on the use of natural metrics, that is metrics derived from conservation laws \cite{PhysRevB.89.115137}. In addition to adiabaticity, this method has been successfully applied to characterize properties of quantum many-body systems in a magnetic field  \cite{PhysRevA.92.032509,JMMM.400.99-102}, density functional theory 
\cite{PhysRevLett.106.050401,PhysRevA.94.062509}, fermionic correlations   \cite{PhysRevMaterials.1.043801}, and quantum phase transitions \cite{BrazJPhys.48.472476,PhysRevResearch.2.033167}.

In \cite{PhysRevA.98.012104, AdvQuantumTechnol.3.1900139,BrazJPhys.48.467471}   Bures \cite{HUBNER1992239, Wu2020}, trace \cite{wilde_2013} and the natural metric for the local particle density \cite{PhysRevLett.106.050401}  are proposed to measure how far from the adiabatic a quantum system evolution is. The authors successfully test their method on dynamics of systems confined by smooth random  \cite{BrazJPhys.48.467471,PhysRevA.98.012104} or harmonic potentials  \cite{PhysRevA.98.012104}, and on driven non-homogeneous Hubbard chains \cite{AdvQuantumTechnol.3.1900139}.
They also show how these metrics overcome pitfalls of the QAC, including accounting for non-Markovian behavior.

Computing the time-dependent wave function of a many-body system is usually challenging. In contrast, calculating the local density is typically easier, especially for calculations performed using electronic density approaches such as Density Functional Theory (DFT) \cite{PhysRev.140.A1133}. Indeed, the Runge–Gross theorem \cite{PhysRevLett.52.997}  states that the information provided by the time-dependent wave function and by the local particle density is the same. This suggests that an appropriate local density metric may be used to characterize the level of adiabaticity in a dynamical system. The analytical and numerical results in  \cite{AdvQuantumTechnol.3.1900139,PhysRevA.98.012104,BrazJPhys.48.467471}, strongly support the use of the  local-density natural metric for identifying dynamical regimes. In particular, in \cite{AdvQuantumTechnol.3.1900139} the authors propose a semi-analytical upper bound on the adiabatic threshold for the  local-density natural metric in terms of the  Bures' distance  of the corresponding quantum states, but recognize that a tighter bound would be auspicable. 
Clearly, further developments of the method, e.g. in relation to the bound, would be welcome, as well as additional tests on challenging dynamics. 

This motivates the present work. First we derive an explicit upper bound for the evolution of the local-density natural metric in terms of the evolution of the corresponding trace metric which is valid for a generic dynamics of any system starting in the ground state. Then, we test this and the metrics' performance in the context of the x-ray photoemission.   
We think that X-ray photoemission in metals\cite{PhysRev.163.612, PhysRev.178.1097,Doniach_1970,PhysRevB.24.4863, PhysRevB.71.045326,J.Phys.019710032011-12091300,RevModPhys.62.929} can provide an acid test to the method: it is a strongly out-of-equilibrium phenomenon in which the Anderson's orthogonality catastrophe  emerges \cite{PhysRevLett.18.1049}. Furthemore, real materials typically have $\sim 10^{23}$ atoms, resulting in a metallic band with a continuum energy spectrum. The metrics allow us study continuum spectra and spectra approaching the continuum limit, generally not possible using the QAC.

This article comprises the following sections. In Section II,  we review the use of metrics to track adiabaticity in many-body problems, provide a geometric visualization for the metrics, which helps to explain why these distances can be used to measure adiabaticity, and present our result for an upper bound to the local-density natural metric. In Section III, we  detail some aspects of the x-ray photoemission problem, present the model we are considering, and the strategy to find its time evolution. This includes the derivation of a simpler analytical solution which allows us to determine the parameter region for adiabaticity as we approach the continuum limit. In Section IV we discuss our analytical and numerical results. Finally, Section V contains summary and conclusion. Details of our derivations are contained in the Appendices.

\section{Metrics to track Adiabaticity}

In this section, we highlight the relationship between the Bures, trace and natural local-density metrics and the instantaneous ground state transition probability. In addition, we provide a geometric visualization of the distances and discuss their connection to adiabaticity.

We will consider a dynamics starting from the ground state and evolving under the Hamiltonian $H(t)$. At zero temperature, adiabatic evolution takes place when the system evolves slowly enough to always remain in its instantaneous ground state, so it is convenient for us to perform all calculations on the instantaneous basis $\{\ket{\varphi_k (t)}\}$ which diagonalizes the Hamiltonian at each instant of time $t$, with $E_k(t)$ the instantaneous eigenenergies.  The expansion of the time-dependent quantum state on the instantaneous basis is 
\begin{equation}\label{Wave_function_expantion}
    \ket{\Psi(t)} = \sum_{k} c_k(t) \ket{\varphi_k (t)},
\end{equation}
with the instantaneous ground state transition probability given by 
\begin{equation}\label{c_0_sq}
    |c_0(t)|^2 = |\bra{\Psi(t)}\ket{\varphi_0(t)}|^2.
\end{equation}
For pure states, both the trace and the Bures distances can be written in terms of the overlap between the two states, so that, in the case of interest, the trace distance can be written as \cite{wilde_2013}
\begin{eqnarray}\label{TR}
D_T (\ket{\Psi(t)},\ket{\varphi_0(t)}) = \sqrt{1 - |c_0(t)|^2},
\end{eqnarray}
and the Bures distance as \cite{PhysRevLett.106.050401}
\begin{eqnarray}\label{BR}
&D_B(\ket{\Psi(t)},\ket{\varphi_0(t)}) = \sqrt{2\left(1- |c_0(t)| \right)}.
\end{eqnarray}
From Eq. \eqref{BR} and \eqref{TR}, both  distances carry the same information.


\subsection{Visualization of quantum state metrics }
We can conceptualize $\ket{\Psi(t)}$ as a time-dependent vector existing within a vector space with basis $\{\ket{\varphi_n (t)}\}$. Since $\ket{\Psi(t)}$ is normalized, it can be geometrically visualized as a hypersphere surface, defined by the condition $\sum_{k} |c_k(t)|^2= 1$. It is possible to simplify this perspective by focusing solely on the instantaneous ground state $\ket{\varphi_0(t)}$, and a linear combination of all the excited states $\ket{\Tilde\varphi (t)}$, writing $\ket{\Psi(t)} = c_0(t)\ket{\varphi_0(t)} + {c}_e(t) \ket{\Tilde\varphi(t)}$, where $\ket{\Tilde\varphi (t)}=\sum_{k>0}  \left({{c}_e(t)}\right)^{-1} {c_k(t)} \ket{\varphi_k(t)}$ and  
\begin{equation}\label{c_tilde}
    {c}_e (t) = \sqrt{\sum_{k>0} |c_k(t)|^2}.
\end{equation}

\begin{figure}[hbt!]
    \centering
    \includegraphics[scale=0.30]{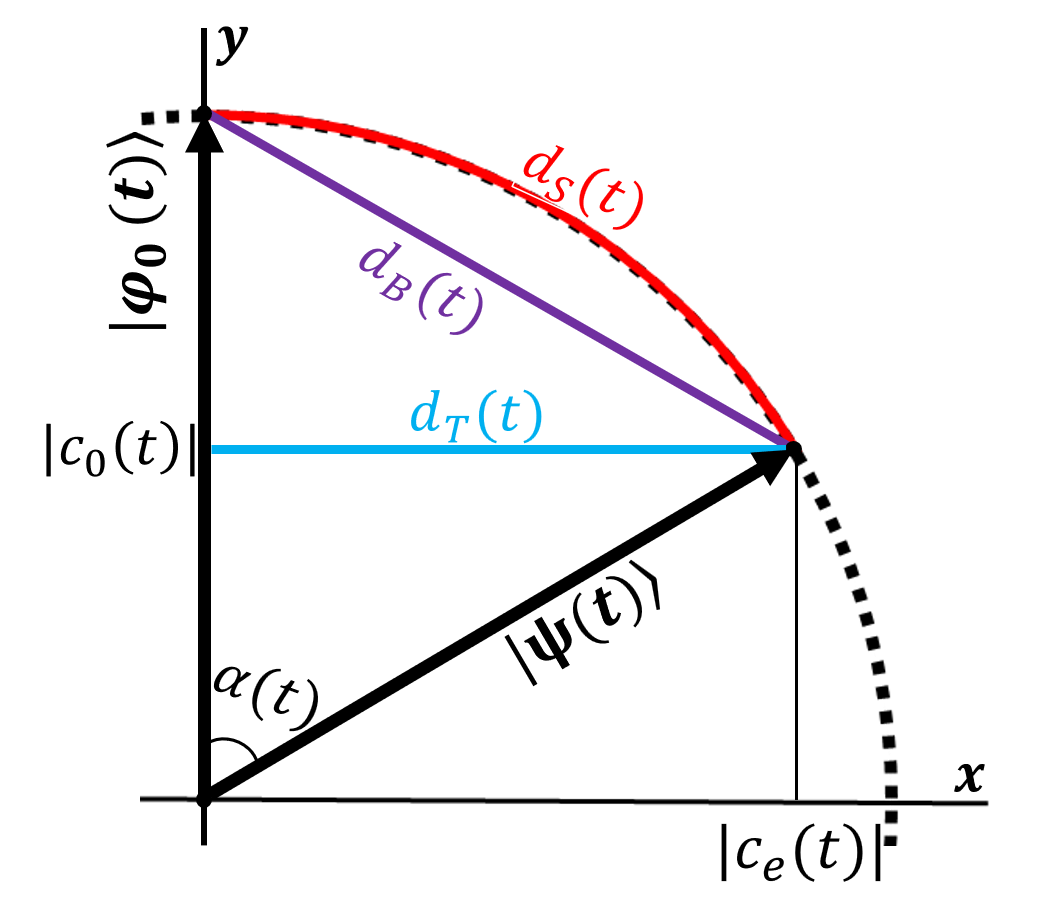}
    \caption{Geometric visualization of the quantum state as a vector in a circle and of the distances $d_T$, $d_S$ and $d_B$ between $\ket{\Psi(t)}$ and $\ket{\varphi_0(t)}$. Here, $|c_0(t)|^2$ represents the probability of finding the quantum state in the instantaneous ground state and $|c_e(t)|^2$ represents the probability of finding it in one of the excited states.}
    \label{fig:sphere}
\end{figure}

After this simplification, and the alignment of $\ket{\Psi(0)}$ with the $y$-axis, $\ket{\Psi(t)}$ lives in the first quadrant of the circle defined by $|c_0(t)|^2 + |{c}_e(t)|^2 = 1$ as in Fig. \ref{fig:sphere}. This construction is equivalent to the metric space geometry for wave-functions described in \cite{PhysRevLett.106.050401}, when this is restricted to the distance between $\ket{\Psi(t)}$ and its initial state.
Once we consider $\ket{\Psi(0)} = \ket{\varphi_0(0)}$, then at $t=0$ the distance between these two vectors is null. As the Hamiltonian changes, the probability of the system be the instantaneous excited states manifold $\ket{\tilde{\varphi} (t)}$ can increase.

There exist numerous methods for measuring the distance between vectors $\ket{\Psi(t)}$ and $\ket{\varphi_0(t)}$ in the Fig. \ref{fig:sphere}. However, the three most intuitive approaches include measuring the direct Euclidean distance $d_B$ between the two vectors, calculating the projection $d_T$ of $\ket{\Psi(t)}$ onto $\ket{\tilde{\varphi}(t)}$, and determining the arc length $d_S$ between them. 

The first corresponds to the Bures distance \footnote{See also \cite{PhysRevLett.106.050401}, and \cite{G_Trindade}}, so $d_B(t) = D_B(\ket{\Psi(t)},\ket{\varphi_0(t)})$.
The trace distance is equivalent here to measure the projection of $\ket{\Psi}$ onto the $\ket{\Tilde\varphi}$ direction, so
 $d_T(t) = D_T(\ket{\Psi(t)},\ket{\varphi_0(t)})$.
The last distance is associated with the Fisher metric \cite{PhysRevX.6.021031, G_Trindade}, since this metric can be used to compute the distances in a hypersphere shell. Measuring the arc length between $\ket{\Psi(t)}$ and $\ket{\varphi_0(t)}$, we obtain $d_S(t) = \rm{acos}|c_0(t)| = \rm{asin}\sqrt{1-|c_0(t)|^2}$.

\subsection{Relation to adiabaticity}
These visualizations improves our understanding of why these metrics can track adiabaticity. If $H(t)$ changes rapidly, the probability amplitudes of the excited states increases, resulting in a decrease in $|c_0(t)|$ and an increase in the distance between $\ket{\Psi(t)}$ and $\ket{\varphi_0(t)}$. Conversely, if the Hamiltonian changes slowly, the reduction in $|c_0(t)|$ is small, and the distance is also small. 

The Adiabatic Criterion estimates the value of $|c_0(t)|$ using time-dependent perturbation theory: it use the expression $|c_0(t)|^2 = 1 - | c_e (t)|^2$  and seeks (perturbative) conditions for $|c_e(t)|$ to be small \cite{Born,JPSJ.5.435, PhysRevLett.98.150402}. On the other hand, the metrics we discussed directly employ $|c_e(t)| = \sqrt{1 - |c_0 (t)|^2}$, without approximations. 

These are the reason the metrics Eqs.~(\ref{BR}) and (\ref{TR}) can be used to track adiabaticity and work so effectively for any dynamics and physical system.

\subsection{Adiabatic threshold and upper bound for adiabatic dynamics of the local density metric}\label{AT}
The  natural local density (NLD) metric \cite{PhysRevLett.106.050401}, when designed for tracking adiabaticity of systems defined on a grid, is expressed as
\begin{equation}\label{ODM}
\begin{aligned}
&D_{n}(n(t),n^A(t)) = \frac{1}{N_e} \sum_{i} \left|n_i(t) - n_i^A (t) \right|.
\end{aligned}
\end{equation}
Here, $n(t)=\{n_i(t)\}$ and $n^A(t)=\{n_i^A(t)\}$, with $n_i(t)$ the occupation of  site $i$ at time $t$ and  $n_i^A(t)$ the  corresponding occupation for the instantaneous ground state.
The geometric interpretation of the NLD  distance is discussed in \cite{PhysRevLett.106.050401}. 

In \cite{AdvQuantumTechnol.3.1900139} the concept of `adiabatic threshold' was introduced: two states would be considered adiabatic (for the purposes at hand) if their Bures distance would be less than a certain fraction of the Bures distance's maximum value (chosen to be 10\% in \cite{AdvQuantumTechnol.3.1900139}).  A similar threshold was defined for the NLD metric, which, crucially, was connected to the adiabatic threshold for the quantum state through an upper bound. However this bound was partly based on inference from numerical results and it was unclear how tight the bound actually was. It is then desirable to derive an analytical expression that can quantify the relationship between NLD and quantum state metrics when evaluating adiabaticity. 

In this paper, we derive a superior limit for the NLD distance \eqref{ODM} which, similar to \eqref{BR} and \eqref{TR}, only depends on $|c_0(t)|^2$. This reads  
\begin{eqnarray}\label{NR}
&D_n(n(t),n^A(t)) \leq 2\sqrt{1-|c_0(t)|^2},
\end{eqnarray}
and is valid for systems with a fixed number of particles.
The inequality \eqref{NR} is derived by computing the occupation numbers $\{n_i\}$ using the density matrices corresponding to $\ket{\Psi(t)}$ and to the instantaneous ground state $\ket{\varphi_0(t)}$. Details of the proof are given in Appendix \ref{SLDD}.

Using Eqs \eqref{NR} and \eqref{TR} we then obtain
\begin{eqnarray}\label{NTR}
&D_n(n(t),n^A(t)) \leq 2 D_T (\ket{\Psi(t)},\ket{\varphi_0(t)}).
\end{eqnarray}

\section{The x-ray photoemission problem}\label{XPP}
In this section, we discuss our model for x-ray photoemission, the underlying physics of this phenomenon, and the strategy to compute the time evolution.

When an x-ray photon strikes a metal, it can interact with the inner-shell electrons, causing their excitation and eventual emission. This constitutes the phenomenon of x-ray photoemission in metals . This strongly out-of-equilibrium phenomenon exhibits numerous interesting properties. As the x-ray photon ejects one deep core electron, the electrons in the metallic band perceive this hole as a localized attractive scattering potential $K(t)$. Under these conditions, the Anderson's orthogonality catastrophe  emerges \cite{PhysRevLett.18.1049}. These characteristics present additional challenges for tracking the level of (non) adiabaticity of the process.
This attractive scattering potential $K(t)$ `shakes' the conduction electrons and changes the energy levels, creating many pairs of particle-hole excitations and giving rise to the physics observed experimentally \cite{PhysRev.178.1097, Doniach_1970}. To follow this phenomenon, we will set the system initially in the ground state, and compute the time evolution generated by the Hamiltonian with the time-dependent scattering potential.

First, we describe the metal using a spin-independent tight-binding model for a 1D chain with half-filled band and a localized potential coupled with the first site. This reads 
\begin{equation}\label{H_t}
\begin{aligned}
\mathcal{H}(t) = \sum_{n=0}^{N-1}\tau \left( a_n^{\dagger}a_{n+1}+\rm{h.c.}\right) + \tau K(t) a_0^{\dagger}a_0,
\end{aligned}
\end{equation}
where $a_n^{\dagger}$ ($a_n$) represents the electron creation (annihilation) operator at site $n$, and $N$ is the number of sites. Here, $\tau$ is the hopping parameter, to which we can associate the characteristic  time
\begin{equation}\label{T_tau}
    T_\tau = \frac{\hbar}{\tau}.
\end{equation}

The tight-binding energy, with reference to the Fermi energy $\epsilon_F$, is $\varepsilon_k = 2\tau \sin\left(\frac{\pi}{N} k\right)$, with $k$ an integer such that $-N/2 \leq k \leq N/2$. Near the Fermi energy, the energy differences between successive levels are given by  $|\Delta\varepsilon_k| = 2 \tau \frac{\pi}{N} \cos\left(\frac{\pi}{N}k\right) \approx \frac{2\pi \tau}{N}$.
To this minimal energy gap
\begin{equation}\label{GAP}
    \Delta\varepsilon = \frac{2\pi \tau}{N}
\end{equation}
 we can associate the characteristic time
 \begin{equation}\label{T_eps}
     T_{\Delta\varepsilon}=\frac{\hbar}{\Delta\varepsilon}.
 \end{equation}

 The gap $\Delta\varepsilon$ vanishes in the limit as $N \rightarrow \infty$: We can approach this limit by increasing the system size. Then, according to the Heisenberg uncertainty principle, our simulations will faithfully reproduce the behavior  of a gapless metallic band up to the corresponding longer and longer time scales $T_{\Delta\varepsilon} (N)$. Recent reports of transient behaviour due to photoemission can be found, e.g., in \cite{Cui2014, PhysRevX.9.011044, PhysRevB.59.10935, PhysRevLett.109.087401}. 
Simulating the x-ray photoemission process by approaching the gapless limit is ideal to test the metrics method for identifying adiabatic and non-adiabatic behaviour:  in fact, the QAC criterion would not allow the study of systems with an energy gap spectrum approaching the continuum limit.

For a half-filled band, i.e. $N_e = N/2$, the number of levels below the Fermi level is given by
\begin{equation}\label{N_e}
    N_e = \frac{\pi\tau}{\Delta\varepsilon}=\pi\frac{ T_{\Delta\varepsilon}}{T_\tau}.
\end{equation}
In this case, Anderson \cite{PhysRevLett.18.1049} showed that the overlap of the ground states, before and after the onset of the localized scattering potential, depends on the number of electrons $N_e$ of the system as
\begin{eqnarray}\label{AOC}
&\left|\bra{\varphi_0(0^-)}\ket{\varphi_0(t)}\right| \sim  N_e^{-\left({\frac{\delta(t)}{\pi}}\right)^2}
\end{eqnarray}
where $\ket{\varphi_0(0^-)}$ indicate the ground state immediately before the potential onset.
Here, 
\begin{equation}
\delta(t) = \rm{atan} \left(-{K(t)}\right)
\label{phase}\end{equation}is called the phase shift. From Eq.~(\ref{AOC}), if we increase $N_e$ the overlap between the ground states decreases, given raise to the Anderson's orthogonality catastrophe.

\subsection{Sudden-quench limit}
The time scale of the interaction between the x-ray and the inner-shell electrons, and consequently the ejection of deep core electrons, is extremely short. Hence, the appearance of the scattering potential  is usually simulated as a sudden quench.
In this case, 
\begin{equation}
    K(t) = \bar{K}\Theta(t),
\end{equation}
 where $\Theta(t)$ is the Heaviside step function, and we can expand the quantum state $\ket{\Psi(t)}$ in the instantaneous basis for $t > 0$  as
\begin{eqnarray}
\ket{\Psi(t)} = \sum_n \bra{\varphi_n(t)} \ket{\varphi_0(0^-)} \exp\left(-i \frac{E_n(t)t}{\hbar} \right) \ket{\varphi_n(t)}. \nonumber
\end{eqnarray}
 Consequently
\begin{equation}
  |\bra{\varphi_0(t)}\ket{\Psi(t)}|^2 = |\bra{\varphi_0(t)}\ket{\varphi_0(0^-)}|^2.  
  \label{c_0_SQ}
\end{equation}
 Then, from Eqs. (\ref{N_e}), (\ref{AOC}), and (\ref{c_0_SQ})  we  obtain
\begin{equation}\label{AOCII}
    |\bra{\varphi_0(t)}\ket{\Psi(t)}|^2\sim{ N_e ^{-2\left({\frac{\delta}{\pi}}\right)^2}}.
\end{equation}

Following the idea of an adiabatic threshold  (see Section \ref{AT}), we consider the dynamics to be adiabatic when distances between states are a percentage $\eta$ of their maximum, typically $\eta = 10\%$. Hence, using the Eq. \eqref{TR} and $\mathrm{max}[D_T(\ket{\Psi(t)},\ket{\varphi_0(t)})] = 1$, we require $D_T(\ket{\Psi(t)},\ket{\varphi_0(t)}) \le \eta$. Then combining this with Eq. \eqref{AOCII}, we can conclude that, even in the sudden-quench picture, de-facto adiabaticity occurs when
\begin{eqnarray}\label{DA:SQC}
    {|{\bar{K}}|} \leq \frac{{\pi \eta}}{ \sqrt{2\ln\left(
    N_e
    \right) }}.
\end{eqnarray}  The derivation of Eq.~(\ref{DA:SQC})  is in Appendix \ref{DEXP14}. From Eqs.~(\ref{DA:SQC}) and (\ref{N_e}), the smaller the gap, the smaller $\bar{K}$ (and hence the  phase shift) must be to satisfy this condition.

\subsection{Beyond the sudden-quench limit}
 Here, we will consider the finite-time dynamics 
 \begin{equation}\label{RAMPUP_K}
  K(t) = \frac{\bar{K}}{T}t,   
 \end{equation}
 where the applied potential grows linearly up to a maximum value $\bar{K}=K(T)$, on a time scale $T \leq T_{\Delta\varepsilon}$.
 When $T=0$, the sudden quench is recovered. This time-dependency will allow us to study all dynamical regimes, from non-adiabatic to adiabatic.
Therefore, two important time scales are present in this problem: $T_{\Delta\varepsilon}$,  the time window in which the system is observed, and the ramp-up time $T$. These time scales define the number of electronic levels in the tight-binding chain and the velocity at which the scattering potential grows, respectively.

Our numerical solution is based on the discretization of the metallic band, followed by implementation of the Real-Space Numerical Renormalization Group (eNRG) method to compute the time evolution. Although it yields accurate results, the computational cost can be relatively high, as finding the wave function for a large number of sites is a challenging task. To address this, we also derive an analytical solution for the problem: a simple approximate solution in the regime in which $K(t)$ is a continuous function of $t$, and only single-particle hole excitations are crucial for describing the time evolution. This is computationally much cheaper but still produces accurate results in a large range of parameters.

\subsection{Method for the Numerical Solution}

\begin{figure}[hbt!]
    \centering
    \includegraphics[scale=0.32]{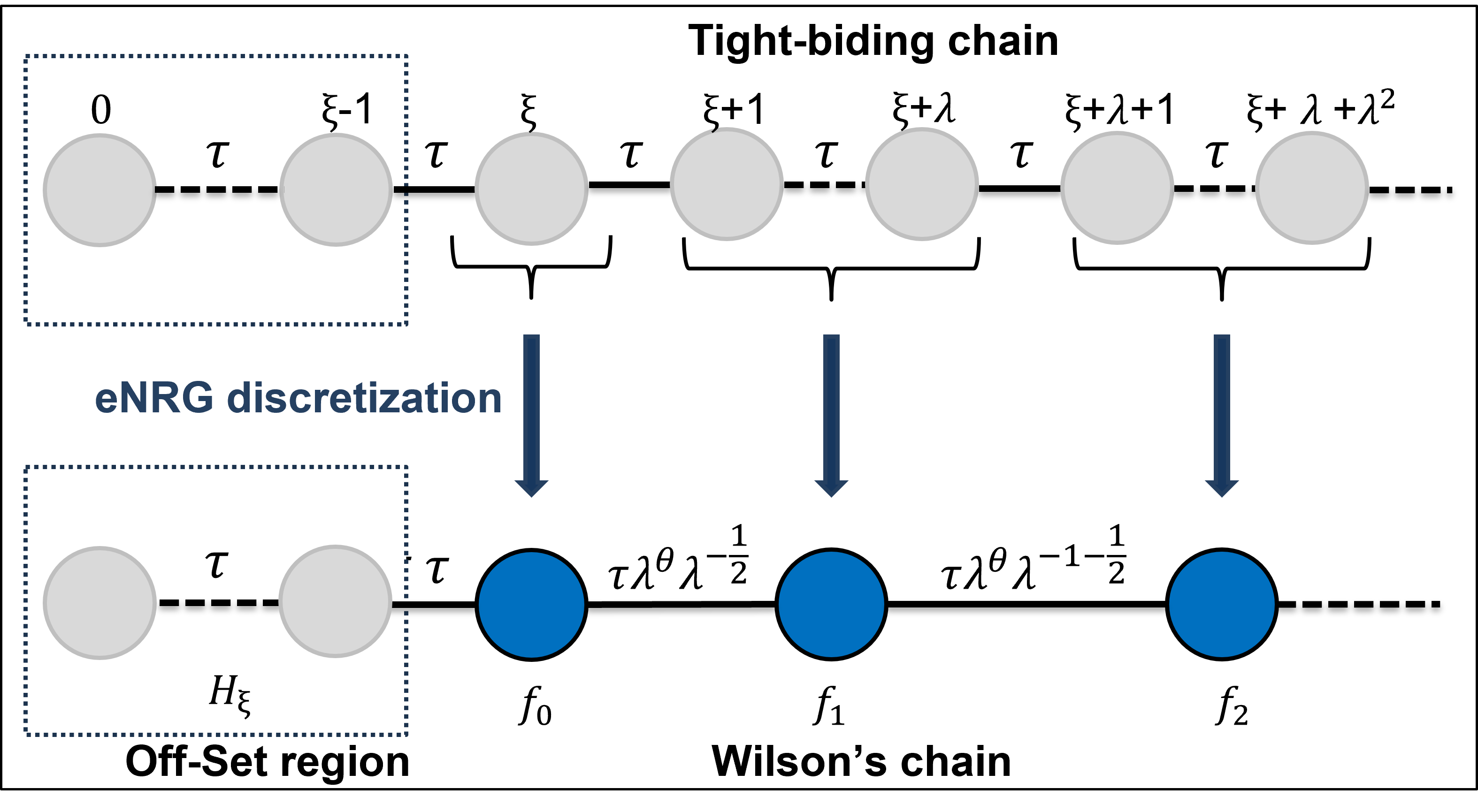}
    \caption{The eNRG procedure initiates with a 1D tight-binding chain (gray circles). Then, it involves grouping $\lambda^n$ sites $(n=0,1,2...)$, where $\lambda>1$, into a single operator $f_n$ (blue circles), and the new hopping parameter becomes $\tau_n=\tau\lambda^{\theta-n-1/2}$. A second parameter $\xi$ (offset) is introduced, where the initial $\xi$ sites are treated individually.}
    \label{fig:eNRG}
\end{figure}

To solve the dynamics generated by Eqs.~\eqref{H_t} and \eqref{RAMPUP_K} numerically, we have implemented the eNRG method \cite{PhysRevB.106.075129}. This organizes the tight-binding sites into powers of the dimensionless discretization parameter $\lambda > 1$, grouping together $\lambda^n$ tight-binding site operators into a single operator $f_n$, except for the first $\xi \ge 0$ (offset) sites (see Fig. \ref{fig:eNRG}). After this procedure, the system is described by the following Hamiltonian
\begin{equation}\label{H}
\begin{aligned}
\mathcal{H}_\mathcal{N}(t) = &\sum_{n=0}^{\xi-1} \tau \left( a_n^{\dagger}a_{n+1}+\rm{H.c.}\right) 
+ \sum_{n=0}^{\mathcal N-1}\tau_n\left( f_n^{\dagger}f_{n+1}+\rm{H.c.}\right)\\
    &+ \tau K(t)a_0^{\dagger}a_0.
\end{aligned}
\end{equation}
This new tight-binding model is written in terms of the eNRG basis such that $\tau_n=\tau\lambda^{\theta-n-1/2}$ is the new hopping parameter, which decreases exponentially with $n$, $\theta$ can assume any value in the range $-1 \le \theta \le 1$, and $\mathcal N$ is the Wilson chain size \cite{RevModPhys.80.395}. Here $f_n^{\dagger}$ ($f_n$) represents the creation (annihilation) operator at the $n$-th eNRG site, with the condition $f_0^\dagger = a_\xi^\dagger$. After this discretization, the smallest energy gap is $\Delta\varepsilon \approx 2\tau\lambda^{-\mathcal N+1/2}$.

The eNRG procedure shares similarities with the traditional Numerical Renormalization Group (NRG) \cite{RevModPhys.47.773}, but it directly uses the real-space tight-binding sites to build \eqref{H}. The exponential grouping of these sites results in a logarithmic discretization in the energy spectra, similar to the NRG \cite{PhysRevB.106.075129}. Recently, a smoothing procedure has been developed for the eNRG method \cite{Picoli} designed to eliminate non-physical oscillations in the time-dependent results arising from the discretization. In essence, this procedure involves integrating $\theta$ over a uniform distribution $\theta\in[-1,1]$, analogous to $z$-averaging \cite{PhysRevB.49.11986}, and over two successive $\xi$. Here we used the smoothing procedure with  $\xi=6$ and $7$ and $100$ values of $\theta$.

~

To compute the time evolution of the system for $t<T_{\Delta\varepsilon}$ under the effect of the scattering potential, we have employed the Crank-Nicolson method \cite{crank_nicolson_1947}. We begin with $\ket{\Psi(0)}=\ket{\varphi_0(0)}$ and use this method to calculate $\ket{\Psi(\Delta t)}$. The time step must be chosen with the condition $ \left(\frac{\bar{K}}{T} \Delta t\right)^2 \ll 1$ to ensure accurate results. Then, we diagonalize the eNRG model to find the eigenstates and eigenenergies. Combined with the smoothing procedure, we compute the wave function and the metrics. The application of this method progressively serves as the time evolution operator, allowing us to determine $\ket{\Psi(t)}$.

\subsection{Proposed Analytical Method}

Eq. \eqref{H_t} is equivalent to a Fermi-gas Hamiltonian with a localized scattering potential and it can be diagonalized analytically \cite{PhysRevLett.18.1049, PhysRevB.71.045326} (see Appendix \ref{Annex_Analytical_Diagonalization} for more details). Using this analytical diagonalization, it can be shown that for continuous $K(t)$, only many-body states differing by a single particle-hole excitation are directly coupled to each other (details in Appendix \ref{DA}).

After the transformation
\begin{equation}\label{transformation}
    \tilde{c}_{n}(t)= {c}_{n}(t)e^{\frac{i}{\hbar}\int_0^t dt' E_{n}(t')},
\end{equation}
 Eq. \eqref{Wave_function_expantion} becomes 
\begin{equation}
    \ket{\Psi(t) } = \sum_n \tilde{c}_n(t) e^{-\frac{i}{\hbar}\int_0^t dt' E_{n}(t')} \ket{\varphi_n(t)},
\end{equation}
and appendix \ref{DA} shows that the evolution of the coefficients $\tilde c_n(t)$ for each many-body state $\ket{\varphi_n(t)}$  is described by
\begin{eqnarray}\label{SEFM}
\frac{d\tilde{c}_n}{dt} & = & -\frac{1}{\pi}\frac{d\delta}{dt}\sum_{p,h\neq p} \frac{\Delta\varepsilon}{\varepsilon_p - \varepsilon_h} \tilde{c}_{n,p,h}(t) e^{-i(\varepsilon_p - \varepsilon_h)\frac{t}{\hbar}}
\\
& = & \frac{1}{\pi} \frac{1}{1+[K(t)]^2}\frac{dK(t)}{dt}\times\nonumber \\
&& \sum_{p,h\neq p}  \frac{1}{p - h} \tilde{c}_{n,p,h}(t) e^{-i\Delta\varepsilon(p - h)\frac{t}{\hbar}}. \label{SEFM2} 
\end{eqnarray}
Here, $\tilde {c}_{n,p,h}(t)$ is the coefficient of $\ket{\varphi_{n,p,h}(t)}$ $=$ $ g_p^\dagger (t) g_h (t)\ket{\varphi_n (t)}$ and $g_k^\dagger$(t) is the creation operator of single-particle eigenstates of $\mathcal{H}(t)$. In deriving Eq. (\ref{SEFM2}) we have used Eq. (\ref{phase}) and that $\varepsilon_k = 2\tau \sin\left(\frac{\pi}{N} k\right) \approx  k\Delta\varepsilon$,
more details in Appendix \ref{DA}. Note that the transformation in Eq. \eqref{transformation} preserves the modulus, i.e., $ |c_n(t)|^2 = |\tilde{c}_n(t)|^2 $.

At $t=0 $, the system is in the ground state. From Eq. \eqref{SEFM2}, large $d K(t)/dt$ values increase the probability of finding the system in the single-particle-hole excited states. The single-particle-hole excited states, in turn, feed the two-particle-hole excitations. This diffusion-like process rapidly reduces the probability to find the system in the ground state, which amounts to non-adiabatic behavior. In addition, the smaller the gap $\Delta\varepsilon$, the larger is the creation of particle-hole pairs, decreasing the occupation probability of the ground state.

If $ K(t) $ changes slowly enough, so that $ \left(\frac{dK}{dt} \right)^2 \ll 1 $, the probability of finding the system in a state with two or more particle-hole excitations is very low, meaning that only single-particle-hole excitations states will be significant. 
In this picture, we can express the set of differential equations in Eq. \eqref{SEFM2} as,
\begin{equation}\label{EDM}
\frac{d\tilde{c}_0(t)}{dt} =  \frac{\frac{1}{\pi} \frac{d K(t)}{dt}}{1+[K(t)]^2}  \sum_{p=0}^{{N_e}} \sum_{h=1}^{N_e} \frac{e^{-i{\Delta\varepsilon(p+h)}\frac{t}{\hbar} }}{p+h}\tilde{c}_{0,p,h}(t),
\end{equation}
with $\tilde c_0(t)$ is the coefficient for the instantaneous ground state, and
\begin{equation}
\frac{d\tilde{c}_{0,p,h}(t)}{dt} 
\approx \frac{\frac{1}{\pi}\frac{dK(t)}{dt}}{1+[K(t)]^2} \frac{e^{i\Delta\varepsilon(p + h)\frac{t}{\hbar} }}{p + h} \tilde{c}_{0}(t).
\end{equation}
with $\tilde c_{0,p,h}(t)$ is the one for $\ket{\varphi_{p,h} (t)} = g_p(t)^\dagger g_h(t) \ket{\varphi_0(t)}$.

\section{Numerical and Analytical Results}

In this section, we present our numerical results. Here, all time-related parameters and variables are expressed in units of $  T_\tau$, and the energy-related parameters and variables are expressed in units of ${\tau}$. 
For clarity, we have divided this section into three parts: in (A) we compare the analytical solution (AS) with the numerical solution obtained from the eNRG method; in (B) we discuss adiabaticity for x-ray photoemission in metals; and in (C) we compare the Bures distance and the local density distance and verify if both give similar information.

\subsection{Analytical solution versus eNRG numerical results}

Here, we compare the results for the time evolution of the model discussed in Section \ref{XPP} using both the analytical solution in \eqref{EDM} and the eNRG method.  To compare methods, we use the eNRG gap $\Delta\varepsilon \approx 2\tau\lambda^{-\mathcal N+1/2}$ as input for the analytical calculations. The other parameters are kept the same for both methods.

\begin{figure}[ht!]
    \centering
    \includegraphics[scale=0.535]{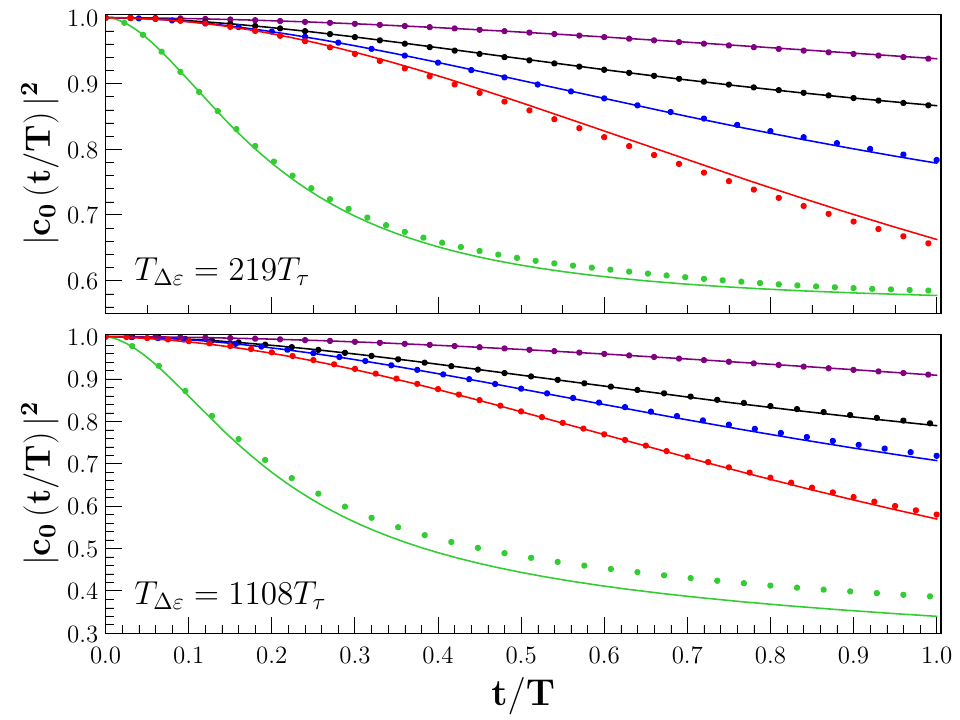}
    \caption{ $|c_0(t/T)|^2 =  |\tilde c_0(t/T)|^2 $  computed with  Eq. \eqref{EDM} (solid lines) and the eNRG (circular dots) method for different combinations of $\bar{K}$ and $T$: purple ($\bar{K}=-0.5$; $T=50 T_\tau$),  black ($\bar{K}=-1.0$; $T=100 T_\tau$), blue ($\bar{K}=-1.0$; $T=20 T_\tau$), green ($\bar{K}=-5.0$; $T=100 T_\tau$) and red ($\bar{K}=-1.0$; $T=1 T_\tau$). {The top panel uses $\lambda=1.5$ and $\mathcal N=15$ ($T_{\Delta\varepsilon}= 219 T_\tau$), while the bottom panel uses $\lambda=1.5$ and $\mathcal N=19$ ($T_{\Delta\varepsilon}=1108 T_\tau$).}}
    \label{fig:RPT}
\end{figure}

As the wave function metrics only depend on $|c_0(t)|^2 =|\bra{\Psi(t)}\ket{\varphi_0(t)}|^2$,  this quantity over time will be the main focus of the discussions. In Fig. \ref{fig:RPT}, we show $|c_0(t)|^2$ over time using Eq. \eqref{EDM} (solid lines) and eNRG (circular dots) for different combinations of $\bar{K}$ and $T$, with $\lambda = 1.5$ and $\mathcal N=15$ (upper panel) and $\mathcal N= 19$ (lower panel). These results show that combinations of $(\bar{K},T)$ with the same ratio $R=\bar{K}/T$ exhibit very different behaviors. For example, each of the pair of lines (black, purple) and (blue, green) corresponds to the same ratio $R$, but, as the magnitude of $|\bar{K}|$ increases, the system becomes less adiabatic, that is $|c_0(t)|^2$ decreases more rapidly, and quite dramatically so for the largest value of $|\bar{K}|$ here considered (green line). In contrast to the usual expectation, we find that the amplitude of $\bar{K}$ is more critical, for tracking adiabaticity, then the ratio $R$. However, with a fixed $\bar{K}$, a longer time scale $T$ leads indeed to a more adiabatic dynamic.

We can also observe, by comparing the upper ($T_{\Delta\varepsilon}= 219 T_\tau$) and bottom ($T_{\Delta\varepsilon}=1108 T_\tau$) panels for the same pair of parameters $(T;\bar{K})$ and time $t/T$, that a smaller gap (longer $T_{\Delta\varepsilon}$ ) is correlated with  smaller values of $|c_0(t)|^2$ (note the different $y-$axis scale in the two panels). This result implies that indeed the energy gap $\Delta\varepsilon$ is very important for the dynamics, since the smaller the gap, the easier it is to create excitations, and less adiabatic is the dynamics.

Fig. \ref{fig:RPT} shows that results from AS and eNRG methods align not just near adiabaticity, when $|c_0(t/T)|^2 \geq 0.90$, but, remarkably, even when the evolution is far from adiabaticity, up to $|c_0(t/T)|^2 \approx 0.6$. The reason for this is that Eq. \eqref{EDM} includes transitions from the ground state to all of the single-particle-hole excited states. When two-particle-hole excited states become significant in the dynamics, they will affect the coefficients $c_0(t)$ and $c_{p,h}(t)$, and AS will become less accurate, as is observed  for the green lines of Fig. \ref{fig:RPT} and $|c_0(t/T)|^2 \lesssim 0.6$. Then, corrections to the AS should be considered if high accuracy is required.

\begin{figure}[hbt!]
    \centering
    \includegraphics[scale=0.54]{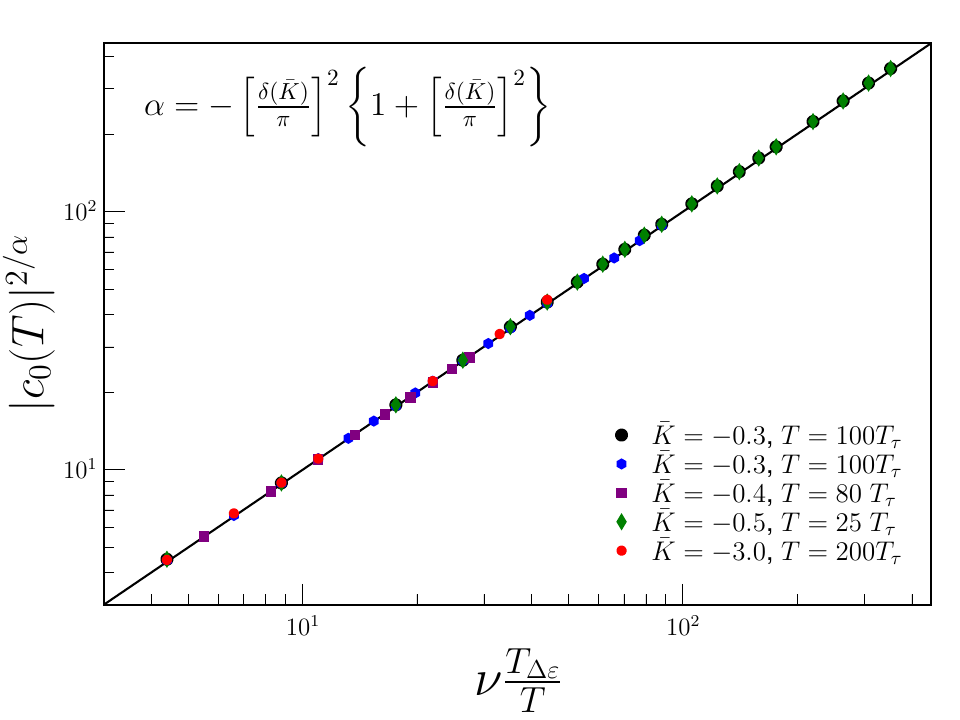}
    \caption{Universal Behavior of $|c_0(T)|^2$ computed using the Eq. \eqref{EDM} for various combinations of $\bar{K}$, $T$, and $\Delta\varepsilon$ (see text). {Here, $\nu = 4.4$: it was chosen to fit the numerical results.}}
    \label{fig:RPTxN}
\end{figure}

The AS method is computationally cheaper and gives very good results for a large range of parameters. Hence, we used it to compute $|c_0(t=T)|^2$  for different combinations of $\bar{K}$, $T$, and $\Delta\varepsilon$. These parameters were varied within the ranges $-3 \leq \bar{K}<0 $, $10 T_\tau \le T \le 200 T_\tau$ and $ 10^{2} T_\tau \le T_{\Delta\varepsilon} \le 10^{4} T_\tau$. The results are shown in Fig. \ref{fig:RPTxN}, where they are plotted with rescaled axis $ y = |c_0(T)|^{2/\alpha} $ against $ x =  \nu\frac{T_{\Delta\varepsilon }}{T}  $, with $ \alpha = -\left[\frac{{\delta}(\bar{K})}{\pi}\right]^2 \left\{ 1 + \left[\frac{{\delta}(\bar{K})}{\pi}\right]^2 \right\} $ and $\nu = 4.4$.\footnote{The parameter $\nu$ helps accounting for the number of electrons participating in the dynamics, see discussion below.} They demonstrate the universal behaviour  
\begin{equation}\label{AS}
\begin{aligned}
|c_0(T)|^2 = \left( \nu \frac{T_{\Delta\varepsilon }}{T} \right)^{- \left[\frac{1}{\pi} {\delta}(\bar{K})\right]^2\left\{ 1 + \left[\frac{1}{\pi} {\delta}(\bar{K}) \right]^2 \right\}}.
\end{aligned}
\end{equation}
We derive this universal behavior to the leading order  $|c_0(T)|^2 \propto \left(\frac{T_{\Delta\varepsilon }}{T} \right)^{- \left[\frac{\delta(\bar K)}{\pi}\right]^2} $ in Appendix \ref{|C_0|}, for $T \gg T_\tau$.

The expected sudden-quench limit for Eq. \ref{AS} is given in Eq. \eqref{AOCII}. Indeed, as $T\to 0$,  $\hbar/T $ becomes comparable to the energy $N_e \tau$, needed to excite all particles to single-particle states above the Fermi level. Substituting $T = T_\tau/ N_e $  into the right-hand side of Eq. \eqref{AS} and using Eq.~(\ref{N_e}) yield
\begin{align}
    \left|c_0\left(T = \frac{T_\tau}{ N_e} \right)\right|^2 &\approx&  &\left({\nu}{\pi}\left(\frac{T_{\Delta\varepsilon }}{T_\tau}\right)^2 \right)^{- \left[\frac{1}{\pi} {\delta}(\bar{K})\right]^2\left\{ 1 + \left[\frac{1}{\pi} {\delta}(\bar{K}) \right]^2 \right\}} \nonumber \\ 
    &=& &\left(\sqrt{\frac{\nu}{\pi}} N_e\right)^{-2\left[\frac{1}{\pi} {\delta}(\bar{K})\right]^2\left\{ 1 + \left[\frac{1}{\pi} {\delta}(\bar{K}) \right]^2 \right\}},
\end{align}
recovering to first order in $\left(\frac{\delta}{\pi}\right)^2$ the expression for the sudden quench case in Eq. \eqref{AOCII}. The correction is due to $T>0$ in the derivation of AS.

One interesting property of Eq. \eqref{AS} is that it depends only on the maximum scattering potential $ \bar{K} $ and the rescaled time $ \frac{T}{T_{\Delta\varepsilon}} $. This means that for a fixed $ \bar{K} $ and constant $ \frac{T}{T_{\Delta\varepsilon}} $, the value of $ |c_0(T)| $ remains unchanged. For a fixed $ \bar{K} $, a larger rescaled time $ \frac{T}{T_{\Delta\varepsilon}} $ results in a more adiabatic behavior, and vice versa. 
If $ T = T_{\Delta\varepsilon}  $, only very low-energy excitations are significant. On the other hand, if $ T = \frac{T_\tau}{N_e } $, close to the sudden quench case, excitations across all energy levels will appear.
Indeed, the  energy scale $ \frac{\hbar}{T} $ acts as a cutoff for possible excitations; hence, the fraction of electrons participating in the dynamics is, approximately, $ \frac{T_ {\Delta\varepsilon}}{T} $, with Eq. \eqref{AS} providing the more precise estimate of $ \nu\frac{T_ {\Delta\varepsilon}}{T} $. 
We can now identify two regimes depending on the value of $ T $. If $ T \leq \frac{T_\tau}{ N_e} $, the system behaves as in a sudden quench, where all possible electronic transitions are influenced by the scattering potential, resulting in a strongly non-adiabatic response. On the other hand, if $ \frac{T_\tau}{ N_e} < T \leq T_{\Delta\varepsilon} $, only a fraction of the possible particle-hole excitations is affected by the increasing potential. As $T$ increases, this fraction decreases, leading to fewer particle-hole excitations and a more adiabatic evolution of the system.

\subsection{Tracking Adiabaticity}
Here we aim to track the dynamical regime, and in particular to check if the system is adiabatic, by using the trace distance Eq. \eqref{TR}. Following the concept of adiabatic threshold \cite{AdvQuantumTechnol.3.1900139}, we define a system to be adiabatic when $D_T(\ket{\Psi(t)},\ket{\varphi_0(t)}) \le \eta$, with $\eta$ a suitably small number and $\mathrm{max}(D_T(\ket{\Psi(t)},\ket{\varphi_0(t)})) = 1$. Except where otherwise stated, in our calculations we use as adiabatic threshold the value $\eta = 0.1$, which corresponds to $|c_0(t)|^2 \ge 0.99$.

As observed, near adiabaticity Eq. \eqref{AS} works well. Then, we can use it for small $\eta$, and we write the adiabaticity condition for $t=T$ as
\begin{eqnarray}\label{DA:RUC}
 {|\bar{K}|}\leq {\pi \eta}\left[\ln\left(\nu\frac{T_{\Delta\varepsilon }}{T} \right)\right]^{-1/2}.
\end{eqnarray}

To derive the expression \eqref{DA:RUC} we used Eqs. \eqref{phase}, \eqref{TR} and \eqref{AS} (see Appendix \ref{ADEXP} for details). It is important to mention that Eq. (\ref{DA:RUC}) is only valid for $T>0$; for $T=0$ (sudden quench) one should use Eq. \eqref{DA:SQC}.

\begin{figure}[hbt!]
    \centering
    \includegraphics[scale=0.535]{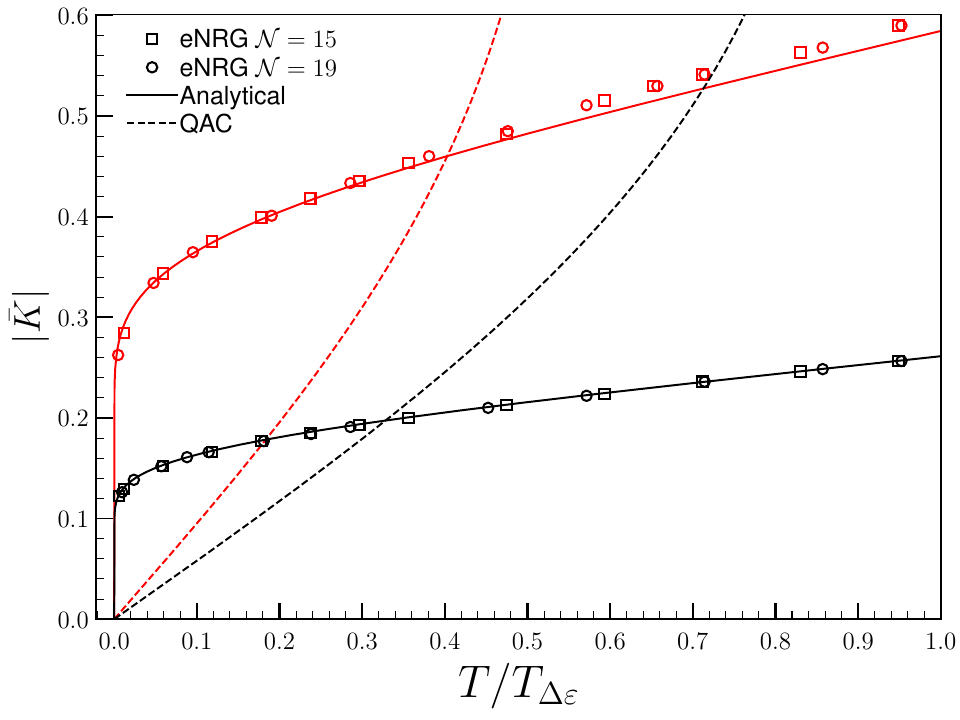}
    \caption{Adiabatic phase diagram for $\mathcal N=19$, $\lambda=1.5$ and $T_{\Delta\varepsilon} = 1108 T_\tau$ (black and red squares), and $\mathcal N=15$ , $\lambda=1.5$ and $T_{\Delta\varepsilon} = 219 T_\tau$ (black and red circles). The  $(\bar{K},T/T_{\Delta\varepsilon})$ curves represent the adiabaticity condition for the trace distance at $t=T$ for adiabatic thresholds $\eta = 0.1$ (black curve) and $\eta = 0.2236$ (red curve). The corresponding QAC results are represented by the dashed black and red lines.}
    \label{fig:Diagram}
\end{figure}

\subsubsection{Adiabatic phase diagram}
In Fig. \ref{fig:Diagram} the region in $\{(\bar{K}, T/T_{\Delta\varepsilon})\}$ below the black solid line satisfies  Eq. \eqref{DA:RUC} with $\eta = 0.1$, corresponding to the condition 
$D_T(\ket{\Psi(T)},\ket{\varphi_0(T)}) \le \eta = 0.1$. 
For this problem  and the range of parameters explored $|c_0(t)|^2$ decreases monotonically (see Fig. \ref{fig:RPT}), hence $D_T(\ket{\Psi(t)},\ket{\varphi_0(t)}) \le D_T(\ket{\Psi(T)},\ket{\varphi_0(T)})$. This implies that dynamics corresponding to the parameter region below the solid black curve will satisfy the adiabatic condition at all times.
Conversely, dynamics corresponding to the parameter region above will always contain time windows of non-adiabatic behavior (according to the chosen threshold), and in particular it will be non-adiabatic at $t=T$. We will refer to this representation as the adiabatic phase diagram.

 \subsubsection{Comparison between the analytic solution and eNRG numerical results at $t=T$}
 In Fig. \ref{fig:Diagram} we compare predictions from Eq. \eqref{DA:RUC} with $\eta = 0.1$ (black solid line) with two corresponding sets of eNRG numerical results: $\lambda=1.5$ and $\mathcal N=15$, consistent with a minimum gap of $\Delta\varepsilon = {1}/{219} ~\tau$ (black circles), and $\mathcal N=19$, $\lambda=1.5$ ($\Delta\varepsilon=1/1108 ~\tau$, black squares). They are in excellent agreement, confirming that the functional dependence of Eqs.\eqref{AS} and \eqref{DA:RUC} on the parameters $\bar K$, $T$, and $T_{\Delta\varepsilon}$ is correct.

Further, we consider $\eta = 0.2236$ (red solid line, corresponding to $|c_0|^2 \ge 0.95$). For this higher threshold, the agreement between the two sets of eNRG results demonstrates that a functional dependence on ${T}/{T_{\Delta\varepsilon}}$ persists, albeit with small corrections with respect to the dependence shown in Eq. \eqref{DA:RUC}. In fact, as ${T}/{T_{\Delta\varepsilon}}$ increases, the agreement with the analytical results remains very good, but not excellent.

 Comparison with our numerical eNRG results suggests that Eq. \eqref{DA:RUC} accurately provides the adiabatic regime for any combinations of $\bar{K}$, $T$ and $T_{\Delta\varepsilon}$. This is reinforced by the universal shape of the curves in Fig. \ref{fig:Diagram}. Then, we expect that this expression holds even for small values of $\Delta\varepsilon$, approaching the gapless spectrum.
 
 \subsubsection{Comparison with the Quantum Adiabatic Criterion for $t\le T$}
 To calculate  the QAC corresponding to an adiabatic threshold $\eta$, we approximate Eq. \eqref{QAC} by using only the first excited state $\ket{\varphi_1(T)}$, with $|E_1(T) - E_0(T)| = \Delta\varepsilon$ (single particle-hole excitation with smaller possible energy) and $\dot H = \tau \frac{\bar{K}}{T} a_0^\dagger a_0$ from  Eq \eqref{H}. The result is 

 \begin{equation}
    |\bar{K}| \frac{T_{\Delta\varepsilon}}{T} \frac{\tau |\bra{\varphi_0(t)} a_0^\dagger a_0 \ket{\varphi_1(t)}|}{\Delta\varepsilon }  \leq \eta . \label{QACtr1}
 \end{equation}
 It can be shown, see Appendix \ref{Matrix:QAC} that 
 \begin{equation} \label{matrix01}
   |\bra{\varphi_0(t)} a_0^\dagger a_0 \ket{\varphi_1(t)}| \approx  \frac{\cos^2 (\delta(t))  }{2\pi}  \frac{T_\tau}{T_{\Delta\varepsilon}}.   
 \end{equation}
 From Eqs. (\ref{phase}) and (\ref{matrix01}) it can be seen that, in this case, the QAC Eq. (\ref{QACtr1}) is better satisfied as time increases. This is in contrast with the dynamics of the system, where the higher eigenstates' population grows with time, as demonstrated by the increase with time of the trace distance. 

At $t=T$ the QAC adiabatic condition 
corresponds
 to the areas below the dashed curves in Fig. \ref{fig:Diagram}. These are calculated using Eqs.~ (\ref{QACtr1}) and (\ref{matrix01}) and display a quasi-linear behaviour up to intermediate values of  $T/T_{\Delta\varepsilon}$: in fact,
 from Eqs. (\ref{phase}) and (\ref{matrix01}), $|\bra{\varphi_0(T)} a_0^\dagger a_0 \ket{\varphi_1(T)}| {\approx} \frac{1  }{2\pi}  \frac{T_\tau}{T_{\Delta\varepsilon}}$ up to second order in $|\bar K|$.
This functional dependence  is  very different from the one of Eq. \eqref{DA:RUC}, and, from Fig. \ref{fig:Diagram}, clearly inconsistent with eNRG results (compare symbols with dashed lines). Also, by construction, the QAC does not account for memory, which is relevant at finite values of $T$. As a result, for small $\frac{T}{T_{\Delta\varepsilon}}$, the QAC strongly underestimates the parametric region corresponding to adiabaticity.
On the other hand, for  $\frac{T}{T_{\Delta\varepsilon}}\gtrsim 0.3$, it estimates as adiabatic larger and larger parameter sets leading to non-adiabatic dynamics. Clearly, the QAC fails to track adiabaticity for this problem.


\subsubsection{Gapless band limit}
We aim to discuss the gapless spectrum, but for practical purposes, our calculations are performed using a finite gap $\Delta \varepsilon$. Hence, according to the time-energy uncertainty principle, our calculations can accurately describe metallic behavior if $ t \le T_{\Delta\varepsilon}$. Alternatively, if in an experiment the measurement time for observing metallic behavior is chosen to be $T_{\Delta\varepsilon}$, the size of the maximum fictitious energy gap that could be allowed in a faithful simulation of the experiment would be $ \Delta \varepsilon = {\hbar}/{T_{\Delta\varepsilon}}$.

Eq. \eqref{DA:RUC} is derived assuming a finite minimum energy gap $ \Delta \varepsilon$, so it is valid only for $ T \le T_{\Delta\varepsilon}$. For $\Delta\varepsilon\to 0$ and any $T$ finite, the adiabaticity condition \eqref{DA:RUC} is not satisfied as the dynamics would produce excitations with energy larger than the infinitesimal gap. For any finite $\Delta\varepsilon$ and $ T = T_{\Delta\varepsilon}$, Eq. \eqref{DA:RUC} reduces to 
\begin{eqnarray}\label{DA:RUC3}
 {|\bar{K}|}\leq \frac{{\pi\eta}}{\sqrt{\ln\left(\nu\right)}}.
\end{eqnarray}
This is a condition on the applied external potential: given the finite gap $\Delta\varepsilon$ and the slowest dynamics compatible with {\it not} discriminating this gap, the applied potential has to be smaller than a certain threshold in order to fulfill adiabaticity.

\subsection{Comparing predictions from quantum state and local density metrics}

In Refs. \cite{AdvQuantumTechnol.3.1900139,PhysRevA.98.012104} it was shown that the NLD distance could witness (non) adiabatic behaviour. Here we will check if this method is robust enough to faithfully describe, qualitatively and quantitatively,  the challeging dynamics of x-ray photoemission. From Eqs.~(\ref{TR})~and (\ref{BR}), Bure and trace distances contain the same information, so, following Refs. \cite{AdvQuantumTechnol.3.1900139,PhysRevA.98.012104},  we will now concentrate on the Bures and NLD distances.

 In Fig, \ref{fig:Results1} we compare the numerical results obtained from eNRG calculations for  $D_B(t) \equiv D_B (\ket{\Psi(t)},\ket{\varphi_0(t)}) $ and the local density distance $D_n(t) \equiv D_n(n(t),n^A(t))$. 

It can be seen that overall both $D_B(t)$ and $D_n(t)$ convey similar information. In cases involving faster evolution (smaller $T$) and/or stronger potentials (greater $\bar{K}$), the values of these metrics generally increase, suggesting that excitations become important. In contrast, for weak and slowly varying potentials, the values remain small, indicating that, when $\ket{\Psi(t)}$ is closer to $\ket{\varphi_0(t)}$,   $n(t)$ remains closer to $n^A(t)$.

Throughout all our numerical calculations, we observe that $D_n(t) \le D_B(t) $, compare, e.g., the panels in the first and second columns of Fig. \ref{fig:Results1}. From Eq. \eqref{TR} and \eqref{BR} the inequality $D_B(t) \leq \sqrt{2} D_T(t)$ holds. The above implies that the NLD distance is indeed bounded by $2D_T(t)$, satisfying Eq. \eqref{NTR}.

\subsubsection{Adiabatic threshold for the natural local density metric}

\begin{figure*}[ht!]
    \centering
    \begin{tabular}{lll}
        \includegraphics[scale=0.362]{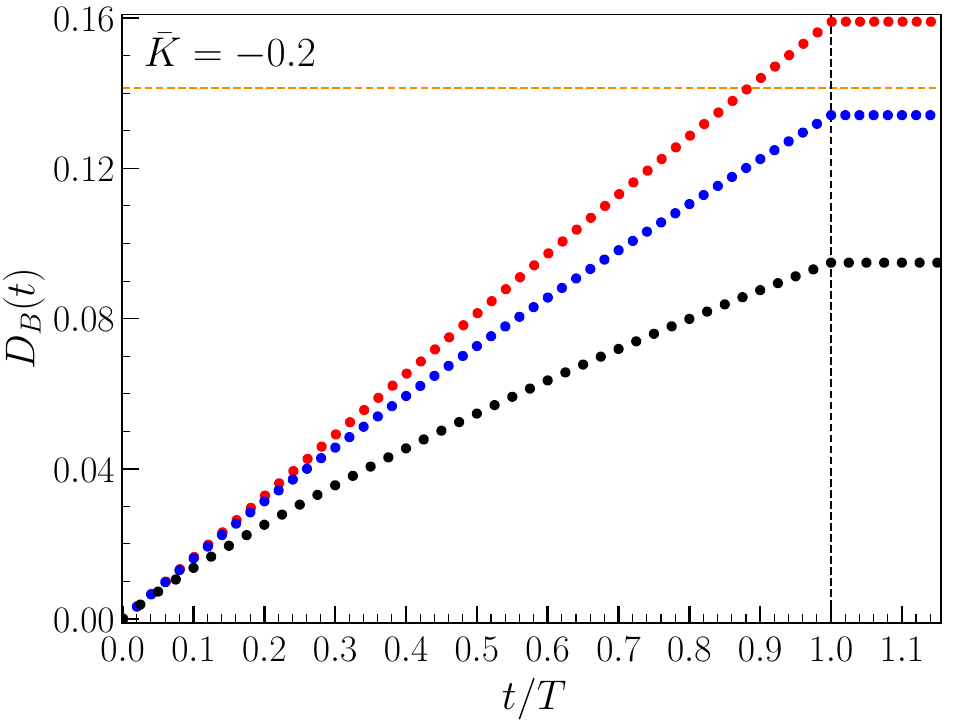}
        &
        \includegraphics[scale=0.362]{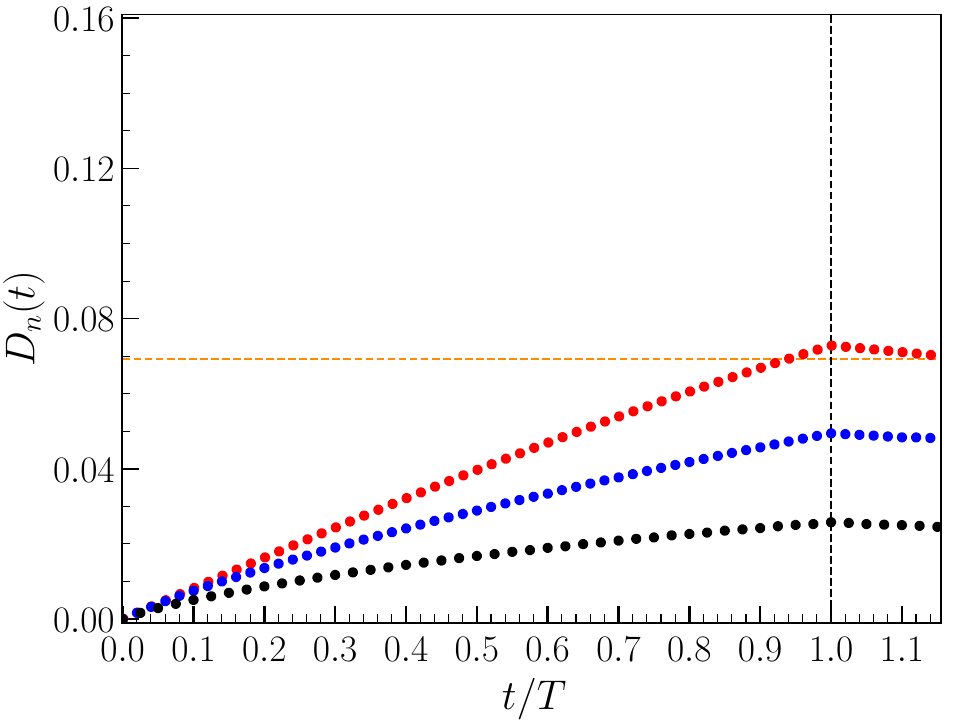}
        &
        \includegraphics[scale=0.362]{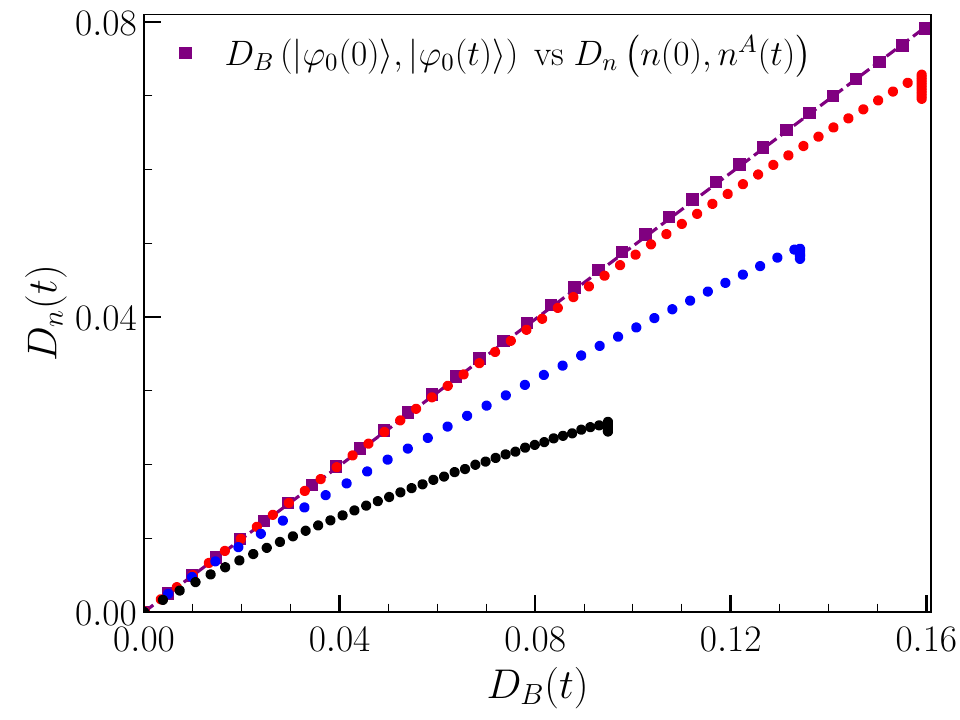}
    \end{tabular}
    \begin{tabular}{lll}
        \includegraphics[scale=0.362]{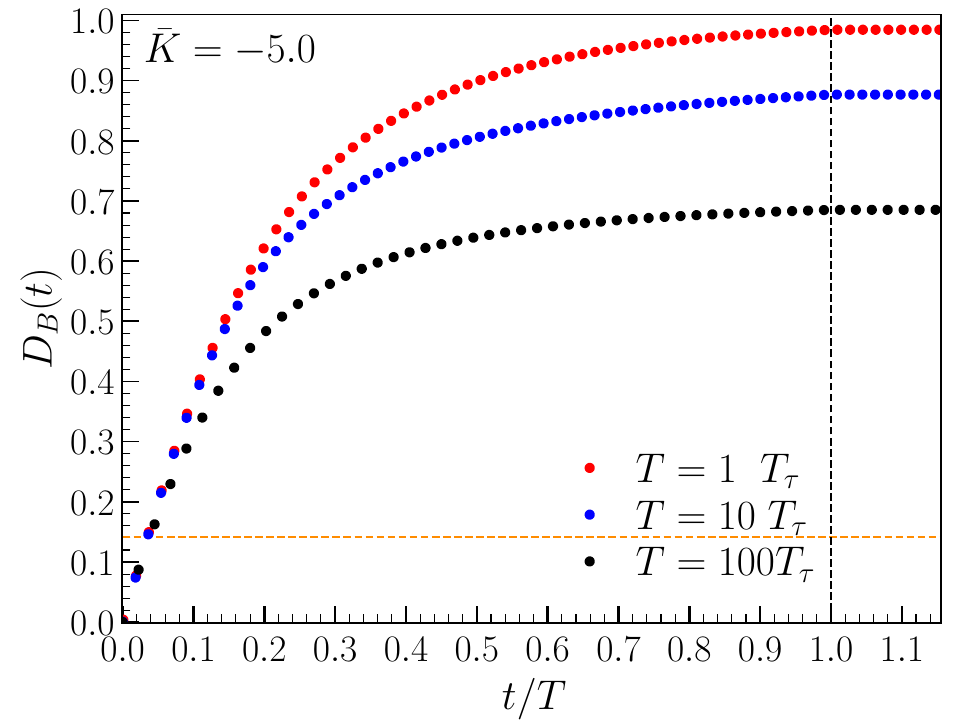}
        &
        \includegraphics[scale=0.362]{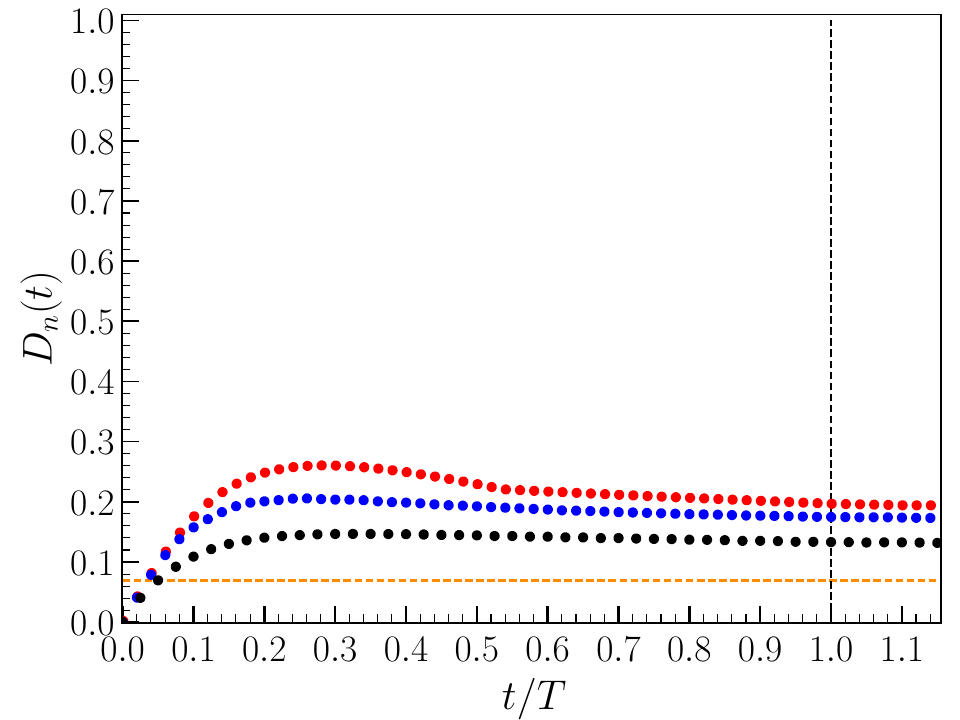}
        &
        \includegraphics[scale=0.362]{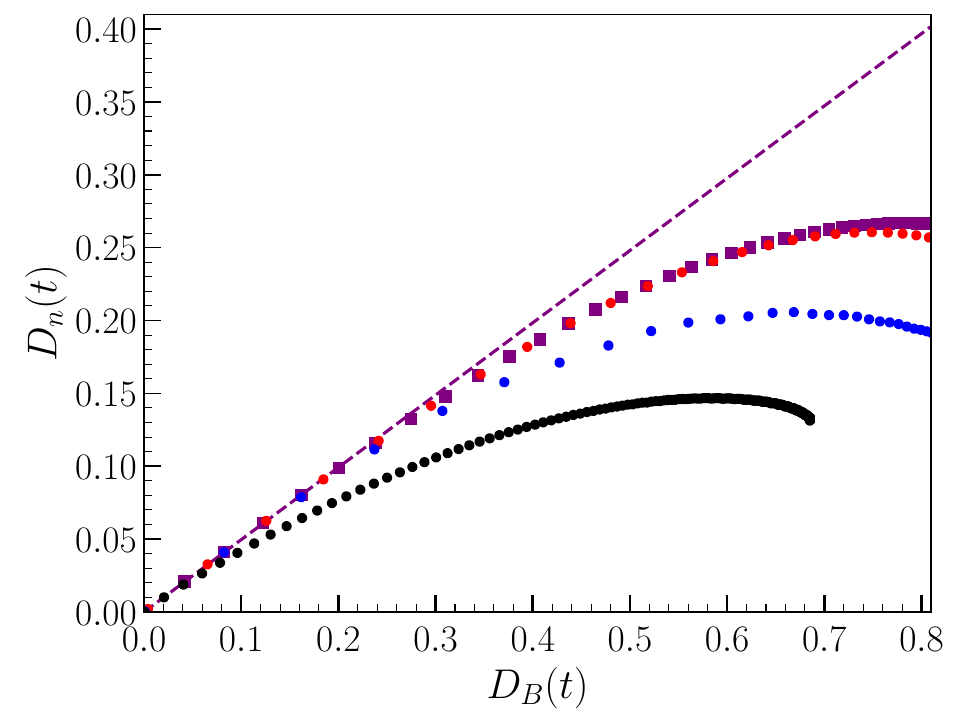}
    \end{tabular}
    \caption{Time evolution of $D_B(t)$ (first column) and $D_n(t)$ (second column)  computed by the eNRG method, for $\bar{K} = -0.2$ (Top panels), and for $\bar{K} = -5.0$ (Bottom panels) for different time scales $T$. For this plot we use $\lambda=1.5$ and $\mathcal N=15$, corresponding to $T_{\Delta\varepsilon} = {219} T_\tau$. The vertical dashed line indicates the point where the amplitude of the scattering potential reaches its maximum. The horizontal dashed orange line represents the adiabatic threshold. The square purple dots in the third column show the Bures distance $D_B(\ket{\varphi_0(0)},\ket{\varphi_0(t)})$ between the instantaneous ground state and the initial ground state versus the local density distance $D_n(n(0),n^A(t))$. It is shown that, up to medium Bures distances, $D_n(n(0),n^A(t)) \approx m D_B(\ket{\varphi_0(0)},\ket{\varphi_0(t)})$ with $m = 0.50 $, value found following the procedure described in Ref. \cite{AdvQuantumTechnol.3.1900139}. The purple dashed line shows $y = m \cdot x$. The third column shows also that, for small values of $D_B(t)$, $D_n(t)$ versus $D_B(t)$ behaves linearly, and according to the same  $m = 0.50 $ linear coefficient.}
    \label{fig:Results1}
\end{figure*}

For a quantitative analysis, we calculate the adiabatic threshold for the NLD metric, following the method proposed in   \cite{AdvQuantumTechnol.3.1900139}.
In detail, we plot $ D_B(\ket{\varphi_0(0)},\ket{\varphi_0(t)}) $, i.e., the Bures distance  between the initial ground state and the instantaneous ground state,  against the corresponding NLD distance $ D_n(n(0),n^A(t)) $ for $ \bar{K} = -0.2 $ and $ \bar{K} = -5.0 $ (purple squares in the third column of Fig.  \ref{fig:Results1}).
We then apply a linear fit to the region where $ D_B(\ket{\varphi_0(0)},\ket{\varphi_0(t)}) \le 0.16 $ and find, in both cases, the angular coefficient $ m = 0.50 $\footnote{As the applied potential is increased linearly, the distances related to $K=-2$ maximum potential are part of the $K=-5$-related distances' set}, see purple dashed line in the third column of Fig.  \ref{fig:Results1}. The adiabatic threshold for the NLD metric is then estimated as  $m\eta\cdot\mathrm{max}(D_B) =0.050\sqrt{2}$ \cite{AdvQuantumTechnol.3.1900139}, which is plotted as a yellow dashed line in the second-column panels of  Fig.  \ref{fig:Results1}. 

In agreement with \cite{AdvQuantumTechnol.3.1900139}, our results show that this estimate for the NLD metric adiabatic threshold is an upper bound, as visually demonstrated by all $D_n(t)$ versus  $D_B(t)$ curves in the third column lying being below the dashed purple lines. However, this bound is close enough to provide quantitative agreement between the regions of adiabatic dynamics estimated using  $D_B(t)$ or $D_n(t)$.

Indeed, results in the first and second column of Fig. \ref{fig:Results1} concur that for $T=100T_\tau$ (black dots) and  $T=10T_\tau$ (blue dots) the dynamics are always adiabatic; however,  they both show that for $T=T_\tau$ and $ \bar{K} = -5.0 $ (red dots, lower panels) adiabaticity is maintain only up to $t/T \approx 0.04$, while for $T=T_\tau$ and $ \bar{K} = -2.0 $ (red dots, upper panels) the dynamics is adiabatic up to $t/T\approx 0.9$.

\subsubsection{Out-of-equilibrium dynamical response}

 For $t>T$, we observe that $D_n(t)$ decreases, while $D_B(t)$  remains constant. In this region, the Hamiltonian does not change anymore, leading to a constant $|c_0(t>T)|^2$ and hence, from Eq.~\eqref{BR}, to a constant $D_B(t)$.

\begin{figure}[ht!]
		\centering
        \includegraphics[scale=0.53]{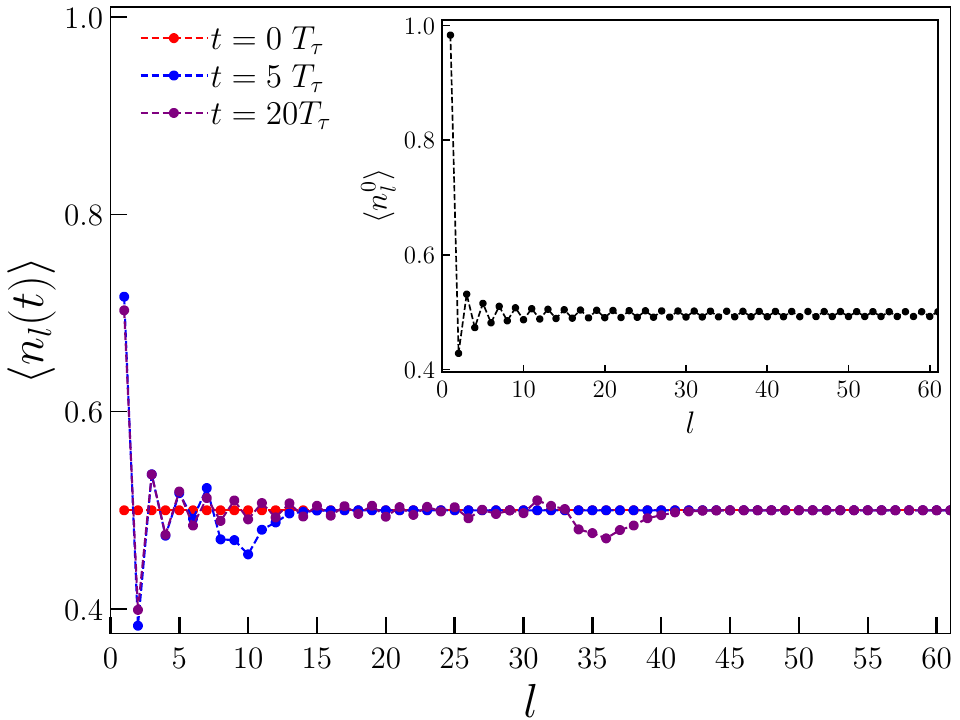}
		\caption{Local density dynamics in response to a sudden quench from $ K=0$ to $ K=-5$ as a function of the site number $l$. At $t=0$ (red dots) the local density configuration is $\langle n_l \rangle = 0.5$ for all sites. At $t=5 T_\tau$ the crest of the density wave arrives at site 10 (blue dots), and at $t=20 T_\tau$ it reaches site 35 (purple dots). Here, we consider a tight-binding chain with $N=120$. Inset: local density of the ground state for $ K=-5$ and $N=120$. }
	\label{Occupation}
\end{figure}

 To understand the decrease of $D_n(t)$  for $t > T$, we can refer to Fig. \ref{Occupation}, where the local density configuration of a tight-binding chain after a sudden quench with $\bar K = -5$ is plotted (main panel). At $t=0^+$, the system is in the $K=0$ ground state, with $\langle n_l (0^+)\rangle = 1/2$ (red dots, half-filling). However, for $t>0$, this state is (strongly) out-of-equilibrium with respect to the instantaneous ground state, where the attractive localized scattering potential $\bar{K}<0$ makes electrons preferentially occupy the first site (black dots, inset).
 This localized scattering potential triggers a dynamical density wave, which propagates along the chain (main panel, blue and purple dots). By reorganizing the occupation of each site closer to the instantaneous ground state configuration, this wave tends to decrease the local density distance. This behavior is associated with Friedel oscillations \cite{friedel1952,Picoli}.

We note that, for finite-time dynamics and $0<t<T$, this equilibrating effect competes with the response to the continuous change in the local density of the instantaneous ground state triggered by \eqref{RAMPUP_K}, which supports an increase in $D_n(t)$. As a consequence, depending on the rate of change of $K(t)$ and the response strength allowed by the hopping parameter $\tau$, $D_n(t)$ can increase or decrease, as seen in the central column of Fig. \ref{fig:Results1}. 

Our results show that, near adiabaticity, the NLD distance contains the same information as the wave function metrics, which is related to $|c_0(t)|^2$. However, if the system substantially deviates from the adiabatic regime, the NLD distance maintains at least some of the local information contained in the $\{|c_{j\ne0}(t)|^2\}$ set \footnote{Through their dependence on $|c_0(t)|$, trace and Bures distance both contain information about the global property $\sum_{j\ne0}|c_{j}(t)|^2 = 1-|c_0(t)|^2$.  }, where this set describes the out-of-equilibrium dynamics that the system state and the corresponding local density undertake in the effort to adjust to the instantaneous equilibrium state.

\section{Summary and Conclusions}

In this paper, we analyzed the dynamics of the Fermi gas subject to a time-dependent localized scattering potential as a severe test for the use of quantum state and local density metrics to measure adiabaticity. 

In this context, we discuss the dependence of the quantum-state metrics on the instantaneous ground-state occupation $|c_0(t)|$, and demonstrate a related upper bound for the local-density metrics. We derive an analytical method to calculate the key quantity $|c_0(t)|$, which discards excitations beyond single particle-holes. Its accuracy is tested against results from the recently introduced Real-Space NRG method. Our results show that for $|c_0(t)|\lesssim 0.6$ the analytical results are virtually-exact to very accurate, offering a fast and reliable method which is valid well-beyond the adiabatic regime. 

Further, we consider $|c_0(T)|$, the instantaneous ground-state occupation at the end of the ramp-up of the applied potential, and show that it displays a universal behavior, determined by the interplay between the maximum applied potential, the minimum energy gap, and the fraction of electrons excited by the process. Using the concept of adiabatic threshold\cite{AdvQuantumTechnol.3.1900139}, this behaviour allows us to construct an adiabatic phase diagram in terms of these parameters and to compare results with predictions from the most-commonly used form of the quantum adiabatic criterion. The latter is shown to fail in predicting adiabatic behaviour, both qualitatively and quantitatively. For this problem, according to the quantum adiabatic criterion, the adiabatic condition should depend on $\bar{K}\cdot T_\tau/T$ for small $|\bar K|$. Instead, we find that it depends on the more complex expression Eq. \eqref{DA:RUC}, which is not a function of the combination $\bar{K}\cdot T_\tau/T$ even for small $|\bar K|$. This discordance becomes more pronounced as $|\bar K|$ increase.

In the last part of the paper, we compare information contained within the Bures and the local density distances. Previous findings suggested that the local density metric can be used to investigate adiabaticity and  our results demonstrate that this is the case even for the tough problem of x-ray photoemission, where the Anderson's catastrophe occurs. 

Moreover, we find that the local-density distance not only correctly predicts and tracks adiabaticity, but also contains information on out-of-equilibrium dynamics that may occur in the absence of applied external fields. In the case at hand, after the application of the external potential, the system attempts to equilibrate via Friedel oscillations and redistribution of charge. This effect is captured by the local density distance between dynamical and equilibrium states, which, accordingly, decreases with time. In contrast, the Bures and trace distances cannot capture this dynamics, as they depend only on the projection on the ground state, which is a constant of motion for a closed system with a time-independent Hamiltonian. 

This suggests that the local density distance -- typically cheaper and faster to accurately compute than the Bures or trace distances -- is not only a trustworthy way to 
assess adiabaticity, but also a probe for general out-of-equilibrium dynamics which can be more sensitive than the related quantum state distances.  

~

\textbf{Acknowledgments} - The authors thank Gabriel dos Reis Trindade for fruitful discussions. This research received financial support from FAPESP (grant 2022/05198-2), the CNPq (grant 311689/2023-0) and the Coordenação de Aperfeiçoamento de Pessoal de Nível Superior – Brasil (CAPES) – Finance Code 001. G. D. acknowledges financial support from the Coordenação de Aperfeiçoamento de Pessoal de Nível Superior (CAPES - 88887.495890/2020-00). F. D. P. acknowledges financial support from FAPESP (grants 2022/09312-4 and 2024/05637-1).  I.D. thanks the Instituto de Fısica de S\~ao Carlos, University  of S\~ao Paulo, Brazil, for the kind hospitality. We gratefully acknowledge the support and access to  computational resources offered by the Center for Mathematical Sciences Applied to Industry (CeMEAI), funded by FAPESP (grant 2013/07375-0).

\nocite{*}
\bibliography{main} 

\begin{thebibliography}{76}%
\makeatletter
\providecommand \@ifxundefined [1]{%
 \@ifx{#1\undefined}
}%
\providecommand \@ifnum [1]{%
 \ifnum #1\expandafter \@firstoftwo
 \else \expandafter \@secondoftwo
 \fi
}%
\providecommand \@ifx [1]{%
 \ifx #1\expandafter \@firstoftwo
 \else \expandafter \@secondoftwo
 \fi
}%
\providecommand \natexlab [1]{#1}%
\providecommand \enquote  [1]{``#1''}%
\providecommand \bibnamefont  [1]{#1}%
\providecommand \bibfnamefont [1]{#1}%
\providecommand \citenamefont [1]{#1}%
\providecommand \href@noop [0]{\@secondoftwo}%
\providecommand \href [0]{\begingroup \@sanitize@url \@href}%
\providecommand \@href[1]{\@@startlink{#1}\@@href}%
\providecommand \@@href[1]{\endgroup#1\@@endlink}%
\providecommand \@sanitize@url [0]{\catcode `\\12\catcode `\$12\catcode
  `\&12\catcode `\#12\catcode `\^12\catcode `\_12\catcode `\%12\relax}%
\providecommand \@@startlink[1]{}%
\providecommand \@@endlink[0]{}%
\providecommand \url  [0]{\begingroup\@sanitize@url \@url }%
\providecommand \@url [1]{\endgroup\@href {#1}{\urlprefix }}%
\providecommand \urlprefix  [0]{URL }%
\providecommand \Eprint [0]{\href }%
\providecommand \doibase [0]{https://doi.org/}%
\providecommand \selectlanguage [0]{\@gobble}%
\providecommand \bibinfo  [0]{\@secondoftwo}%
\providecommand \bibfield  [0]{\@secondoftwo}%
\providecommand \translation [1]{[#1]}%
\providecommand \BibitemOpen [0]{}%
\providecommand \bibitemStop [0]{}%
\providecommand \bibitemNoStop [0]{.\EOS\space}%
\providecommand \EOS [0]{\spacefactor3000\relax}%
\providecommand \BibitemShut  [1]{\csname bibitem#1\endcsname}%
\let\auto@bib@innerbib\@empty
\bibitem [{\citenamefont {Farhi}\ \emph {et~al.}(2002)\citenamefont {Farhi},
  \citenamefont {Goldstone},\ and\ \citenamefont {Gutmann}}]{farhi2002quantum}%
  \BibitemOpen
  \bibfield  {author} {\bibinfo {author} {\bibfnamefont {E.}~\bibnamefont
  {Farhi}}, \bibinfo {author} {\bibfnamefont {J.}~\bibnamefont {Goldstone}},\
  and\ \bibinfo {author} {\bibfnamefont {S.}~\bibnamefont {Gutmann}},\
  }\href@noop {} {\bibinfo {title} {Quantum adiabatic evolution algorithms with
  different paths}} (\bibinfo {year} {2002}),\ \Eprint
  {https://arxiv.org/abs/quant-ph/0208135} {arXiv:quant-ph/0208135 [quant-ph]}
  \BibitemShut {NoStop}%
\bibitem [{\citenamefont {Albash}\ and\ \citenamefont
  {Lidar}(2018)}]{RevModPhys.90.015002}%
  \BibitemOpen
  \bibfield  {author} {\bibinfo {author} {\bibfnamefont {T.}~\bibnamefont
  {Albash}}\ and\ \bibinfo {author} {\bibfnamefont {D.~A.}\ \bibnamefont
  {Lidar}},\ }\href {https://doi.org/10.1103/RevModPhys.90.015002} {\bibfield
  {journal} {\bibinfo  {journal} {Rev. Mod. Phys.}\ }\textbf {\bibinfo {volume}
  {90}},\ \bibinfo {pages} {015002} (\bibinfo {year} {2018})}\BibitemShut
  {NoStop}%
\bibitem [{\citenamefont {Fiusa}\ \emph {et~al.}(2023)\citenamefont {Fiusa},
  \citenamefont {Soares-Pinto},\ and\ \citenamefont
  {Pires}}]{PhysRevA.107.032422}%
  \BibitemOpen
  \bibfield  {author} {\bibinfo {author} {\bibfnamefont {G.}~\bibnamefont
  {Fiusa}}, \bibinfo {author} {\bibfnamefont {D.~O.}\ \bibnamefont
  {Soares-Pinto}},\ and\ \bibinfo {author} {\bibfnamefont {D.~P.}\ \bibnamefont
  {Pires}},\ }\href {https://doi.org/10.1103/PhysRevA.107.032422} {\bibfield
  {journal} {\bibinfo  {journal} {Phys. Rev. A}\ }\textbf {\bibinfo {volume}
  {107}},\ \bibinfo {pages} {032422} (\bibinfo {year} {2023})}\BibitemShut
  {NoStop}%
\bibitem [{\citenamefont {Gu\'ery-Odelin}\ \emph {et~al.}(2019)\citenamefont
  {Gu\'ery-Odelin}, \citenamefont {Ruschhaupt}, \citenamefont {Kiely},
  \citenamefont {Torrontegui}, \citenamefont {Mart\'{\i}nez-Garaot},\ and\
  \citenamefont {Muga}}]{RevModPhys.91.045001}%
  \BibitemOpen
  \bibfield  {author} {\bibinfo {author} {\bibfnamefont {D.}~\bibnamefont
  {Gu\'ery-Odelin}}, \bibinfo {author} {\bibfnamefont {A.}~\bibnamefont
  {Ruschhaupt}}, \bibinfo {author} {\bibfnamefont {A.}~\bibnamefont {Kiely}},
  \bibinfo {author} {\bibfnamefont {E.}~\bibnamefont {Torrontegui}}, \bibinfo
  {author} {\bibfnamefont {S.}~\bibnamefont {Mart\'{\i}nez-Garaot}},\ and\
  \bibinfo {author} {\bibfnamefont {J.~G.}\ \bibnamefont {Muga}},\ }\href
  {https://doi.org/10.1103/RevModPhys.91.045001} {\bibfield  {journal}
  {\bibinfo  {journal} {Rev. Mod. Phys.}\ }\textbf {\bibinfo {volume} {91}},\
  \bibinfo {pages} {045001} (\bibinfo {year} {2019})}\BibitemShut {NoStop}%
\bibitem [{\citenamefont {Torrontegui}\ \emph {et~al.}(2013)\citenamefont
  {Torrontegui}, \citenamefont {Ibáñez}, \citenamefont {Martínez-Garaot},
  \citenamefont {Modugno}, \citenamefont {{del Campo}}, \citenamefont
  {Guéry-Odelin}, \citenamefont {Ruschhaupt}, \citenamefont {Chen},\ and\
  \citenamefont {Muga}}]{TORRONTEGUI2013117}%
  \BibitemOpen
  \bibfield  {author} {\bibinfo {author} {\bibfnamefont {E.}~\bibnamefont
  {Torrontegui}}, \bibinfo {author} {\bibfnamefont {S.}~\bibnamefont
  {Ibáñez}}, \bibinfo {author} {\bibfnamefont {S.}~\bibnamefont
  {Martínez-Garaot}}, \bibinfo {author} {\bibfnamefont {M.}~\bibnamefont
  {Modugno}}, \bibinfo {author} {\bibfnamefont {A.}~\bibnamefont {{del
  Campo}}}, \bibinfo {author} {\bibfnamefont {D.}~\bibnamefont
  {Guéry-Odelin}}, \bibinfo {author} {\bibfnamefont {A.}~\bibnamefont
  {Ruschhaupt}}, \bibinfo {author} {\bibfnamefont {X.}~\bibnamefont {Chen}},\
  and\ \bibinfo {author} {\bibfnamefont {J.~G.}\ \bibnamefont {Muga}},\ }in\
  \href {https://doi.org/https://doi.org/10.1016/B978-0-12-408090-4.00002-5}
  {\emph {\bibinfo {booktitle} {Advances in Atomic, Molecular, and Optical
  Physics}}},\ Vol.~\bibinfo {volume} {62},\ \bibinfo {editor} {edited by\
  \bibinfo {editor} {\bibfnamefont {E.}~\bibnamefont {Arimondo}}, \bibinfo
  {editor} {\bibfnamefont {P.~R.}\ \bibnamefont {Berman}},\ and\ \bibinfo
  {editor} {\bibfnamefont {C.~C.}\ \bibnamefont {Lin}}}\ (\bibinfo  {publisher}
  {Academic Press},\ \bibinfo {year} {2013})\ pp.\ \bibinfo {pages}
  {117--169}\BibitemShut {NoStop}%
\bibitem [{\citenamefont {Grace}\ \emph {et~al.}(2010)\citenamefont {Grace},
  \citenamefont {Dominy}, \citenamefont {Kosut}, \citenamefont {Brif},\ and\
  \citenamefont {Rabitz}}]{Grace_2010}%
  \BibitemOpen
  \bibfield  {author} {\bibinfo {author} {\bibfnamefont {M.~D.}\ \bibnamefont
  {Grace}}, \bibinfo {author} {\bibfnamefont {J.}~\bibnamefont {Dominy}},
  \bibinfo {author} {\bibfnamefont {R.~L.}\ \bibnamefont {Kosut}}, \bibinfo
  {author} {\bibfnamefont {C.}~\bibnamefont {Brif}},\ and\ \bibinfo {author}
  {\bibfnamefont {H.}~\bibnamefont {Rabitz}},\ }\href
  {https://doi.org/10.1088/1367-2630/12/1/015001} {\bibfield  {journal}
  {\bibinfo  {journal} {New Journal of Physics}\ }\textbf {\bibinfo {volume}
  {12}},\ \bibinfo {pages} {015001} (\bibinfo {year} {2010})}\BibitemShut
  {NoStop}%
\bibitem [{\citenamefont {Su}\ \emph {et~al.}(2018)\citenamefont {Su},
  \citenamefont {Chen}, \citenamefont {Ma}, \citenamefont {Chen},\ and\
  \citenamefont {Sun}}]{Su_2018}%
  \BibitemOpen
  \bibfield  {author} {\bibinfo {author} {\bibfnamefont {S.}~\bibnamefont
  {Su}}, \bibinfo {author} {\bibfnamefont {J.}~\bibnamefont {Chen}}, \bibinfo
  {author} {\bibfnamefont {Y.}~\bibnamefont {Ma}}, \bibinfo {author}
  {\bibfnamefont {J.}~\bibnamefont {Chen}},\ and\ \bibinfo {author}
  {\bibfnamefont {C.}~\bibnamefont {Sun}},\ }\href
  {https://doi.org/10.1088/1674-1056/27/6/060502} {\bibfield  {journal}
  {\bibinfo  {journal} {Chinese Physics B}\ }\textbf {\bibinfo {volume} {27}},\
  \bibinfo {pages} {060502} (\bibinfo {year} {2018})}\BibitemShut {NoStop}%
\bibitem [{\citenamefont {Quan}(2009)}]{PhysRevE.79.041129}%
  \BibitemOpen
  \bibfield  {author} {\bibinfo {author} {\bibfnamefont {H.~T.}\ \bibnamefont
  {Quan}},\ }\href {https://doi.org/10.1103/PhysRevE.79.041129} {\bibfield
  {journal} {\bibinfo  {journal} {Phys. Rev. E}\ }\textbf {\bibinfo {volume}
  {79}},\ \bibinfo {pages} {041129} (\bibinfo {year} {2009})}\BibitemShut
  {NoStop}%
\bibitem [{\citenamefont {Deffner}\ and\ \citenamefont
  {Lutz}(2013)}]{PhysRevE.87.022143}%
  \BibitemOpen
  \bibfield  {author} {\bibinfo {author} {\bibfnamefont {S.}~\bibnamefont
  {Deffner}}\ and\ \bibinfo {author} {\bibfnamefont {E.}~\bibnamefont {Lutz}},\
  }\href {https://doi.org/10.1103/PhysRevE.87.022143} {\bibfield  {journal}
  {\bibinfo  {journal} {Phys. Rev. E}\ }\textbf {\bibinfo {volume} {87}},\
  \bibinfo {pages} {022143} (\bibinfo {year} {2013})}\BibitemShut {NoStop}%
\bibitem [{\citenamefont {Skelt}\ \emph {et~al.}(2019)\citenamefont {Skelt},
  \citenamefont {Zawadzki},\ and\ \citenamefont {D’Amico}}]{Skelt_2019}%
  \BibitemOpen
  \bibfield  {author} {\bibinfo {author} {\bibfnamefont {A.~H.}\ \bibnamefont
  {Skelt}}, \bibinfo {author} {\bibfnamefont {K.}~\bibnamefont {Zawadzki}},\
  and\ \bibinfo {author} {\bibfnamefont {I.}~\bibnamefont {D’Amico}},\ }\href
  {https://doi.org/10.1088/1751-8121/ab4fb6} {\bibfield  {journal} {\bibinfo
  {journal} {Journal of Physics A: Mathematical and Theoretical}\ }\textbf
  {\bibinfo {volume} {52}},\ \bibinfo {pages} {485304} (\bibinfo {year}
  {2019})}\BibitemShut {NoStop}%
\bibitem [{\citenamefont {Funo}\ \emph {et~al.}(2017)\citenamefont {Funo},
  \citenamefont {Zhang}, \citenamefont {Chatou}, \citenamefont {Kim},
  \citenamefont {Ueda},\ and\ \citenamefont {del
  Campo}}]{PhysRevLett.118.100602}%
  \BibitemOpen
  \bibfield  {author} {\bibinfo {author} {\bibfnamefont {K.}~\bibnamefont
  {Funo}}, \bibinfo {author} {\bibfnamefont {J.-N.}\ \bibnamefont {Zhang}},
  \bibinfo {author} {\bibfnamefont {C.}~\bibnamefont {Chatou}}, \bibinfo
  {author} {\bibfnamefont {K.}~\bibnamefont {Kim}}, \bibinfo {author}
  {\bibfnamefont {M.}~\bibnamefont {Ueda}},\ and\ \bibinfo {author}
  {\bibfnamefont {A.}~\bibnamefont {del Campo}},\ }\href
  {https://doi.org/10.1103/PhysRevLett.118.100602} {\bibfield  {journal}
  {\bibinfo  {journal} {Phys. Rev. Lett.}\ }\textbf {\bibinfo {volume} {118}},\
  \bibinfo {pages} {100602} (\bibinfo {year} {2017})}\BibitemShut {NoStop}%
\bibitem [{\citenamefont {Venuti}\ \emph {et~al.}(2016)\citenamefont {Venuti},
  \citenamefont {Albash}, \citenamefont {Lidar},\ and\ \citenamefont
  {Zanardi}}]{PhysRevA.93.032118}%
  \BibitemOpen
  \bibfield  {author} {\bibinfo {author} {\bibfnamefont {L.~C.}\ \bibnamefont
  {Venuti}}, \bibinfo {author} {\bibfnamefont {T.}~\bibnamefont {Albash}},
  \bibinfo {author} {\bibfnamefont {D.~A.}\ \bibnamefont {Lidar}},\ and\
  \bibinfo {author} {\bibfnamefont {P.}~\bibnamefont {Zanardi}},\ }\href
  {https://doi.org/10.1103/PhysRevA.93.032118} {\bibfield  {journal} {\bibinfo
  {journal} {Phys. Rev. A}\ }\textbf {\bibinfo {volume} {93}},\ \bibinfo
  {pages} {032118} (\bibinfo {year} {2016})}\BibitemShut {NoStop}%
\bibitem [{\citenamefont {Patan\`e}\ \emph {et~al.}(2008)\citenamefont
  {Patan\`e}, \citenamefont {Silva}, \citenamefont {Amico}, \citenamefont
  {Fazio},\ and\ \citenamefont {Santoro}}]{PhysRevLett.101.175701}%
  \BibitemOpen
  \bibfield  {author} {\bibinfo {author} {\bibfnamefont {D.}~\bibnamefont
  {Patan\`e}}, \bibinfo {author} {\bibfnamefont {A.}~\bibnamefont {Silva}},
  \bibinfo {author} {\bibfnamefont {L.}~\bibnamefont {Amico}}, \bibinfo
  {author} {\bibfnamefont {R.}~\bibnamefont {Fazio}},\ and\ \bibinfo {author}
  {\bibfnamefont {G.~E.}\ \bibnamefont {Santoro}},\ }\href
  {https://doi.org/10.1103/PhysRevLett.101.175701} {\bibfield  {journal}
  {\bibinfo  {journal} {Phys. Rev. Lett.}\ }\textbf {\bibinfo {volume} {101}},\
  \bibinfo {pages} {175701} (\bibinfo {year} {2008})}\BibitemShut {NoStop}%
\bibitem [{\citenamefont {Eisert~J.}(2015)}]{Eisert2015}%
  \BibitemOpen
  \bibfield  {author} {\bibinfo {author} {\bibfnamefont {G.~C.}\ \bibnamefont
  {Eisert~J.}, \bibfnamefont {Friesdorf~M.}},\ }\href
  {https://doi.org/10.1038/nphys3215} {\bibfield  {journal} {\bibinfo
  {journal} {Phys. Rev. Lett.}\ }\textbf {\bibinfo {volume} {11}},\ \bibinfo
  {pages} {124} (\bibinfo {year} {2015})}\BibitemShut {NoStop}%
\bibitem [{\citenamefont {Bl\"ochl}\ and\ \citenamefont
  {Parrinello}(1992)}]{PhysRevB.45.9413}%
  \BibitemOpen
  \bibfield  {author} {\bibinfo {author} {\bibfnamefont {P.~E.}\ \bibnamefont
  {Bl\"ochl}}\ and\ \bibinfo {author} {\bibfnamefont {M.}~\bibnamefont
  {Parrinello}},\ }\href {https://doi.org/10.1103/PhysRevB.45.9413} {\bibfield
  {journal} {\bibinfo  {journal} {Phys. Rev. B}\ }\textbf {\bibinfo {volume}
  {45}},\ \bibinfo {pages} {9413} (\bibinfo {year} {1992})}\BibitemShut
  {NoStop}%
\bibitem [{\citenamefont {Born}\ and\ \citenamefont {Fock}(1928)}]{Born}%
  \BibitemOpen
  \bibfield  {author} {\bibinfo {author} {\bibfnamefont {M.}~\bibnamefont
  {Born}}\ and\ \bibinfo {author} {\bibfnamefont {V.}~\bibnamefont {Fock}},\
  }\href {https://doi.org/https://doi.org/10.1007/BF01343193} {\bibfield
  {journal} {\bibinfo  {journal} {Reviews of Modern Physics}\ }\textbf
  {\bibinfo {volume} {51}},\ \bibinfo {pages} {165} (\bibinfo {year}
  {1928})}\BibitemShut {NoStop}%
\bibitem [{Note1()}]{Note1}%
  \BibitemOpen
  \bibinfo {note} {See e.g. J. E. Avron and A. Elgart, Communications in
  Mathematical Physics 203, 445 (1999) and references therein.}\BibitemShut
  {Stop}%
\bibitem [{\citenamefont {Kato}(1950)}]{JPSJ.5.435}%
  \BibitemOpen
  \bibfield  {author} {\bibinfo {author} {\bibfnamefont {T.}~\bibnamefont
  {Kato}},\ }\href {https://doi.org/10.1143/JPSJ.5.435} {\bibfield  {journal}
  {\bibinfo  {journal} {Journal of the Physical Society of Japan}\ }\textbf
  {\bibinfo {volume} {5}},\ \bibinfo {pages} {435} (\bibinfo {year}
  {1950})}\BibitemShut {NoStop}%
\bibitem [{\citenamefont {Nenciu}(1980)}]{JPhysAMath.13.L15}%
  \BibitemOpen
  \bibfield  {author} {\bibinfo {author} {\bibfnamefont {G.}~\bibnamefont
  {Nenciu}},\ }\href {https://doi.org/10.1088/0305-4470/13/2/002} {\bibfield
  {journal} {\bibinfo  {journal} {Journal of Physics A: Mathematical and
  General}\ }\textbf {\bibinfo {volume} {13}},\ \bibinfo {pages} {L15}
  (\bibinfo {year} {1980})}\BibitemShut {NoStop}%
\bibitem [{\citenamefont {Avron}\ \emph {et~al.}(1987)\citenamefont {Avron},
  \citenamefont {Seiler},\ and\ \citenamefont
  {Yaffe}}]{CommunMathPhys.110.3349}%
  \BibitemOpen
  \bibfield  {author} {\bibinfo {author} {\bibfnamefont {J.~E.}\ \bibnamefont
  {Avron}}, \bibinfo {author} {\bibfnamefont {R.}~\bibnamefont {Seiler}},\ and\
  \bibinfo {author} {\bibfnamefont {L.~G.}\ \bibnamefont {Yaffe}},\ }\href@noop
  {} {\bibfield  {journal} {\bibinfo  {journal} {Commun. Math. Phys.}\ }\textbf
  {\bibinfo {volume} {110}},\ \bibinfo {pages} {33} (\bibinfo {year}
  {1987})}\BibitemShut {NoStop}%
\bibitem [{\citenamefont {Avron}\ \emph {et~al.}(1993)\citenamefont {Avron},
  \citenamefont {Seiler},\ and\ \citenamefont
  {Yaffe}}]{CommunMathPhys.3.649650}%
  \BibitemOpen
  \bibfield  {author} {\bibinfo {author} {\bibfnamefont {J.~E.}\ \bibnamefont
  {Avron}}, \bibinfo {author} {\bibfnamefont {R.}~\bibnamefont {Seiler}},\ and\
  \bibinfo {author} {\bibfnamefont {L.~G.}\ \bibnamefont {Yaffe}},\ }\href@noop
  {} {\bibfield  {journal} {\bibinfo  {journal} {Commun. Math. Phys.}\ }\textbf
  {\bibinfo {volume} {156}},\ \bibinfo {pages} {649} (\bibinfo {year}
  {1993})}\BibitemShut {NoStop}%
\bibitem [{\citenamefont {Avron}\ and\ \citenamefont
  {Elgart}(1999)}]{CommMathPhys.203.445463}%
  \BibitemOpen
  \bibfield  {author} {\bibinfo {author} {\bibfnamefont {J.~E.}\ \bibnamefont
  {Avron}}\ and\ \bibinfo {author} {\bibfnamefont {A.}~\bibnamefont {Elgart}},\
  }\href {https://doi.org/https://doi.org/10.1007/s002200050620} {\bibfield
  {journal} {\bibinfo  {journal} {Communications in Mathematical Physics}\
  }\textbf {\bibinfo {volume} {203}},\ \bibinfo {pages} {445} (\bibinfo {year}
  {1999})}\BibitemShut {NoStop}%
\bibitem [{\citenamefont {Skelt}\ and\ \citenamefont
  {D'Amico}(2020)}]{AdvQuantumTechnol.3.1900139}%
  \BibitemOpen
  \bibfield  {author} {\bibinfo {author} {\bibfnamefont {A.~H.}\ \bibnamefont
  {Skelt}}\ and\ \bibinfo {author} {\bibfnamefont {I.}~\bibnamefont
  {D'Amico}},\ }\href {https://doi.org/https://doi.org/10.1002/qute.201900139}
  {\bibfield  {journal} {\bibinfo  {journal} {Advanced Quantum Technologies}\
  }\textbf {\bibinfo {volume} {3}},\ \bibinfo {pages} {1900139} (\bibinfo
  {year} {2020})}\BibitemShut {NoStop}%
\bibitem [{\citenamefont {Du}\ \emph {et~al.}(2008)\citenamefont {Du},
  \citenamefont {Hu}, \citenamefont {Wang}, \citenamefont {Wu}, \citenamefont
  {Zhao},\ and\ \citenamefont {Suter}}]{PhysRevLett.101.060403}%
  \BibitemOpen
  \bibfield  {author} {\bibinfo {author} {\bibfnamefont {J.}~\bibnamefont
  {Du}}, \bibinfo {author} {\bibfnamefont {L.}~\bibnamefont {Hu}}, \bibinfo
  {author} {\bibfnamefont {Y.}~\bibnamefont {Wang}}, \bibinfo {author}
  {\bibfnamefont {J.}~\bibnamefont {Wu}}, \bibinfo {author} {\bibfnamefont
  {M.}~\bibnamefont {Zhao}},\ and\ \bibinfo {author} {\bibfnamefont
  {D.}~\bibnamefont {Suter}},\ }\href
  {https://doi.org/10.1103/PhysRevLett.101.060403} {\bibfield  {journal}
  {\bibinfo  {journal} {Phys. Rev. Lett.}\ }\textbf {\bibinfo {volume} {101}},\
  \bibinfo {pages} {060403} (\bibinfo {year} {2008})}\BibitemShut {NoStop}%
\bibitem [{\citenamefont {Tong}\ \emph {et~al.}(2005)\citenamefont {Tong},
  \citenamefont {Singh}, \citenamefont {Kwek},\ and\ \citenamefont
  {Oh}}]{PhysRevLett.95.110407}%
  \BibitemOpen
  \bibfield  {author} {\bibinfo {author} {\bibfnamefont {D.~M.}\ \bibnamefont
  {Tong}}, \bibinfo {author} {\bibfnamefont {K.}~\bibnamefont {Singh}},
  \bibinfo {author} {\bibfnamefont {L.~C.}\ \bibnamefont {Kwek}},\ and\
  \bibinfo {author} {\bibfnamefont {C.~H.}\ \bibnamefont {Oh}},\ }\href
  {https://doi.org/10.1103/PhysRevLett.95.110407} {\bibfield  {journal}
  {\bibinfo  {journal} {Phys. Rev. Lett.}\ }\textbf {\bibinfo {volume} {95}},\
  \bibinfo {pages} {110407} (\bibinfo {year} {2005})}\BibitemShut {NoStop}%
\bibitem [{\citenamefont {Comparat}(2009)}]{PhysRevA.80.012106}%
  \BibitemOpen
  \bibfield  {author} {\bibinfo {author} {\bibfnamefont {D.}~\bibnamefont
  {Comparat}},\ }\href {https://doi.org/10.1103/PhysRevA.80.012106} {\bibfield
  {journal} {\bibinfo  {journal} {Phys. Rev. A}\ }\textbf {\bibinfo {volume}
  {80}},\ \bibinfo {pages} {012106} (\bibinfo {year} {2009})}\BibitemShut
  {NoStop}%
\bibitem [{\citenamefont {Marzlin}\ and\ \citenamefont
  {Sanders}(2004)}]{PhysRevLett.93.160408}%
  \BibitemOpen
  \bibfield  {author} {\bibinfo {author} {\bibfnamefont {K.-P.}\ \bibnamefont
  {Marzlin}}\ and\ \bibinfo {author} {\bibfnamefont {B.~C.}\ \bibnamefont
  {Sanders}},\ }\href {https://doi.org/10.1103/PhysRevLett.93.160408}
  {\bibfield  {journal} {\bibinfo  {journal} {Phys. Rev. Lett.}\ }\textbf
  {\bibinfo {volume} {93}},\ \bibinfo {pages} {160408} (\bibinfo {year}
  {2004})}\BibitemShut {NoStop}%
\bibitem [{\citenamefont {Ortigoso}(2012)}]{PhysRevA.86.032121}%
  \BibitemOpen
  \bibfield  {author} {\bibinfo {author} {\bibfnamefont {J.}~\bibnamefont
  {Ortigoso}},\ }\href {https://doi.org/10.1103/PhysRevA.86.032121} {\bibfield
  {journal} {\bibinfo  {journal} {Phys. Rev. A}\ }\textbf {\bibinfo {volume}
  {86}},\ \bibinfo {pages} {032121} (\bibinfo {year} {2012})}\BibitemShut
  {NoStop}%
\bibitem [{\citenamefont {Amin}(2009)}]{PhysRevLett.102.220401}%
  \BibitemOpen
  \bibfield  {author} {\bibinfo {author} {\bibfnamefont {M.~H.~S.}\
  \bibnamefont {Amin}},\ }\href
  {https://doi.org/10.1103/PhysRevLett.102.220401} {\bibfield  {journal}
  {\bibinfo  {journal} {Phys. Rev. Lett.}\ }\textbf {\bibinfo {volume} {102}},\
  \bibinfo {pages} {220401} (\bibinfo {year} {2009})}\BibitemShut {NoStop}%
\bibitem [{\citenamefont {Jansen}\ \emph {et~al.}(2007)\citenamefont {Jansen},
  \citenamefont {Ruskai},\ and\ \citenamefont {Seiler}}]{JMathPhys.48.102111}%
  \BibitemOpen
  \bibfield  {author} {\bibinfo {author} {\bibfnamefont {S.}~\bibnamefont
  {Jansen}}, \bibinfo {author} {\bibfnamefont {M.-B.}\ \bibnamefont {Ruskai}},\
  and\ \bibinfo {author} {\bibfnamefont {R.}~\bibnamefont {Seiler}},\ }\href
  {https://doi.org/10.1063/1.2798382} {\bibfield  {journal} {\bibinfo
  {journal} {Journal of Mathematical Physics}\ }\textbf {\bibinfo {volume}
  {48}},\ \bibinfo {pages} {102111} (\bibinfo {year} {2007})}\BibitemShut
  {NoStop}%
\bibitem [{\citenamefont {Lychkovskiy}\ \emph {et~al.}(2018)\citenamefont
  {Lychkovskiy}, \citenamefont {Gamayun},\ and\ \citenamefont
  {Cheianov}}]{PhysRevB.98.024307}%
  \BibitemOpen
  \bibfield  {author} {\bibinfo {author} {\bibfnamefont {O.}~\bibnamefont
  {Lychkovskiy}}, \bibinfo {author} {\bibfnamefont {O.}~\bibnamefont
  {Gamayun}},\ and\ \bibinfo {author} {\bibfnamefont {V.}~\bibnamefont
  {Cheianov}},\ }\href {https://doi.org/10.1103/PhysRevB.98.024307} {\bibfield
  {journal} {\bibinfo  {journal} {Phys. Rev. B}\ }\textbf {\bibinfo {volume}
  {98}},\ \bibinfo {pages} {024307} (\bibinfo {year} {2018})}\BibitemShut
  {NoStop}%
\bibitem [{\citenamefont {Il`in}\ \emph {et~al.}(2021)\citenamefont {Il`in},
  \citenamefont {Aristova},\ and\ \citenamefont
  {Lychkovskiy}}]{PhysRevA.104.L030202}%
  \BibitemOpen
  \bibfield  {author} {\bibinfo {author} {\bibfnamefont {N.}~\bibnamefont
  {Il`in}}, \bibinfo {author} {\bibfnamefont {A.}~\bibnamefont {Aristova}},\
  and\ \bibinfo {author} {\bibfnamefont {O.}~\bibnamefont {Lychkovskiy}},\
  }\href {https://doi.org/10.1103/PhysRevA.104.L030202} {\bibfield  {journal}
  {\bibinfo  {journal} {Phys. Rev. A}\ }\textbf {\bibinfo {volume} {104}},\
  \bibinfo {pages} {L030202} (\bibinfo {year} {2021})}\BibitemShut {NoStop}%
\bibitem [{\citenamefont {Skelt}\ \emph
  {et~al.}(2018{\natexlab{a}})\citenamefont {Skelt}, \citenamefont {Godby},\
  and\ \citenamefont {D'Amico}}]{PhysRevA.98.012104}%
  \BibitemOpen
  \bibfield  {author} {\bibinfo {author} {\bibfnamefont {A.~H.}\ \bibnamefont
  {Skelt}}, \bibinfo {author} {\bibfnamefont {R.~W.}\ \bibnamefont {Godby}},\
  and\ \bibinfo {author} {\bibfnamefont {I.}~\bibnamefont {D'Amico}},\ }\href
  {https://doi.org/10.1103/PhysRevA.98.012104} {\bibfield  {journal} {\bibinfo
  {journal} {Phys. Rev. A}\ }\textbf {\bibinfo {volume} {98}},\ \bibinfo
  {pages} {012104} (\bibinfo {year} {2018}{\natexlab{a}})}\BibitemShut
  {NoStop}%
\bibitem [{\citenamefont {Ilyin}(2022)}]{TheorMathPhys.211.545}%
  \BibitemOpen
  \bibfield  {author} {\bibinfo {author} {\bibfnamefont {N.~B.}\ \bibnamefont
  {Ilyin}},\ }\href@noop {} {\bibfield  {journal} {\bibinfo  {journal} {Theor.
  Math. Phys.}\ }\textbf {\bibinfo {volume} {211}},\ \bibinfo {pages} {545}
  (\bibinfo {year} {2022})}\BibitemShut {NoStop}%
\bibitem [{\citenamefont {Skelt}\ \emph
  {et~al.}(2018{\natexlab{b}})\citenamefont {Skelt}, \citenamefont {Godby},\
  and\ \citenamefont {D'Amico}}]{BrazJPhys.48.467471}%
  \BibitemOpen
  \bibfield  {author} {\bibinfo {author} {\bibfnamefont {A.}~\bibnamefont
  {Skelt}}, \bibinfo {author} {\bibfnamefont {R.}~\bibnamefont {Godby}},\ and\
  \bibinfo {author} {\bibfnamefont {I.}~\bibnamefont {D'Amico}},\ }\href
  {https://doi.org/10.1007/s13538-018-0589-1} {\bibfield  {journal} {\bibinfo
  {journal} {Brazilian Journal of Physics}\ }\textbf {\bibinfo {volume} {48}},\
  \bibinfo {pages} {467} (\bibinfo {year} {2018}{\natexlab{b}})}\BibitemShut
  {NoStop}%
\bibitem [{\citenamefont {D'Amico}\ \emph {et~al.}(2011)\citenamefont
  {D'Amico}, \citenamefont {Coe}, \citenamefont {Fran\ifmmode~\mbox{\c{c}}\else
  \c{c}\fi{}a},\ and\ \citenamefont {Capelle}}]{PhysRevLett.106.050401}%
  \BibitemOpen
  \bibfield  {author} {\bibinfo {author} {\bibfnamefont {I.}~\bibnamefont
  {D'Amico}}, \bibinfo {author} {\bibfnamefont {J.~P.}\ \bibnamefont {Coe}},
  \bibinfo {author} {\bibfnamefont {V.~V.}\ \bibnamefont
  {Fran\ifmmode~\mbox{\c{c}}\else \c{c}\fi{}a}},\ and\ \bibinfo {author}
  {\bibfnamefont {K.}~\bibnamefont {Capelle}},\ }\href
  {https://doi.org/10.1103/PhysRevLett.106.050401} {\bibfield  {journal}
  {\bibinfo  {journal} {Phys. Rev. Lett.}\ }\textbf {\bibinfo {volume} {106}},\
  \bibinfo {pages} {050401} (\bibinfo {year} {2011})}\BibitemShut {NoStop}%
\bibitem [{\citenamefont {Sharp}\ and\ \citenamefont
  {D'Amico}(2014)}]{PhysRevB.89.115137}%
  \BibitemOpen
  \bibfield  {author} {\bibinfo {author} {\bibfnamefont {P.~M.}\ \bibnamefont
  {Sharp}}\ and\ \bibinfo {author} {\bibfnamefont {I.}~\bibnamefont
  {D'Amico}},\ }\href {https://doi.org/10.1103/PhysRevB.89.115137} {\bibfield
  {journal} {\bibinfo  {journal} {Phys. Rev. B}\ }\textbf {\bibinfo {volume}
  {89}},\ \bibinfo {pages} {115137} (\bibinfo {year} {2014})}\BibitemShut
  {NoStop}%
\bibitem [{\citenamefont {Sharp}\ and\ \citenamefont
  {D'Amico}(2015)}]{PhysRevA.92.032509}%
  \BibitemOpen
  \bibfield  {author} {\bibinfo {author} {\bibfnamefont {P.~M.}\ \bibnamefont
  {Sharp}}\ and\ \bibinfo {author} {\bibfnamefont {I.}~\bibnamefont
  {D'Amico}},\ }\href {https://doi.org/10.1103/PhysRevA.92.032509} {\bibfield
  {journal} {\bibinfo  {journal} {Phys. Rev. A}\ }\textbf {\bibinfo {volume}
  {92}},\ \bibinfo {pages} {032509} (\bibinfo {year} {2015})}\BibitemShut
  {NoStop}%
\bibitem [{\citenamefont {Sharp}\ and\ \citenamefont
  {D'Amico}(2016{\natexlab{a}})}]{JMMM.400.99-102}%
  \BibitemOpen
  \bibfield  {author} {\bibinfo {author} {\bibfnamefont {P.}~\bibnamefont
  {Sharp}}\ and\ \bibinfo {author} {\bibfnamefont {I.}~\bibnamefont
  {D'Amico}},\ }\href
  {https://doi.org/https://doi.org/10.1016/j.jmmm.2015.08.061} {\bibfield
  {journal} {\bibinfo  {journal} {Journal of Magnetism and Magnetic Materials}\
  }\textbf {\bibinfo {volume} {400}},\ \bibinfo {pages} {99} (\bibinfo {year}
  {2016}{\natexlab{a}})},\ \bibinfo {note} {proceedings of the 20th
  International Conference on Magnetism (Barcelona) 5-10 July 2015}\BibitemShut
  {NoStop}%
\bibitem [{\citenamefont {Sharp}\ and\ \citenamefont
  {D'Amico}(2016{\natexlab{b}})}]{PhysRevA.94.062509}%
  \BibitemOpen
  \bibfield  {author} {\bibinfo {author} {\bibfnamefont {P.~M.}\ \bibnamefont
  {Sharp}}\ and\ \bibinfo {author} {\bibfnamefont {I.}~\bibnamefont
  {D'Amico}},\ }\href {https://doi.org/10.1103/PhysRevA.94.062509} {\bibfield
  {journal} {\bibinfo  {journal} {Phys. Rev. A}\ }\textbf {\bibinfo {volume}
  {94}},\ \bibinfo {pages} {062509} (\bibinfo {year}
  {2016}{\natexlab{b}})}\BibitemShut {NoStop}%
\bibitem [{\citenamefont {Marocchi}\ \emph {et~al.}(2017)\citenamefont
  {Marocchi}, \citenamefont {Pittalis},\ and\ \citenamefont
  {D'Amico}}]{PhysRevMaterials.1.043801}%
  \BibitemOpen
  \bibfield  {author} {\bibinfo {author} {\bibfnamefont {S.}~\bibnamefont
  {Marocchi}}, \bibinfo {author} {\bibfnamefont {S.}~\bibnamefont {Pittalis}},\
  and\ \bibinfo {author} {\bibfnamefont {I.}~\bibnamefont {D'Amico}},\ }\href
  {https://doi.org/10.1103/PhysRevMaterials.1.043801} {\bibfield  {journal}
  {\bibinfo  {journal} {Phys. Rev. Mater.}\ }\textbf {\bibinfo {volume} {1}},\
  \bibinfo {pages} {043801} (\bibinfo {year} {2017})}\BibitemShut {NoStop}%
\bibitem [{\citenamefont {de~Picoli}\ \emph {et~al.}(2018)\citenamefont
  {de~Picoli}, \citenamefont {D'Amico},\ and\ \citenamefont {Fran{\c
  c}a}}]{BrazJPhys.48.472476}%
  \BibitemOpen
  \bibfield  {author} {\bibinfo {author} {\bibfnamefont {T.}~\bibnamefont
  {de~Picoli}}, \bibinfo {author} {\bibfnamefont {I.}~\bibnamefont {D'Amico}},\
  and\ \bibinfo {author} {\bibfnamefont {V.~V.}\ \bibnamefont {Fran{\c c}a}},\
  }\href@noop {} {\bibfield  {journal} {\bibinfo  {journal} {Braz. J. Phys.}\
  }\textbf {\bibinfo {volume} {48}},\ \bibinfo {pages} {472} (\bibinfo {year}
  {2018})}\BibitemShut {NoStop}%
\bibitem [{\citenamefont {Zawadzki}\ \emph {et~al.}(2020)\citenamefont
  {Zawadzki}, \citenamefont {Serra},\ and\ \citenamefont
  {D'Amico}}]{PhysRevResearch.2.033167}%
  \BibitemOpen
  \bibfield  {author} {\bibinfo {author} {\bibfnamefont {K.}~\bibnamefont
  {Zawadzki}}, \bibinfo {author} {\bibfnamefont {R.~M.}\ \bibnamefont
  {Serra}},\ and\ \bibinfo {author} {\bibfnamefont {I.}~\bibnamefont
  {D'Amico}},\ }\href {https://doi.org/10.1103/PhysRevResearch.2.033167}
  {\bibfield  {journal} {\bibinfo  {journal} {Phys. Rev. Res.}\ }\textbf
  {\bibinfo {volume} {2}},\ \bibinfo {pages} {033167} (\bibinfo {year}
  {2020})}\BibitemShut {NoStop}%
\bibitem [{\citenamefont {Hübner}(1992)}]{HUBNER1992239}%
  \BibitemOpen
  \bibfield  {author} {\bibinfo {author} {\bibfnamefont {M.}~\bibnamefont
  {Hübner}},\ }\href
  {https://doi.org/https://doi.org/10.1016/0375-9601(92)91004-B} {\bibfield
  {journal} {\bibinfo  {journal} {Physics Letters A}\ }\textbf {\bibinfo
  {volume} {163}},\ \bibinfo {pages} {239} (\bibinfo {year}
  {1992})}\BibitemShut {NoStop}%
\bibitem [{\citenamefont {Wu}\ and\ \citenamefont {Yu}(2020)}]{Wu2020}%
  \BibitemOpen
  \bibfield  {author} {\bibinfo {author} {\bibfnamefont {S.-x.}\ \bibnamefont
  {Wu}}\ and\ \bibinfo {author} {\bibfnamefont {C.-s.}\ \bibnamefont {Yu}},\
  }\href {https://doi.org/10.1038/s41598-020-62409-w} {\bibfield  {journal}
  {\bibinfo  {journal} {Scientific Reports}\ }\textbf {\bibinfo {volume}
  {10}},\ \bibinfo {pages} {2045} (\bibinfo {year} {2020})}\BibitemShut
  {NoStop}%
\bibitem [{\citenamefont {Wilde}(2013)}]{wilde_2013}%
  \BibitemOpen
  \bibfield  {author} {\bibinfo {author} {\bibfnamefont {M.~M.}\ \bibnamefont
  {Wilde}},\ }\href {https://doi.org/10.1017/CBO9781139525343} {\emph {\bibinfo
  {title} {Quantum Information Theory}}}\ (\bibinfo  {publisher} {Cambridge
  University Press},\ \bibinfo {year} {2013})\BibitemShut {NoStop}%
\bibitem [{\citenamefont {Kohn}\ and\ \citenamefont
  {Sham}(1965)}]{PhysRev.140.A1133}%
  \BibitemOpen
  \bibfield  {author} {\bibinfo {author} {\bibfnamefont {W.}~\bibnamefont
  {Kohn}}\ and\ \bibinfo {author} {\bibfnamefont {L.~J.}\ \bibnamefont
  {Sham}},\ }\href {https://doi.org/10.1103/PhysRev.140.A1133} {\bibfield
  {journal} {\bibinfo  {journal} {Phys. Rev.}\ }\textbf {\bibinfo {volume}
  {140}},\ \bibinfo {pages} {A1133} (\bibinfo {year} {1965})}\BibitemShut
  {NoStop}%
\bibitem [{\citenamefont {Runge}\ and\ \citenamefont
  {Gross}(1984)}]{PhysRevLett.52.997}%
  \BibitemOpen
  \bibfield  {author} {\bibinfo {author} {\bibfnamefont {E.}~\bibnamefont
  {Runge}}\ and\ \bibinfo {author} {\bibfnamefont {E.~K.~U.}\ \bibnamefont
  {Gross}},\ }\href {https://doi.org/10.1103/PhysRevLett.52.997} {\bibfield
  {journal} {\bibinfo  {journal} {Phys. Rev. Lett.}\ }\textbf {\bibinfo
  {volume} {52}},\ \bibinfo {pages} {997} (\bibinfo {year} {1984})}\BibitemShut
  {NoStop}%
\bibitem [{\citenamefont {Mahan}(1967)}]{PhysRev.163.612}%
  \BibitemOpen
  \bibfield  {author} {\bibinfo {author} {\bibfnamefont {G.~D.}\ \bibnamefont
  {Mahan}},\ }\href {https://doi.org/10.1103/PhysRev.163.612} {\bibfield
  {journal} {\bibinfo  {journal} {Phys. Rev.}\ }\textbf {\bibinfo {volume}
  {163}},\ \bibinfo {pages} {612} (\bibinfo {year} {1967})}\BibitemShut
  {NoStop}%
\bibitem [{\citenamefont {Nozi\`eres}\ and\ \citenamefont
  {De~Dominicis}(1969)}]{PhysRev.178.1097}%
  \BibitemOpen
  \bibfield  {author} {\bibinfo {author} {\bibfnamefont {P.}~\bibnamefont
  {Nozi\`eres}}\ and\ \bibinfo {author} {\bibfnamefont {C.~T.}\ \bibnamefont
  {De~Dominicis}},\ }\href {https://doi.org/10.1103/PhysRev.178.1097}
  {\bibfield  {journal} {\bibinfo  {journal} {Phys. Rev.}\ }\textbf {\bibinfo
  {volume} {178}},\ \bibinfo {pages} {1097} (\bibinfo {year}
  {1969})}\BibitemShut {NoStop}%
\bibitem [{\citenamefont {Doniach}\ and\ \citenamefont
  {Sunjic}(1970)}]{Doniach_1970}%
  \BibitemOpen
  \bibfield  {author} {\bibinfo {author} {\bibfnamefont {S.}~\bibnamefont
  {Doniach}}\ and\ \bibinfo {author} {\bibfnamefont {M.}~\bibnamefont
  {Sunjic}},\ }\href {https://doi.org/10.1088/0022-3719/3/2/010} {\bibfield
  {journal} {\bibinfo  {journal} {Journal of Physics C: Solid State Physics}\
  }\textbf {\bibinfo {volume} {3}},\ \bibinfo {pages} {285} (\bibinfo {year}
  {1970})}\BibitemShut {NoStop}%
\bibitem [{\citenamefont {Oliveira}\ and\ \citenamefont
  {Wilkins}(1981)}]{PhysRevB.24.4863}%
  \BibitemOpen
  \bibfield  {author} {\bibinfo {author} {\bibfnamefont {L.~N.}\ \bibnamefont
  {Oliveira}}\ and\ \bibinfo {author} {\bibfnamefont {J.~W.}\ \bibnamefont
  {Wilkins}},\ }\href {https://doi.org/10.1103/PhysRevB.24.4863} {\bibfield
  {journal} {\bibinfo  {journal} {Phys. Rev. B}\ }\textbf {\bibinfo {volume}
  {24}},\ \bibinfo {pages} {4863} (\bibinfo {year} {1981})}\BibitemShut
  {NoStop}%
\bibitem [{\citenamefont {d'Ambrumenil}\ and\ \citenamefont
  {Muzykantskii}(2005)}]{PhysRevB.71.045326}%
  \BibitemOpen
  \bibfield  {author} {\bibinfo {author} {\bibfnamefont {N.}~\bibnamefont
  {d'Ambrumenil}}\ and\ \bibinfo {author} {\bibfnamefont {B.}~\bibnamefont
  {Muzykantskii}},\ }\href {https://doi.org/10.1103/PhysRevB.71.045326}
  {\bibfield  {journal} {\bibinfo  {journal} {Phys. Rev. B}\ }\textbf {\bibinfo
  {volume} {71}},\ \bibinfo {pages} {045326} (\bibinfo {year}
  {2005})}\BibitemShut {NoStop}%
\bibitem [{\citenamefont {{Combescot, M.}}\ and\ \citenamefont {{Nozières,
  P.}}(1971)}]{J.Phys.019710032011-12091300}%
  \BibitemOpen
  \bibfield  {author} {\bibinfo {author} {\bibnamefont {{Combescot, M.}}}\ and\
  \bibinfo {author} {\bibnamefont {{Nozières, P.}}},\ }\href
  {https://doi.org/10.1051/jphys:019710032011-12091300} {\bibfield  {journal}
  {\bibinfo  {journal} {J. Phys. France}\ }\textbf {\bibinfo {volume} {32}},\
  \bibinfo {pages} {913} (\bibinfo {year} {1971})}\BibitemShut {NoStop}%
\bibitem [{\citenamefont {Ohtaka}\ and\ \citenamefont
  {Tanabe}(1990)}]{RevModPhys.62.929}%
  \BibitemOpen
  \bibfield  {author} {\bibinfo {author} {\bibfnamefont {K.}~\bibnamefont
  {Ohtaka}}\ and\ \bibinfo {author} {\bibfnamefont {Y.}~\bibnamefont
  {Tanabe}},\ }\href {https://doi.org/10.1103/RevModPhys.62.929} {\bibfield
  {journal} {\bibinfo  {journal} {Rev. Mod. Phys.}\ }\textbf {\bibinfo {volume}
  {62}},\ \bibinfo {pages} {929} (\bibinfo {year} {1990})}\BibitemShut
  {NoStop}%
\bibitem [{\citenamefont {Anderson}(1967)}]{PhysRevLett.18.1049}%
  \BibitemOpen
  \bibfield  {author} {\bibinfo {author} {\bibfnamefont {P.~W.}\ \bibnamefont
  {Anderson}},\ }\href {https://doi.org/10.1103/PhysRevLett.18.1049} {\bibfield
   {journal} {\bibinfo  {journal} {Phys. Rev. Lett.}\ }\textbf {\bibinfo
  {volume} {18}},\ \bibinfo {pages} {1049} (\bibinfo {year}
  {1967})}\BibitemShut {NoStop}%
\bibitem [{Note2()}]{Note2}%
  \BibitemOpen
  \bibinfo {note} {See also \cite {PhysRevLett.106.050401}, and \cite
  {G_Trindade}}\BibitemShut {NoStop}%
\bibitem [{\citenamefont {Pires}\ \emph {et~al.}(2016)\citenamefont {Pires},
  \citenamefont {Cianciaruso}, \citenamefont {C\'eleri}, \citenamefont
  {Adesso},\ and\ \citenamefont {Soares-Pinto}}]{PhysRevX.6.021031}%
  \BibitemOpen
  \bibfield  {author} {\bibinfo {author} {\bibfnamefont {D.~P.}\ \bibnamefont
  {Pires}}, \bibinfo {author} {\bibfnamefont {M.}~\bibnamefont {Cianciaruso}},
  \bibinfo {author} {\bibfnamefont {L.~C.}\ \bibnamefont {C\'eleri}}, \bibinfo
  {author} {\bibfnamefont {G.}~\bibnamefont {Adesso}},\ and\ \bibinfo {author}
  {\bibfnamefont {D.~O.}\ \bibnamefont {Soares-Pinto}},\ }\href
  {https://doi.org/10.1103/PhysRevX.6.021031} {\bibfield  {journal} {\bibinfo
  {journal} {Phys. Rev. X}\ }\textbf {\bibinfo {volume} {6}},\ \bibinfo {pages}
  {021031} (\bibinfo {year} {2016})}\BibitemShut {NoStop}%
\bibitem [{\citenamefont {Trindade}(2023)}]{G_Trindade}%
  \BibitemOpen
  \bibfield  {author} {\bibinfo {author} {\bibfnamefont {G.~R.}\ \bibnamefont
  {Trindade}},\ }\emph {\bibinfo {title} {From statistical models to
  $\alpha$-connections: an overview of information geometry}},\ \href
  {https://doi.org/https://doi.org/10.11606/D.55.2023.tde-09012024-110256}
  {Ph.D. thesis},\ \bibinfo  {school} {Instituto de Ciências Matemáticas e de
  Computação, Universidade de São Paulo, São Carlos} (\bibinfo {year}
  {2023})\BibitemShut {NoStop}%
\bibitem [{\citenamefont {Tong}\ \emph {et~al.}(2007)\citenamefont {Tong},
  \citenamefont {Singh}, \citenamefont {Kwek},\ and\ \citenamefont
  {Oh}}]{PhysRevLett.98.150402}%
  \BibitemOpen
  \bibfield  {author} {\bibinfo {author} {\bibfnamefont {D.~M.}\ \bibnamefont
  {Tong}}, \bibinfo {author} {\bibfnamefont {K.}~\bibnamefont {Singh}},
  \bibinfo {author} {\bibfnamefont {L.~C.}\ \bibnamefont {Kwek}},\ and\
  \bibinfo {author} {\bibfnamefont {C.~H.}\ \bibnamefont {Oh}},\ }\href
  {https://doi.org/10.1103/PhysRevLett.98.150402} {\bibfield  {journal}
  {\bibinfo  {journal} {Phys. Rev. Lett.}\ }\textbf {\bibinfo {volume} {98}},\
  \bibinfo {pages} {150402} (\bibinfo {year} {2007})}\BibitemShut {NoStop}%
\bibitem [{\citenamefont {Cui}\ \emph {et~al.}(2014)\citenamefont {Cui},
  \citenamefont {Wang}, \citenamefont {Argondizzo}, \citenamefont
  {Garrett-Roe}, \citenamefont {Gumhalter},\ and\ \citenamefont
  {Petek}}]{Cui2014}%
  \BibitemOpen
  \bibfield  {author} {\bibinfo {author} {\bibfnamefont {X.}~\bibnamefont
  {Cui}}, \bibinfo {author} {\bibfnamefont {C.}~\bibnamefont {Wang}}, \bibinfo
  {author} {\bibfnamefont {A.}~\bibnamefont {Argondizzo}}, \bibinfo {author}
  {\bibfnamefont {S.}~\bibnamefont {Garrett-Roe}}, \bibinfo {author}
  {\bibfnamefont {B.}~\bibnamefont {Gumhalter}},\ and\ \bibinfo {author}
  {\bibfnamefont {H.}~\bibnamefont {Petek}},\ }\href
  {https://doi.org/10.1038/nphys2981} {\bibfield  {journal} {\bibinfo
  {journal} {Nature Physics}\ }\textbf {\bibinfo {volume} {10}},\ \bibinfo
  {pages} {505} (\bibinfo {year} {2014})}\BibitemShut {NoStop}%
\bibitem [{\citenamefont {Reutzel}\ \emph {et~al.}(2019)\citenamefont
  {Reutzel}, \citenamefont {Li},\ and\ \citenamefont
  {Petek}}]{PhysRevX.9.011044}%
  \BibitemOpen
  \bibfield  {author} {\bibinfo {author} {\bibfnamefont {M.}~\bibnamefont
  {Reutzel}}, \bibinfo {author} {\bibfnamefont {A.}~\bibnamefont {Li}},\ and\
  \bibinfo {author} {\bibfnamefont {H.}~\bibnamefont {Petek}},\ }\href
  {https://doi.org/10.1103/PhysRevX.9.011044} {\bibfield  {journal} {\bibinfo
  {journal} {Phys. Rev. X}\ }\textbf {\bibinfo {volume} {9}},\ \bibinfo {pages}
  {011044} (\bibinfo {year} {2019})}\BibitemShut {NoStop}%
\bibitem [{\citenamefont {Borisov}\ \emph {et~al.}(1999)\citenamefont
  {Borisov}, \citenamefont {Kazansky},\ and\ \citenamefont
  {Gauyacq}}]{PhysRevB.59.10935}%
  \BibitemOpen
  \bibfield  {author} {\bibinfo {author} {\bibfnamefont {A.~G.}\ \bibnamefont
  {Borisov}}, \bibinfo {author} {\bibfnamefont {A.~K.}\ \bibnamefont
  {Kazansky}},\ and\ \bibinfo {author} {\bibfnamefont {J.~P.}\ \bibnamefont
  {Gauyacq}},\ }\href {https://doi.org/10.1103/PhysRevB.59.10935} {\bibfield
  {journal} {\bibinfo  {journal} {Phys. Rev. B}\ }\textbf {\bibinfo {volume}
  {59}},\ \bibinfo {pages} {10935} (\bibinfo {year} {1999})}\BibitemShut
  {NoStop}%
\bibitem [{\citenamefont {Neppl}\ \emph {et~al.}(2012)\citenamefont {Neppl},
  \citenamefont {Ernstorfer}, \citenamefont {Bothschafter}, \citenamefont
  {Cavalieri}, \citenamefont {Menzel}, \citenamefont {Barth}, \citenamefont
  {Krausz}, \citenamefont {Kienberger},\ and\ \citenamefont
  {Feulner}}]{PhysRevLett.109.087401}%
  \BibitemOpen
  \bibfield  {author} {\bibinfo {author} {\bibfnamefont {S.}~\bibnamefont
  {Neppl}}, \bibinfo {author} {\bibfnamefont {R.}~\bibnamefont {Ernstorfer}},
  \bibinfo {author} {\bibfnamefont {E.~M.}\ \bibnamefont {Bothschafter}},
  \bibinfo {author} {\bibfnamefont {A.~L.}\ \bibnamefont {Cavalieri}}, \bibinfo
  {author} {\bibfnamefont {D.}~\bibnamefont {Menzel}}, \bibinfo {author}
  {\bibfnamefont {J.~V.}\ \bibnamefont {Barth}}, \bibinfo {author}
  {\bibfnamefont {F.}~\bibnamefont {Krausz}}, \bibinfo {author} {\bibfnamefont
  {R.}~\bibnamefont {Kienberger}},\ and\ \bibinfo {author} {\bibfnamefont
  {P.}~\bibnamefont {Feulner}},\ }\href
  {https://doi.org/10.1103/PhysRevLett.109.087401} {\bibfield  {journal}
  {\bibinfo  {journal} {Phys. Rev. Lett.}\ }\textbf {\bibinfo {volume} {109}},\
  \bibinfo {pages} {087401} (\bibinfo {year} {2012})}\BibitemShut {NoStop}%
\bibitem [{\citenamefont {Ferrari}\ and\ \citenamefont
  {de~Oliveira}(2022)}]{PhysRevB.106.075129}%
  \BibitemOpen
  \bibfield  {author} {\bibinfo {author} {\bibfnamefont {A.~L.}\ \bibnamefont
  {Ferrari}}\ and\ \bibinfo {author} {\bibfnamefont {L.~N.}\ \bibnamefont
  {de~Oliveira}},\ }\href {https://doi.org/10.1103/PhysRevB.106.075129}
  {\bibfield  {journal} {\bibinfo  {journal} {Phys. Rev. B}\ }\textbf {\bibinfo
  {volume} {106}},\ \bibinfo {pages} {075129} (\bibinfo {year}
  {2022})}\BibitemShut {NoStop}%
\bibitem [{\citenamefont {Bulla}\ \emph {et~al.}(2008)\citenamefont {Bulla},
  \citenamefont {Costi},\ and\ \citenamefont {Pruschke}}]{RevModPhys.80.395}%
  \BibitemOpen
  \bibfield  {author} {\bibinfo {author} {\bibfnamefont {R.}~\bibnamefont
  {Bulla}}, \bibinfo {author} {\bibfnamefont {T.~A.}\ \bibnamefont {Costi}},\
  and\ \bibinfo {author} {\bibfnamefont {T.}~\bibnamefont {Pruschke}},\ }\href
  {https://doi.org/10.1103/RevModPhys.80.395} {\bibfield  {journal} {\bibinfo
  {journal} {Rev. Mod. Phys.}\ }\textbf {\bibinfo {volume} {80}},\ \bibinfo
  {pages} {395} (\bibinfo {year} {2008})}\BibitemShut {NoStop}%
\bibitem [{\citenamefont {Wilson}(1975)}]{RevModPhys.47.773}%
  \BibitemOpen
  \bibfield  {author} {\bibinfo {author} {\bibfnamefont {K.~G.}\ \bibnamefont
  {Wilson}},\ }\href {https://doi.org/10.1103/RevModPhys.47.773} {\bibfield
  {journal} {\bibinfo  {journal} {Rev. Mod. Phys.}\ }\textbf {\bibinfo {volume}
  {47}},\ \bibinfo {pages} {773} (\bibinfo {year} {1975})}\BibitemShut
  {NoStop}%
\bibitem [{\citenamefont {Picoli}\ \emph {et~al.}(2024)\citenamefont {Picoli},
  \citenamefont {Diniz}, \citenamefont {Oliveira},\ and\ \citenamefont
  {D'Amico}}]{Picoli}%
  \BibitemOpen
  \bibfield  {author} {\bibinfo {author} {\bibfnamefont {F.~D.}\ \bibnamefont
  {Picoli}}, \bibinfo {author} {\bibfnamefont {G.}~\bibnamefont {Diniz}},
  \bibinfo {author} {\bibfnamefont {L.~N.}\ \bibnamefont {Oliveira}},\ and\
  \bibinfo {author} {\bibfnamefont {I.}~\bibnamefont {D'Amico}},\ }\href@noop
  {} {\bibfield  {journal} {\bibinfo  {journal} {Unpublished}\ } (\bibinfo
  {year} {2024})}\BibitemShut {NoStop}%
\bibitem [{\citenamefont {Oliveira}\ and\ \citenamefont
  {Oliveira}(1994)}]{PhysRevB.49.11986}%
  \BibitemOpen
  \bibfield  {author} {\bibinfo {author} {\bibfnamefont {W.~C.}\ \bibnamefont
  {Oliveira}}\ and\ \bibinfo {author} {\bibfnamefont {L.~N.}\ \bibnamefont
  {Oliveira}},\ }\href {https://doi.org/10.1103/PhysRevB.49.11986} {\bibfield
  {journal} {\bibinfo  {journal} {Phys. Rev. B}\ }\textbf {\bibinfo {volume}
  {49}},\ \bibinfo {pages} {11986} (\bibinfo {year} {1994})}\BibitemShut
  {NoStop}%
\bibitem [{\citenamefont {Crank}\ and\ \citenamefont
  {Nicolson}(1947)}]{crank_nicolson_1947}%
  \BibitemOpen
  \bibfield  {author} {\bibinfo {author} {\bibfnamefont {J.}~\bibnamefont
  {Crank}}\ and\ \bibinfo {author} {\bibfnamefont {P.}~\bibnamefont
  {Nicolson}},\ }\href {https://doi.org/10.1017/S0305004100023197} {\bibfield
  {journal} {\bibinfo  {journal} {Mathematical Proceedings of the Cambridge
  Philosophical Society}\ }\textbf {\bibinfo {volume} {43}},\ \bibinfo {pages}
  {50–67} (\bibinfo {year} {1947})}\BibitemShut {NoStop}%
\bibitem [{Note3()}]{Note3}%
  \BibitemOpen
  \bibinfo {note} {The parameter $\nu $ helps accounting for the number of
  electrons participating in the dynamics, see discussion below.}\BibitemShut
  {Stop}%
\bibitem [{Note4()}]{Note4}%
  \BibitemOpen
  \bibinfo {note} {As the applied potential is increased linearly, the
  distances related to $K=-2$ maximum potential are part of the $K=-5$-related
  distances' set}\BibitemShut {NoStop}%
\bibitem [{\citenamefont {Friedel}(1952)}]{friedel1952}%
  \BibitemOpen
  \bibfield  {author} {\bibinfo {author} {\bibfnamefont {J.}~\bibnamefont
  {Friedel}},\ }\href@noop {} {\bibfield  {journal} {\bibinfo  {journal} {The
  London, Edinburgh, and Dublin Philosophical Magazine and Journal of Science}\
  }\textbf {\bibinfo {volume} {43}},\ \bibinfo {pages} {153} (\bibinfo {year}
  {1952})}\BibitemShut {NoStop}%
\bibitem [{Note5()}]{Note5}%
  \BibitemOpen
  \bibinfo {note} {Through their dependence on $|c_0(t)|$, trace and Bures
  distance both contain information about the global property $\DOTSB \sum@
  \slimits@ _{j\protect \ne 0}|c_{j}(t)|^2 = 1-|c_0(t)|^2$.}\BibitemShut
  {Stop}%
\bibitem [{\citenamefont {Sommerfeld}(1949)}]{sommerfeld1949partial}%
  \BibitemOpen
  \bibfield  {author} {\bibinfo {author} {\bibfnamefont {A.}~\bibnamefont
  {Sommerfeld}},\ }\href {https://books.google.com.br/books?id=iQU1zwEACAAJ}
  {\emph {\bibinfo {title} {Partial Differential Equations in Physics}}},\
  \bibinfo {series} {Lectures on theoretical physics}\ No.\ \bibinfo {number}
  {v. 1,pt. 1}\ (\bibinfo  {publisher} {Academic Press},\ \bibinfo {year}
  {1949})\BibitemShut {NoStop}%
\bibitem [{\citenamefont {Mahan}(2010)}]{Mahan2010-xj}%
  \BibitemOpen
  \bibfield  {author} {\bibinfo {author} {\bibfnamefont {G.~D.}\ \bibnamefont
  {Mahan}},\ }\href@noop {} {\emph {\bibinfo {title} {Many-particle
  physics}}},\ Physics of solids and liquids\ (\bibinfo  {publisher}
  {Springer},\ \bibinfo {address} {New York, NY},\ \bibinfo {year}
  {2010})\BibitemShut {NoStop}%
\end{thebibliography}%

\newpage
\appendix
\section{Superior Limit of the Local Density Distance}\label{SLDD}

Here, we detail the derivation of inequality~(\eqref{NR}), which establishes an upper bound for the local-density metric $D_{n}(n(t),n^A(t))$ at time $t$.

To perform the derivation, we will need to express the operators using two basis, and explicitly use at times the matrix notation (indicated by a double underline). The first basis is the time-dependent site occupation basis
\begin{equation}
    \{\ket{n, j} \} : = \left\{ \ket{n_1^j (t); n_2^j (t); ...;n_i^j (t);...} \right\}
\end{equation}
where, for any $j$ and $t$, the set of site-occupations $\{n_i^j(t)\}$ and $n =\sum_i^L n_i^j(t)$, with $n=N_e$ in the main of this paper. The site-occupation operator $\hat{n}_i$ is diagonal by construction in this basis
\begin{equation}
    {\hat{n}_i} \ket{n,j} = n_i^j(t) \ket{n,j}.
\end{equation}
The second basis is the instantaneous eigenstates' basis $\{\ket{\varphi_k(t)} \}$. 

We also define the change of basis transformation between these two representations $\hat V(t)$. These transformations are given by
\begin{eqnarray}
    \ket{\varphi_k(t)} = \hat V (t) \ket{n,j}
\end{eqnarray}
for the basis states, while for the operators, in matrix notation,  we have
\begin{equation} \label{ch_bas}
    \uu{O}^{\varphi}  = \uu{ V}^{\dagger}(t)~\uu{ O} ~\uu{ V} (t).
\end{equation}
Here the $\varphi$ index denotes the instantaneous basis and without the index the site-occupation basis. Since we are performing the calculations for the most general system possible, we do not have the elements of the change of basis transformation matrix $\uu{V}(t)$. We are only assuming knowledge of the system's wave function $\ket{\Psi(t)}$ in the instantaneous basis.

At zero temperature and for a closed system, we can express the density matrix operator in the instantaneous basis as
\begin{eqnarray}\label{RD}
{\hat{\rho}}(t) &=& \ket{\Psi (t)}\bra{\Psi (t)} \nonumber \\ &=& \sum_{k,k'}c_k (t) c_{k'}^{*}(t) \ket{\varphi_k(t)}\bra{\varphi_{k'}(t)}.
\end{eqnarray}
If the evolution is adiabatic, then the system remains in the instantaneous ground state and the density matrix is
\begin{eqnarray}\label{RA}
&\hat\rho_A (t)=  \ket{\varphi_0(t)}\bra{\varphi_0(t)}. 
\end{eqnarray} 

To compute the local density distance Eq. (\ref{ODM}), we can calculate the site occupation ${n}_i^j$ for each site $i$ corresponding to the time-evolved quantum state $\ket{\Psi(t)}$ through
\begin{eqnarray}\label{EV}
&{n_i^j(t)} = \langle \bm\hat {n}_i \rangle = \rm{Tr} \{\bm\hat\rho(t)\bm\hat{n}_i\},
\end{eqnarray} 
and the site occupation
corresponding to the instantaneous ground state $\ket{\varphi_0(t)}$ with the corresponding equation for $\rho_A (t)$.

The local density $n_i^j(t)$ is an observable and hence basis-independent, but its calculation through Eq. \eqref{EV} requires knowledge of the elements of both the site operator $\hat{n}_i$ and the density matrix $\hat{\rho}(t)$ in the same basis. A significant challenge arises, since in general $[\hat{n}_i,\hat{\rho}(t)]\ne 0$ and the two operators cannot be easily expressed in the same basis.

Therefore, our first task is to find a way to relate the density matrices in the instantaneous basis and the site-occupation basis.

It is convenient to use the site-occupation basis first, since the occupation operator is diagonal. To simplify the notation, we will omit the time dependence. Although we do not know the specific elements $ \rho_{(n,j)(n',j')}$, we can formally express 
the density matrix into the site-occupation basis
\begin{eqnarray}\label{GD_AUX_I}
&\hat \rho = \sum_{{(n,j)}} \sum_{{(n',j')}} \rho_{{ }_{(n,j)(n',j')}} \ket{n,j} \bra{n',j'}.
\end{eqnarray}


Then using the Eq. \eqref{GD_AUX_I}, as the site occupation operator is diagonal, we get 
\begin{equation*}
\hat\rho  \hat n_i =\sum_{(n,j)}  \sum_{(n',j')} \rho_{(n,j);(n',j')} n_i(n;j)  \ket{n,j} \bra{n',j'}
\end{equation*}
and the local density of each site becomes
\begin{equation}\label{GD_AUX_II}
 n_i^j = {\rm Tr} \{\hat\rho\hat n_i\} = \sum_{(n,j)}\rho_{(n,j)(n,j)} n_i(n,j).
\end{equation}

By using the same procedure to compute the local density of each site for the instantaneous ground state, we find
\begin{equation}\label{GD_AUX_III}
 n_i^A = {\rm Tr} \{\hat\rho_A\hat n_i\} = \sum_{(n,j)}\rho_{(n,j)(n,j)}^A n_i(n,j).
\end{equation}

To find the local density distance we need to compute $\left| n_i^{A} - n_i \right|$, which can be obtained using Eq. \eqref{GD_AUX_II} and \eqref{GD_AUX_III} by computing 
\begin{eqnarray*}
  \left| n_i^{A} - n_i \right| = \left| \sum_{(n,j)} \rho^{A}_{(n,j)}n_i(n,j)-\sum_{(n,j)}\rho_{(n,j)}n_i(n,j) \right|.  
\end{eqnarray*}

Now,  the occupation of each site is a positive 
number, so, by using the modular properties and the triangular inequality it is obtained
\begin{eqnarray}
    \left| \sum_{(n,j)}  \left(\rho^{A}_{(n,j)} - \rho_{(n,j)}\right) n_i(n,j) \right| \nonumber \\
    \le \sum_{(n,j)} \left| \rho^{A}_{(n,j)} - \rho_{(n,j)} \right| n_i(n,j).
\end{eqnarray}

As, in the site occupation representation,
\begin{equation*}
    \sum_{(n,j)} \left| \rho^{A}_{(n,j)} - \rho_{(n,j)} \right| n_i(n,j) = \mathrm{Tr} \left\{ \left|\hat\rho_{A} - \hat\rho \right| \hat n_i  \right\}, 
\end{equation*}
 we can write
\begin{equation}\label{Occupation Ineq.}
\left| n_i^{A}(t) - n_i (t) \right| \leq  \mathrm{Tr} \left\{ \left| \hat\rho_{A}(t) -  \hat\rho (t) \right|  \hat{n}_i  \right\}.
\end{equation}

On first look, it does not seem like we have found a good connection, as we do not know anything about the matrix elements of $\left|\hat\rho_{A} - \hat\rho\right|$ in the site-occupation basis. However, since the trace is basis agnostic, we can now turn to the instantaneous eigenstates' basis  to find an expression for $ |\hat\rho-\hat \rho_A| =\sqrt{\left[\hat\rho - \hat \rho_A\right]^{\dagger} \left[\hat\rho-\hat\rho_A\right]}$.

We can start it with the definitions of density matrices by Eq. \eqref{RA} and \eqref{RD} and write $\hat\rho-\hat \rho_A$ as
\begin{equation*}
    \hat\rho -\hat\rho_A  = \sum_{k,k'} c_k c_{k'}^{*} \ket{\varphi_k}\bra{\varphi_{k'}} -\ket{\varphi_0}\bra{\varphi_0},
\end{equation*}
and it is not difficult to show that
\begin{equation}
|\hat\rho-\hat\rho_A|^2  = \ket{\varphi_0}\bra{\varphi_0} +\sum_{k,k'} c_k c_{k'}^{*}\left[1 - \delta_{k,0}-\delta_{k',0}\right] \ket{\varphi_k}\bra{\varphi_k'}.
\label{rhophysq}\end{equation}
The above calculations only use linear algebra and the proprieties of the closed system density matrices like $\hat\rho ^ 2 = \hat\rho$ and $\hat\rho^\dagger = \hat\rho$.

In matrix form, (\ref{rhophysq}) becomes
\begin{eqnarray}
\uu{|\hat\rho^{\varphi}-\hat\rho_A^{\varphi}|^2}
&=&\left(\begin{array}{ccccccc}
1-|c_0|^2&0&0&...\\
0&|c_1|^2&c_1c_2^*&...\\
0&c_2c_1^*&|c_{2}|^2&...\\
\vdots&\vdots&\vdots&\ddots\\
\end{array}\right)
\\
&=&\left(\begin{array}{cc}
1-|c_0|^2&0\\
0& \uu{\varrho}\\
\end{array}\right),
\end{eqnarray}
where we have we defined the new Hermitian operator
\begin{equation}\label{varrho}
    \hat\varrho = \sum_{k,k' \neq 0} c_k c_{k'}^{*} \ket{\varphi_k}\bra{\varphi_{k'}},
\end{equation}
which excludes the instantaneous ground state, and with  $\rm{Tr}(\hat \varrho)=\sum_{k\ne0} |c_k|^2 = 1 - |c_0^2|$.

Once the matrix is block diagonal, we can simply apply the mathematical operation in the diagonal blocks and take their square root, as 
\begin{eqnarray}\label{RHO}
\uu{|\hat\rho^{\varphi}-\hat \rho_A^{\varphi}|}=\left(\begin{array}{cc}
\sqrt{1-|c_0|^2}&0\\
0& \uu{\sqrt{ \varrho}}\\
\end{array}\right). 
\end{eqnarray}
Now we only need the expression for the matrix $\uu{\sqrt{\varrho}}$. We can find it by doing the opposite operations, by taking the square of $\bm\hat\varrho$ using Eq. \eqref{varrho} we found that $ \bm\hat\varrho^2 = (1 - |c_0|^2) \bm\hat \varrho$, resulting in 
\begin{equation}
    \uu{\sqrt{ \varrho}} = \frac{1}{\sqrt{1 - |c_0|^2}} \uu{\varrho}, 
    \label{sqvar_mat}
\end{equation} with corresponding operator
\begin{equation}
  \hat{\sqrt{ \varrho}} =   \frac{1}{\sqrt{1 - |c_0|^2}} \sum_{k,k' \neq 0} c_k c_{k'}^{*} \ket{\varphi_k}\bra{\varphi_{k'}},
\label{sqvar}
\end{equation}

We now define the Hermitian  operator
\begin{equation}
    \hat y = c_0^* \sum_{k>0} c_k \ket{\varphi_k}\bra{\varphi_0} +c_0 \sum_{k>0} c_k^* \ket{\varphi_0}\bra{\varphi_k},
\end{equation}
whose corresponding matrix has non-null elements in the first row and first column only, and it is such that  $\rm{Tr} (\hat y (t))  =  0$
Then, utilizing Eq. \eqref{RD} we can write 
\begin{equation}
    \hat\rho = |c_0|^2 \hat \rho_A + \hat\varrho  + \hat y 
    \label{rhoy}
\end{equation}

Hence, by using Eqs.(\ref{RHO}),  and (\ref{sqvar}), 
we can write 
\begin{eqnarray}\label{Connection Rho}
{\left| \bm\hat\rho_A - \bm\hat\rho \right|} & = & \sqrt{1-|c_0|^2} \bm\hat\rho_A + \frac{1}{\sqrt{1-|c_0|^2}}\bm\hat\varrho \\
&=& \sqrt{1-|c_0|^2} \bm\hat\rho_A + \frac{\bm\hat\rho - |c_0|^2 \bm\hat\rho_A - \bm\hat y  }{ \sqrt{1-|c_0|^2}},\label{diff_rho_y} \label{rhodiff_y}
\end{eqnarray}
where the latter expression has used Eq. (\ref{rhoy}).

Then, we can substitute Eq. \ref{rhodiff_y}  into the inequality \eqref{Occupation Ineq.}, resulting in 
\begin{eqnarray}\label{Occupation Metric}
& \left| n_i^A - n_i \right|(t) \leq  \sqrt{1-|c_0(t)|^2} n_i^A (t) + \frac{ n_i (t) - |c_0(t)|^2 n_i^A (t)}{\sqrt{1-|c_0(t)|^2}} \nonumber \\
& \hspace{0.1 cm} - \dfrac{\rm{Tr}\left( \hat y (t)   \hat n_i\right)}{\sqrt{1-|c_0(t)|^2}}.
\end{eqnarray}
The expression \eqref{Occupation Metric} is generic and valid for any closed physical system at zero temperature. However, it is still hard to find the analytical expression for the last term, since the operators within the trace are naturally expressed in different basis and we do not know the explicit generic elements of the change of basis matrix $\uu{V}(t)$ of Eq. \ref{ch_bas}.

However, assuming total number of particle conservation,  then $\sum_i \bm\hat{n}_i = N_e \bm\hat I$. Thus, applying the sum over all the sites in Eq. \eqref{Occupation Metric}, and using $\rm{Tr} (\hat y (t)) =   0$, we finally find 
\begin{eqnarray}
&D_{n}(n(t),n^A(t)) \leq 2\sqrt{1-|c_0(t)|^2}.
\end{eqnarray}
This equation demonstrates that the local density distance has a superior limit that only depends on the instantaneous ground state transition probability $|c_0(t)|^2$.

\section{Analytical diagonalization of the Hamiltonian}\label{Annex_Analytical_Diagonalization}

Let us start with the Hamiltonian in Eq.\eqref{H_t} $H \equiv \mathcal{H}(t')$, for a fixed $t'$ and $K \equiv K(t')$. After the Fourier transformations 
\begin{equation}\label{fourier}
    a_n^\dagger = \frac{1}{\sqrt{N}} \sum_k \tilde{a}_k^\dagger \exp\left({-i \left(\frac{\pi k}{N} + \frac{\pi}{2} \right) n}\right),
\end{equation}
this Hamiltonian becomes:
\begin{equation}
\begin{aligned}
 H = \sum_k \varepsilon_k \tilde{a}_k^\dagger\tilde{a}_k + \frac{\tau K}{N} \sum_{k,q}\tilde{a}_k^\dagger\tilde{a}_{q}.
\end{aligned}
\end{equation}
Here $\varepsilon_k = 2\tau\rm{sin}\left(\frac{\pi k}{N} \right)$, and $q$ is an integer such that $-N/2\le q \le+N/2$.

We can diagonalize this Hamiltonian if we can write it as $H = \sum_{m} \gamma_{m} g_{m}^\dagger g_m$, where the $\{\gamma_m\}$ are the eigenvalues and $\{g_m^\dagger \}$ the single-particle eigenoperators of the Hamiltonian. The operator $g_m$ can be written as a linear combination of the initial operators $\{\tilde a_k\}$ as
\begin{eqnarray}\label{gm}
g_m = \sum_k \alpha_{k,m} \tilde a_{k}.
\end{eqnarray}

Our task now is to determine the eigenvalues $\{\gamma_m\}$ and the "eigenvectors" $\{g_m\}$. To achieve this, we need to compute the eigenvalues $\{\gamma_m\}$ and the coefficients $\{\alpha_{k,m}\}$ to diagonalize the Hamiltonian. The proceeding starts by computing the main commutators $[H,\tilde a_k]$ and $[H,g_m]$. After some manipulations, we arrive at the following expressions:
\begin{eqnarray}\label{[H,Ck]}
[H, \tilde a_k] = (-1) \left[ \epsilon_k \tilde a_k + \frac{K\tau}{N} \sum_q \tilde a_q    \right],
\end{eqnarray}
and
\begin{eqnarray}\label{[H,gm]}
[H,g_m] = -\gamma_{m}g_m,
\end{eqnarray}
where the latter has been derived by using that $g_m$ are the eigenvectors and $\gamma_m$ the eigenvalues of the diagonalized Hamiltonian.  

Using the equality defined in equation \eqref{[H,gm]} by substituting $g_m$ with the linear combination defined in \eqref{gm} and after some algebraic manipulations, we arrive at:
\begin{eqnarray}
\sum_k \left[\varepsilon_k \alpha_{k,m} +\frac{K\tau}{N} \sum_q \alpha_{q,m} \right] \tilde a_k = \sum_k \gamma_m \alpha_{k,m}\tilde a_k.
\end{eqnarray}

As the operators $\tilde a_k$ are linearly independent, each term of the $k$-sum above must satisfy the equation independently, that is
\begin{eqnarray}\label{E__I}
\left( \gamma_m-\varepsilon_k \right) \alpha_{k,m} - \frac{K\tau}{N} \sum_q \alpha_{q,m} = 0.
\end{eqnarray}
After summing over all $k$ states, we find
\begin{eqnarray}\label{E__II}
\sum_k \alpha_{k,m} = K \tau \left( \frac{1}{N} \sum_k \frac{1}{\left(\gamma_m-\varepsilon_k \right)} \right) \sum_q \alpha_{q,m},
\end{eqnarray}
with $\sum_k  \alpha_{k,m}  \ne 0$, from which we obtain
\begin{eqnarray}\label{Ktau}
1 = K \tau \left( \frac{1}{N} \sum_k \frac{1}{\left(\gamma_m-\varepsilon_k \right)} \right).
\end{eqnarray}

We wish now to analyze the sum on the r.h.s. of Eq.~\ref{Ktau}. It is convenient to define a variable $\mathcal{S}_m$ such that:
\begin{eqnarray}
\mathcal{S}_m &= \dfrac{1}{N} \mathlarger{\sum_q} \dfrac{1}{\gamma_m-\varepsilon_q}.
\end{eqnarray}

The scattering potential $K$ in the Hamiltonian $H$ shifts each initial eigenvalue $\varepsilon_m$ by $- \left(\dfrac{\delta_m }{\pi} \right) \Delta\varepsilon$, where $\delta_m$ is the phase shift. In principle, $\delta_m$ can take any value that adjusts the initial eigenvalue to match the eigenvalue $\gamma_m$ in the presence of the scattering potential. In this situation (see ref. \cite{PhysRevLett.18.1049,PhysRev.178.1097,PhysRevB.106.075129, RevModPhys.47.773,PhysRevB.49.11986}), we can write
\begin{eqnarray}\label{eigenvalues}
\gamma_m = \varepsilon_m - \frac{\delta_m}{\pi} \Delta\varepsilon.
\end{eqnarray}
Here $\Delta \varepsilon$ is the difference between two consecutive energy levels near the Fermi energy.

Now, we can write $\mathcal{S}_m$ as
\begin{eqnarray}
\mathcal{S}_m &= \dfrac{1}{N} \mathlarger{\sum_q} \dfrac{1}{\varepsilon_m-\varepsilon_q - \frac{\delta_m}{\pi} \Delta\varepsilon}.
\end{eqnarray}

To simplify the calculations, after this point we will consider a linear energy dispersion  $\varepsilon_q = q \Delta \varepsilon$, and $\Delta \varepsilon_q = \Delta \varepsilon$,  where $\Delta \varepsilon$ is the difference between two consecutive energy levels. The density of states for this linear dispersion is 
\begin{eqnarray}\label{Flatband_dos}
\rho \equiv \dfrac{1}{(N\Delta\varepsilon)},
\end{eqnarray}
and we can write
\begin{eqnarray}\label{AUX_SWSUM_I}
\mathcal{S}_m &  &\approx - \rho \sum_q \dfrac{1}{q - m + \frac{\delta_m}{\pi}}.
\end{eqnarray}

\begin{figure}[htb!]
		\centering
		\includegraphics[scale=0.38]{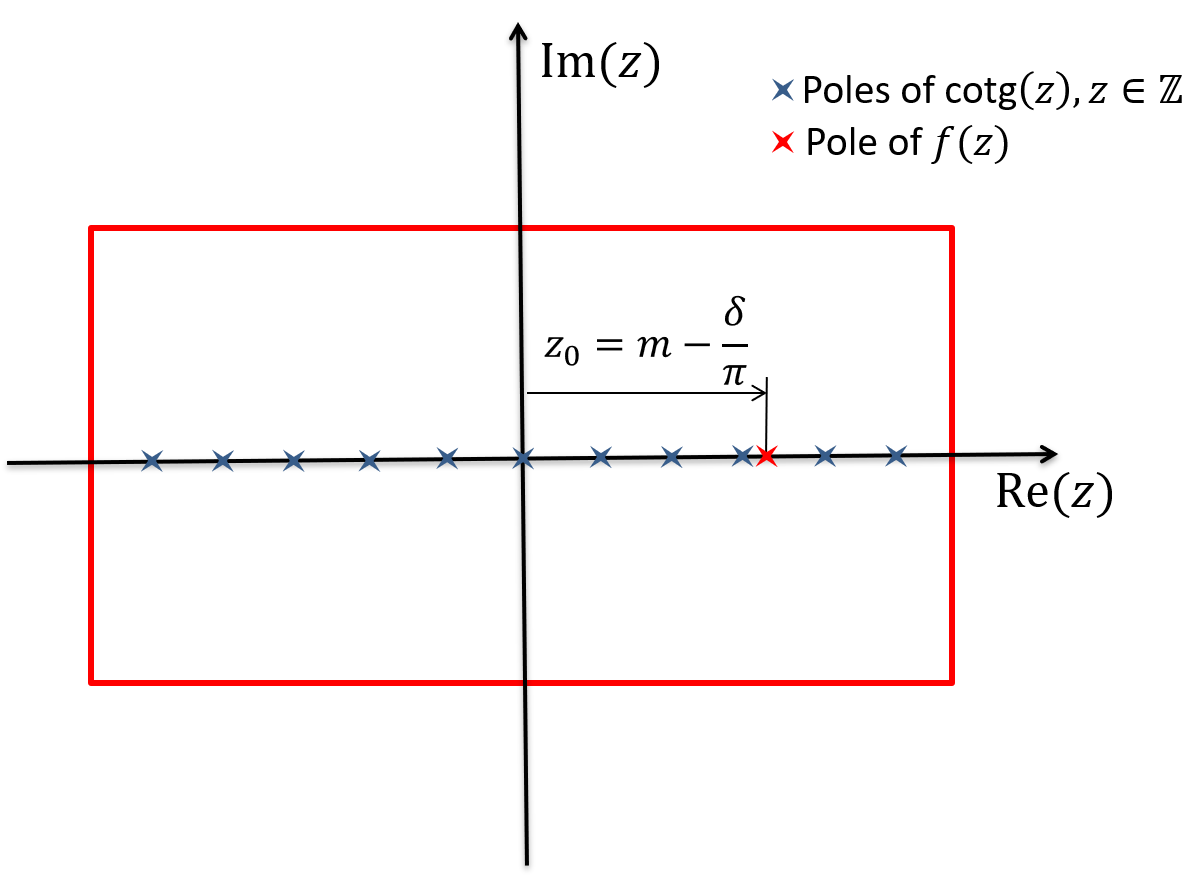}
		\caption{Integration path of \eqref{ComplexIntegral}.}
	\label{Sommerfeld-Watson}
\end{figure}

To solve the sum in Eq. \eqref{AUX_SWSUM_I}, we can use the Sommerfeld-Watson transformation \cite{sommerfeld1949partial}. Considering a sum of the form $\sum_n f(n)$, where $f(z)$ is a function with non-integer poles, we can define a function $F(z) = \pi f(z) \cot (\pi z)$. The function $F(z)$ has simple integer poles $\{n\}$ coming from the term $\sin (\pi z)$ and non integer poles $\{z_{n_i}\}$ coming from the term $f(z)$.  Applied  this transformation to find the value of $\mathcal{S}_m = \rho \sum_n f(n)$, since $f(z)=-\frac{1}{z-\left(m-\frac{\delta_m}{\pi}\right)}$ with a single simple pole $z_0= m-\frac{\delta_m}{\pi}$, after the Sommerfeld-Watson transformation Eq.~(\ref{AUX_SWSUM_I})   results in
\begin{eqnarray}\label{ComplexIntegral}
 \mathcal{S}_m \approx -\frac{1}{2 i}\oint\frac{dz~\rho~\mathrm{cot} (\pi z)}{z-\left(m-\frac{\delta_m}{\pi}\right)}+ \pi \rho \mathrm{cot}  (\pi m - \delta_m).
\end{eqnarray}

The last term can be simplified to $\cot (\pi m - \delta_m) = -\cot (\delta_m)$. Now we only need to find the integral. For this purpose, it is necessary to define a closed path in the domain of the function $F(z)=f(z) \mathrm{cot}  (z)$ as shown in Fig. \ref{Sommerfeld-Watson}. Taking the limit where the box size tends to infinity, and considering $\lim_{(\mathrm{Im} z \rightarrow \pm \infty)}\cot(\mathrm{Re}(z) +i\mathrm{Im} z) = \pm -i$, we can write
\begin{eqnarray}\label{Complex_integral_II}
  \frac{1}{2 i}\oint &dz& \frac{\mathrm{cot} (\pi z)}{z-\left(m-\frac{\delta_m}{\pi}\right)} = 
    \\  -&\frac{1}{2}&  \lim_{y \rightarrow + \infty }  \int_{+\infty}^{-\infty} dx \frac{1}{x+iy-z_0}~
    \\  +&\frac{1}{2}&  \lim_{y \rightarrow - \infty }  \int_{-\infty}^{+\infty} dx \frac{1}{x+iy-z_0}.~
\end{eqnarray}

\begin{figure}[htb!]
		\centering
		\includegraphics[scale=0.39]{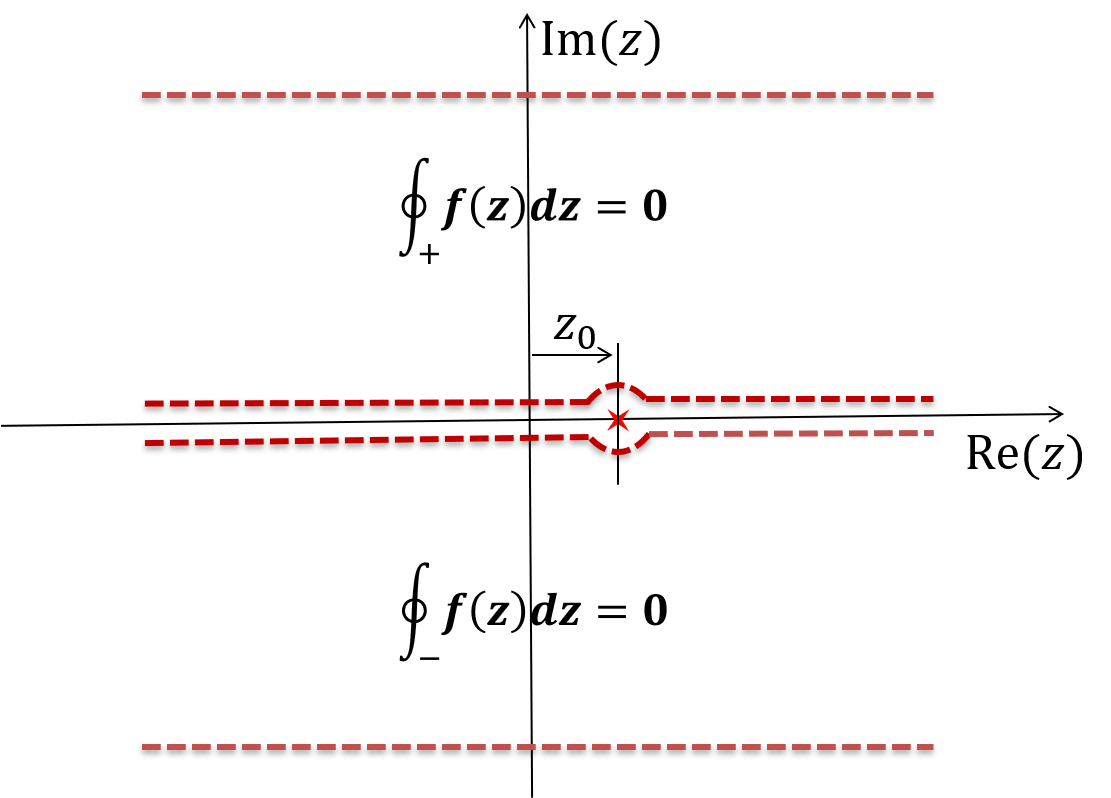}
		\caption{ In the Upper and Bottom boundaries of the integration path $\lim_{(\mathrm{Im} z \rightarrow \pm \infty)}\cot(\mathrm{Re}(z) +i\mathrm{Im} z) = \pm -i$, which results in the expression in Eq. \eqref{soma}.}
	\label{Contorno}
\end{figure}

Using a auxiliary path showed in Fig. \ref{Contorno}, for a function with poles only real axis $\frac{1}{2 i}\oint_\pm f(z) dz  = 0 $, and it is possible to show that
\begin{eqnarray}\label{Complex_integral_III}
\lim_{y \rightarrow \pm \infty } \int_{+\infty}^{-\infty} &dx& \frac{1}{x+iy-z_0} =  \nonumber \\  &-& \lim_{N \rightarrow \infty }\lim_{y \rightarrow  0^{\pm} }  \int_{-N/2}^{+N/2} dx \frac{1}{x+iy-z_0}.~~~~~~~~~
\end{eqnarray}

Finally, with the above results, we found that the contribution from the closed path in Fig. \ref{Sommerfeld-Watson} to $\mathcal{S}_m$ is $ \mathcal{P}\mathlarger{\int}_{-N \Delta\varepsilon/2}^{+N \Delta\varepsilon/2} d\epsilon \dfrac{\rho }{\varepsilon_m - \epsilon}$. Then, the expression in Eq. \eqref{ComplexIntegral} becomes
\begin{eqnarray}\label{soma}
\mathcal{S}_m \approx  - \pi\rho \mathrm{cot}(\delta_m) + \mathcal{P} \mathlarger{\int_{-\frac{N \Delta \varepsilon}{2}}^{+\frac{N \Delta \varepsilon}{2}}} \dfrac{d\epsilon~\rho}{\varepsilon_m - \epsilon}. 
\end{eqnarray}
Here $\mathcal{P}$ represents the Cauchy principal value of the integral. 

~ 

From Eq.~\ref{Ktau} we have
$ K \tau \mathcal{S}_m = 1$.
Then, by using the value of $\mathcal{S}_m$ calculated  in Eq. \eqref{soma}, we can  write 
\begin{eqnarray}\label{phase_shift_equation}
\cot(\delta_m) &\approx& \frac{1}{\pi}\mathcal{P}\int_{-N \Delta \varepsilon /2 }^{+N \Delta \varepsilon /2 }  \frac{1}{\varepsilon_m-\epsilon} d\epsilon  - \dfrac{1}{\pi \rho K \tau},  \nonumber \\
~ \nonumber \\
&\approx& - \dfrac{1}{\pi \rho K \tau} + \frac{1}{\pi}\ln \left( \frac{N \Delta\varepsilon/2 + \varepsilon_m }{N \Delta\varepsilon/2 - \varepsilon_m } \right).~ 
\end{eqnarray}
From this expression, it is possible to find the value of $\delta_m \equiv \delta_m(\varepsilon_m, K)$ for each $\varepsilon_m$ and fixed $K$ and consequently the values of the energy eigenvalues by Eq. \eqref{eigenvalues}.

Isolating the coefficient $\alpha_{k,m}$ in Eq. \eqref{E__I} and squaring both sides and then applying the sum over $k$, we obtain
\begin{eqnarray}
\sum_k |\alpha_{k,m}|^2=\frac{K^2 \tau^2 }{N} \frac{1}{N} \sum_k \frac{1}{\left( \gamma_m-\varepsilon_k \right) ^2} \left( \sum_q \alpha_{q,m} \right)^2. \nonumber
\end{eqnarray}
Where $\frac{1}{N} \sum_k \frac{1}{\left( \gamma_m-\varepsilon_k \right) ^2}$ $=$ $-\frac{d\mathcal{S}_m}{d\gamma_m} \approx \frac{1}{N \Delta \varepsilon^2 }|\frac{\pi}{\sin \delta_m}|^2$ considering only the first term in $\mathcal{S}_m$ and $\sum_k |\alpha_{k,m}|^2 = 1$ since the basis must be normalized. This leads to the expression
\begin{eqnarray}
1={ \rho^2 K^2 \tau^2} \left(\left|\frac{\pi}{\sin \delta_m}\right|^2\right) \left( \sum_q \alpha_{q,m} \right)^2, \nonumber \end{eqnarray}
or after taking the square root 
\begin{eqnarray}\label{sum_coef}
 \sum_q \alpha_{q,m} = \pm \frac{\sin \delta_m}{\pi \rho K \tau }.\end{eqnarray}

Finally, we have found the coefficients $\alpha_{k,m}$ as
\begin{eqnarray}
 \alpha_{k,m} = -\frac{\Delta \varepsilon}{\left( \gamma_m-\varepsilon_k \right)} \frac{\sin \delta_m}{\pi}, \end{eqnarray}
the energy eigenvalues $\gamma_m$ by Eq. \eqref{eigenvalues}
and from Eq. \eqref{phase_shift_equation} the phase shift as
\begin{eqnarray}\label{phase_shift_Aux}
&\tan(\delta_m) \approx - {\pi \rho K\tau}, \mathrm{~for~small~}\varepsilon_m.
\end{eqnarray}
Note that $\ln \left( \frac{N \Delta\varepsilon/2 + \varepsilon_m }{N \Delta\varepsilon/2 - \varepsilon_m } \right) $ is very small if $|m| \ll N/2$.

In the tight-binding model, the effective density of states is $\rho \equiv {1}/{(\pi \tau)}$,  allowing us to express the phase shift in Eq. \eqref{phase_shift_Aux} as
\begin{eqnarray}\label{phase_shift_}
&\tan(\delta_m) \approx - {K}, \mathrm{~for~small~}\varepsilon_m.
\end{eqnarray}

\section{Analytical Determination of Wave function Expansion Coefficients}\label{DA}

As we already discussed in Appendix \ref{Annex_Analytical_Diagonalization} for its time-independent counterpart, the Hamiltonian in Eq.\eqref{H_t} after the Fourier transformation becomes:
\begin{equation}\label{H_f}
\begin{aligned}
\mathcal{H}(t) = \sum_q \varepsilon_q \tilde{a}_q^\dagger\tilde{a}_q + \frac{ \tau K(t)}{N} \sum_{q,q'} \tilde{a}_q^\dagger\tilde{a}_{q'}.
\end{aligned}
\end{equation}
Here $\varepsilon_q = 2\tau\rm{sin}\left(\frac{\pi q}{2N_e} \right)$, $-N/2\le q \le+N/2$, and the phase shift 
\begin{eqnarray}
    {\delta} (t) = \rm{atan} \left(-K(t)\right).
\end{eqnarray}

Once again, for simplicity, it is convenient for us consider the energy spectrum of the Hamiltonian \eqref{H_f} as a linear dispersion $\varepsilon_q = \Delta\varepsilon \cdot q$, since the energy dispersion for the tight-binding can be approximated by $\varepsilon_q = 2\tau\sin\left(\frac{\pi q}{N}\right) \approx \frac{2\pi\tau}{N} q$ for small energies.

The single-particle Hamiltonian in Eq. \eqref{H_f} is well-known for a fermionic gas with a localized scattering potential and it can be diagonalized analytically \cite{PhysRevLett.18.1049} as shown in details in Appendix \ref{Annex_Analytical_Diagonalization}. The eigenstates of a single particle and the eigenenergies of $\mathcal{H}(t)$ \eqref{H_f} are given by the creation operator $g_l^\dagger(t) = \sum_q \alpha_{q,l}(t) \tilde{a}_q^\dagger$ and the energy 
\begin{equation}
    \gamma_l(t)= {\varepsilon}_l -\left(\frac{\delta(t)}{\pi}\right)\Delta\varepsilon.
\end{equation}
Here, the change of basis coefficients are
\begin{equation}\label{t.d.coefficients}
    \alpha_{q,l}(t) = -\frac{\sin{\delta}(t)}{\pi} \frac{\Delta\varepsilon}{{\varepsilon}_l - {\varepsilon}_q - \frac{\delta(t)}{\pi}\Delta\varepsilon }.
\end{equation} Note that the differences in the energy levels of the eigenvalues $\gamma_p(t) - \gamma_{h}(t) = \Delta\varepsilon (p - h)$ do not depend on the phase shift or time.

The Hamiltonian \eqref{H_f} is time-dependent and we need to connect the instantaneous basis at different times. To do it we can re-write $\mathcal{H}(t>t')$ as 
\begin{equation}
\begin{aligned}
\mathcal{H}(t) = \mathcal{H}(t') + \frac{\tau \left( K(t)-K(t') \right)}{N} \sum_{q,q'} \tilde{a}_q^\dagger\tilde{a}_{q'}.
\end{aligned}
\end{equation}
Now, using the Eqs. \eqref{sum_coef}, \eqref{Flatband_dos} and \eqref{phase_shift_Aux} then 
\begin{equation}
    \sum_{q,q'} \tilde{a}_q^\dagger\tilde{a}_{q'} = \cos^2\delta(t') \sum_{q,q'} g_q^\dagger(t') g_{q'}(t'),
\end{equation} and we can rewrite the Hamiltonian:
\begin{equation}\label{H_tt'}
\begin{aligned}
\mathcal{H}(t) = \mathcal{H}(t') + \frac{\tau (K(t)-K(t')) \cos^2\delta(t')}{N} \sum_{q,q'} g_q^\dagger(t') g_{q'}(t'). \nonumber
\end{aligned}
\end{equation}

Then, it is not difficult to show that $g_l^\dagger(t) =$ $\sum_q {\beta}_{l,q}(t,t') {g}_q^\dagger(t')$ 
with 
\begin{equation}\label{beta}
   \beta_{l,q}(t,t') =\frac{\sin\Delta\delta}{\pi}\frac{\Delta\varepsilon  }{\gamma_q(t') - \gamma_l(t') + \frac{\Delta\delta (t,t') }{\pi}\Delta\varepsilon} 
\end{equation}
the energies 
\begin{equation}
\gamma_l(t) = \gamma_l(t') -\left(\frac{\Delta\delta (t,t') }{\pi}\right)\Delta\varepsilon,
\end{equation}
and the difference in the phase shift
\begin{equation}
  \Delta\delta (t,t')=\delta(t) - \delta(t') + \mathcal{O}\left(|\delta(t) - \delta(t')|^3\right).
\end{equation}

To track the time evolution of the many-body system, we need to solve the Schrödinger equation, which, in the instantaneous basis, give to us the time propagation of the coefficients by 
\begin{align}\label{SE}
\frac{d\tilde{c}_m}{dt}=-\sum_n \Big( \tilde{c}_n (t) e^{-\frac{i}{\hbar}\int_0^t [E_n (t') - E_m (t')] dt' }  \nonumber \\
 \times \bra{\varphi_m} {\partial _t} \ket{\varphi_n} \Big),
\end{align}
where $E_m$ are the many-body eigenenergies.
The coefficient 
\begin{equation}
    \tilde{c}_n(t) \equiv c_{n}(t) e^{+\frac{i}{\hbar}\int_0^t dt' E_n(t')},
\end{equation}
contains all the information about the contribution of the many-body eigenstate $\ket{\varphi_n(t)}$.

The time derivative term $\bra{\varphi_n} {\partial _t} \ket{\varphi_m}$ in the differential equation \eqref{SE} can be written explicit in the limit definition of the derivative by
\begin{eqnarray}\label{lim}
\bra{\varphi_m}\partial_t\ket{\varphi_n}=\lim_{dt \rightarrow 0} \frac{\delta_{n,m} - \bra{\varphi_m(t)}\ket{\varphi_n(t^-)} }{dt}.
\end{eqnarray}
Here $t^- = t - dt$.

However, even knowing the general analytical single-particle solution, to build the many-body wave function, we need to deal with many-body projections $\bra{\varphi_n(t^-)}\ket{\varphi_m(t)}$, which can be obtained by the Slater determinants of matrices with elements $\beta_{n,q} \equiv \beta_{n,q}(t,t^-)$. 

Let us start by $\bra{\varphi_n(t)}\ket{\varphi_n(t^-)}$, written as
\begin{align}\label{Slater_I}
&\bra{\varphi_n(t^-)}\ket{\varphi_n(t)} = \nonumber  \\ &~~~~~~~\mathrm{det}
\begin{pmatrix}
\beta_{l_1,l_1} & \beta_{l_1,l_2} & ... &  \beta_{l_1,h}   & ... & \beta_{l_1,l_M} \\
\beta_{l_2,l_1} & \beta_{l_2,l_2} & ... &  \beta_{l_2,h}   & ... & \beta_{l_2,l_M} \\
\vdots          & \vdots          & ... &   \vdots           & ... & \vdots \\
\beta_{l_M,l_1} & \beta_{l_M,l_2} & ... &  \beta_{l_M,h}   & ... & \beta_{l_M,l_M} \\
\end{pmatrix} ~. 
\end{align}

Now, to compute $\bra{\varphi_n(t^-)} g_p^\dagger (t) g_h(t) \ket{\varphi_n(t)}$, it is necessary to modify all elements in the column $h$ by $\beta_{l,p}$ and the determinant becomes
\begin{align}\label{Slater_II}
&\bra{\varphi_n(t^-)} g_p^\dagger (t) g_h(t) \ket{\varphi_n(t)} = \nonumber  \\ &~~~~~~~\mathrm{det}
\begin{pmatrix}
\beta_{l_1,l_1} & \beta_{l_1,l_2} & ... &  \beta_{l_1,p}   & ... & \beta_{l_1,l_M} \\
\beta_{l_2,l_1} & \beta_{l_2,l_2} & ... &  \beta_{l_2,p}   & ... & \beta_{l_2,l_M} \\
\vdots          & \vdots          & ... &   \vdots           & ... & \vdots \\
\beta_{l_M,l_1} & \beta_{l_M,l_2} & ... &  \beta_{l_M,p}   & ... & \beta_{l_M,l_M} \\
\end{pmatrix} ~.
\end{align}

Since $\Delta \delta(t,t^-) \ll 1$ , the diagonal elements of the matrix are dominant. By using Eq. \ref{beta} $\beta_{h,h} =  \frac{\sin(\Delta \delta)}{\Delta \delta} $ and $\beta_{h,p} = \frac{\sin(\Delta \delta)}{\pi} \frac{\Delta \varepsilon}{\varepsilon_p - \varepsilon_h + \frac{\Delta\delta}{\pi }\Delta\varepsilon }$, we can approximate the projections by the diagonal terms as
\begin{align}
    \bra{\varphi_n(t^-)}\ket{\varphi_n(t)} \approx \left(\frac{\sin(\Delta \delta)}{\Delta \delta} \right)^M, \nonumber
\end{align}
and
\begin{align} 
    \bra{\varphi_n(t^-)} & g_p^\dagger (t) g_h(t)  \ket{\varphi_n(t)}  \nonumber \approx \\ &\frac{\Delta \delta}{\pi} \frac{\Delta \varepsilon}{\varepsilon_p - \varepsilon_h + \frac{\Delta\delta}{\pi }\Delta\varepsilon }  \left( \frac{\sin(\Delta \delta)}{\Delta \delta} \right)^{M}. \nonumber
\end{align}

It is straightforwards to show that 
\begin{align} 
    \bra{\varphi_n(t^-)} & g_p^\dagger (t) g_h(t)  \ket{\varphi_n(t)}  \nonumber \approx \\ &\frac{\Delta \delta}{\pi} \frac{\Delta \varepsilon}{\varepsilon_p - \varepsilon_h + \frac{\Delta\delta}{\pi }\Delta\varepsilon }  \bra{\varphi_n(t^-)}\ket{\varphi_n(t)}. \nonumber
\end{align}

We can think about $\ket{\varphi_m(t)}$ as a product of pair particle-hole excitations $ \prod_j g_{p_j}^\dagger (t) g_{h_j}(t)$ of the configuration $\ket{\varphi_n(t)}$ as $\ket{\varphi_m(t)} = \prod_j g_{p_j}^\dagger(t) g_{h_j}(t) \ket{\varphi_n(t)}$. Then, as we discussed above, for small $\Delta\delta$, the projection $\bra{\varphi_n(t^-)}\ket{\varphi_m(t)}$ can be similarly approximate by the product of the diagonal terms as
\begin{eqnarray}\label{Proj}
\frac{\bra{\varphi_n(t^-)}\ket{\varphi_m(t)}}{\bra{\varphi_n(t^-)}\ket{\varphi_n(t)}} \approx \left[ \prod_j \left(\frac{\Delta\delta}{\pi} \right) \frac{\Delta\varepsilon}{\varepsilon_{p_j} - \varepsilon_{h_j} + \frac{\Delta\delta}{\pi} \Delta\varepsilon} \right].~~~~
\end{eqnarray}

Nevertheless, we know from Eq.\eqref{AOC} that the projection $\bra{\varphi_n(t)}\ket{\varphi_n(t^-)} \sim N_e^{-\left(\frac{\Delta\delta}{\pi} \right)^2}$.
Then, combining the Eq.\eqref{Proj} and \eqref{lim} we obtain
\begin{align}\label{derivative}
\bra{\varphi_m}\partial_t\ket{\varphi_n} &=& &\frac{1}{\pi}\frac{d\delta}{dt} \frac{\Delta\varepsilon}{\varepsilon_p - \varepsilon_h} &~&\mathrm{if}~ \ket{\varphi_m} = g_p^\dagger g_h \ket{\varphi_n}, \nonumber \\
    &=& &0 &~&\mathrm{otherwise}.
\end{align}
Eq. \eqref{derivative} is different from zero only if $\ket{\varphi_m} = g_p^\dagger g_h \ket{\varphi_n}$, for $h \neq p$, or, in other words, the many-body state is a particle-hole excitation away from the other one.

Finally, we can re-write the Eq \eqref{SE} as
\begin{equation}\label{SEF}
\begin{aligned}
\frac{d\tilde{c}_n}{dt} = -\frac{1}{\pi} \frac{d\delta}{dt}\sum_p\sum_{h\neq p} \tilde{c}_{n,p,h} \frac{\Delta\varepsilon}{\varepsilon_p - \varepsilon_h} e^{-i(\varepsilon_p - \varepsilon_h)\frac{t}{\hbar}},
\end{aligned}
\end{equation}
where $\tilde{c}_{n,p,h}$ are the coefficients of the instantaneous eigenstate $\ket{\varphi_{n,p,h}} = g_p^\dagger g_h \ket{\varphi_n }$.

This result demonstrates that for this type of system, when the localized scattering potential changes continuously, a direct coupling exists between two many-body instantaneous eigenstates only when it is possible to represent one many-body eigenstate as a particle-hole excitation from the other one.

\section{Estimating the Instantaneous Ground State Probability at the End of the Ramp-Up}\label{|C_0|}

Let us start from the Eq. \eqref{SEFM}, considering $\frac{d\delta}{dt}$ constant, we can estimate the contribution of the single-particle-hole excitations as
\begin{equation}
\begin{aligned}
\tilde{c}_{0,p,h}(t) \approx   \frac{1}{\pi} \frac{d\delta}{dt}  \frac{1}{p + h}  \int_0^t dt' \tilde{c}_{0}(t') e^{+i(p +h)\frac{\Delta\varepsilon t'}{\hbar}}.
\end{aligned}
\end{equation}

Now, assuming $ c_0(t') \approx 1$, we can solve the integral and obtain an approximation for the coefficient of a single-particle-hole excitation as follows:
\begin{equation}\label{1PH}
\begin{aligned}
\tilde{c}_{0,p,h}(t) \approx \frac{1}{\pi} \frac{d\delta}{dt}  \frac{e^{+i\frac{\Delta\varepsilon(p +h)  T}{2\hbar}} }{p + h} \frac{\sin\left( \frac{\Delta\varepsilon (p+h) T}{2\hbar} \right) }{\left( \frac{\Delta\varepsilon(p+h)}{2\hbar} \right)}.
\end{aligned}
\end{equation}

By the normalization, the probability to found the system in the ground state can be obtained by
\begin{equation}
\begin{aligned}
|\tilde{c}_{0}(T)|^2  =  1 - \sum_{p,h} |\tilde{c}_{0,p,h}(T) |^2 - ...,
\end{aligned}
\end{equation}
considering that the probability of finding the system in a state with two or more particle-hole pairs is very small. Note that $|\tilde{c}_{0}(T)|^2 =  |{c}_{0}(T)|^2$ by Eq. \eqref{transformation}. By using Eq. \eqref{1PH}, we find that
\begin{equation}
\begin{aligned}
|{c}_{0}(T)|^2  \approx 1 - \left(\frac{1}{\pi} \frac{d\delta}{dt} \right)^2 \sum_{p,h}   \frac{1}{(p + h)^2} \frac{\sin^2\left( \frac{\Delta\varepsilon (p+h) T}{2\hbar} \right) }{\left( \frac{\Delta\varepsilon(p+h)}{2\hbar} \right)^2}.
\end{aligned}
\end{equation}

By approximating the sum $ \sum_{p>0} \sum_{h \ge 1} $ with the integral $ \int_0^{N_e} \int_1^{N_e} dp \, dh $, once there are no poles within the summation range, we find, after some manipulations, that:
\begin{equation}
\begin{aligned}
|{c}_{0}(T)|^2  \approx 1 - \left(\frac{1}{\pi} \frac{d\delta}{dt} T \right)^2 \int_{\frac{\Delta\varepsilon T}{2\hbar}}^{\frac{\pi \tau T}{2\hbar}} du \frac{\sin^2 u}{u^3}.
\end{aligned}
\end{equation}

The function $ f(x) = \sin^2(x)/x^3 $ behaves as $ 1/x $ for $ x \le 0.1 $ and as $ 1/x^3 $ for $ x > 10$. Therefore, if $ \frac{\pi \tau T}{2\hbar} > 10$  and  $\frac{T_\tau}{T_{\Delta\varepsilon}} \le 0.1$, we can approximately split the integral into three parts as follows:
\begin{equation}
\begin{aligned}
|{c}_{0}(T)|^2  \approx 1 - \left(\frac{1}{\pi} \frac{d\delta}{dt} T \right)^2 \Bigg(\int_{\frac{\Delta\varepsilon T}{2\hbar}}^{0.1}  \frac{du}{u} +\\ \int_{0.1}^{10} du \frac{\sin^2 u}{u^3} +  \int_{10}^{\frac{\pi \tau T}{2\hbar}} \frac{du}{u^3} \Bigg).
\end{aligned}
\end{equation}

Note that the fist integral result in 
\begin{equation}
\begin{aligned}
\int_{\frac{\Delta\varepsilon T}{2\hbar}}^{0.1}  \frac{du}{u}  = \ln\left(0.2\frac{T_{\Delta\varepsilon}}{T}\right).
\end{aligned}
\end{equation}
The second one can be computed numerically and it results in
\begin{equation}
\begin{aligned} 
\int_{0.1}^{10} du \frac{\sin^2 u}{u^3} = 2.53118
\end{aligned}
\end{equation}
The last integral give to us
\begin{equation}
\begin{aligned} 
\int_{10}^{\frac{\pi \tau T}{2\hbar}} du \frac{1}{u^3} = \frac{1}{2} \left( 10^{-2} - \left(\frac{\pi T}{2 T_\tau}\right)^{-2} \right).
\end{aligned}
\end{equation}

Grouping the contributions of each above integral leads to the following result:
\begin{equation}
\begin{aligned}
|{c}_{0}(T)|^2 \approx 1 - &\left(\frac{T}{\pi} \frac{d\delta}{dt} \right)^2\ln\left( 2.53 \frac{T_{\Delta\varepsilon}}{T}\right) + 2\left( \frac{T_\tau}{\pi^2} \frac{d\delta}{dt}  \right)^2.
\end{aligned}
\end{equation}
Here we used that $y = \ln(e^{y})$. The last term contribution is very small if $T \gg T_\tau$.

Now, by using the expansion $ a^{-x} = 1 - x\ln a + \dots $, we can approximate $ |c_0(T)| $ as:
\begin{equation}
\begin{aligned}
|{c}_{0}(T)|^2  \approx \left(\bar \nu \frac{T_{\Delta\varepsilon}}{T}\right)^{-\left(\dfrac{1}{\pi} \dfrac{d\delta}{dt} T \right)^2 }, 
\end{aligned}
\end{equation}
where $\bar \nu = 2.53$.

If $\delta(T)$ is small, $\frac{d \delta }{d t} \approx \frac{\delta(T) }{T}$, and then 
\begin{equation}\label{E_by_PT}
\begin{aligned}
|{c}_{0}(T)|^2  \propto \left(\frac{T_{\Delta\varepsilon}}{T}\right)^{-\left(\frac{\delta(T) }{\pi} \right)^2 }.
\end{aligned}
\end{equation}

The expression in Eq. \eqref{E_by_PT} is obtained from the first iteration by initially assuming $c_0(t) \approx 1$ for $0\le t \le T$, in the first order of $\left(\frac{\delta(T) }{\pi}\right)^2$, and assuming $T_\tau \ll T \ll T_{\Delta\varepsilon}$.

\section{Derivation of Eq. \eqref{DA:SQC} }\label{DEXP14}
From Eq. \eqref{AOCII} and considering $|c_0|^2 \ge 1 - \eta^2$: 
\begin{equation}
|c_0|^2 \approx \left(N_e\right)^{-2\left({\frac{\delta}{\pi}}\right)^2}  \ge 1 - \eta^2, \nonumber
\end{equation}
resulting in 
\begin{equation}
\left(\frac{\delta}{\pi} \right)^2  \le -0.5 \ln(1-\eta^2) \left(\ln \left( N_e \right)\right)^{-1}. \nonumber
\end{equation}

Taking the square root on both sides of the above inequality results in the inequality
\begin{equation}\label{AUXGDI}
\left|\frac{\delta}{\pi} \right|  \le \sqrt{-0.5 \ln(1-\eta^2) \left(\ln \left( N_e  \right)\right)^{-1}}. \nonumber
\end{equation}

Since $u  \le  \tan u,~u~\in~\left[0,\frac{\pi}{2}\right),$ by using the Eq. \eqref{phase}, the Eq.\eqref{AUXGDI} results in
\begin{equation}\label{Cond. II}
 |K|  \le {\pi \eta^*} \left( 2\ln \left( N_e \right)\right)^{-1/2}. 
\end{equation}
Here $\eta^* = \left(-\ln(1-\eta^2)\right)^{1/2} = \eta \left(1 + 0.25\eta^2\right)+ \mathcal{O}\left( \eta^5 \right)$, which for small values of $\eta$ implies $\eta^* \approx \eta$.

\section{Derivation of Eq. \eqref{DA:RUC}}\label{ADEXP}
From Eq. \eqref{AS} and considering $|c_0|^2 \ge 1 - \eta^2$: 
\begin{equation}
|c_0|^2 = \left( \nu \frac{T_{\Delta\varepsilon }}{T} \right)^{- \left(\frac{\delta}{\pi} \right)^2\left( 1 + \left(\frac{{\delta} }{\pi} \right)^2 \right)} \ge 1 - \eta^2. \nonumber
\end{equation}

Now, let us consider that  $|c_0|^2$ is fixed, then:
\begin{equation}
\left(\frac{\delta}{\pi} \right)^2  + \left(\frac{\delta}{\pi} \right)^4  = -\frac{\ln(|c_0|^2)}{\ln\left( \nu \frac{T_{\Delta\varepsilon }}{T} \right)}. \nonumber
\end{equation}


But since $|c_0|^2 \ge 1 - \eta^2$:
\begin{equation}
\left(\frac{\delta}{\pi} \right)^2\left( 1 + \left(\frac{{\delta} }{\pi} \right)^2 \right) \le -\ln(1-\eta^2) \left( \ln \left( \nu \frac{T_{\Delta\varepsilon }}{T} \right)\right)^{-1}. \nonumber
\end{equation}
Since $x^2 \le x^2 + x^4$,
\begin{equation}
\left(\frac{\delta}{\pi} \right)^2  \le -\ln(1-\eta^2) \left( \ln \left( \nu \frac{T_{\Delta\varepsilon }}{T} \right)\right)^{-1}, \nonumber
\end{equation}
taking the square root on both sides of the above inequality results in 
\begin{equation}
\left|\frac{\delta}{\pi} \right|  \le \sqrt{-\ln(1-\eta^2) \left( \ln \left( \nu \frac{T_{\Delta\varepsilon }}{T} \right)\right)^{-1}}. \nonumber
\end{equation}

Since $ u  \le  \tan u,~u~\in~\left[0,\frac{\pi}{2}\right)$, by using the Eq. \eqref{phase}, it results in
\begin{equation}\label{Cond. I}
 |K|  \le \pi \eta \left( \ln \left( \nu \frac{T_{\Delta\varepsilon }}{T} \right)\right)^{-1/2}. 
\end{equation}

\section{The matrix element in the QAC}\label{Matrix:QAC}

We aim here to find analytically the matrix element defined by the expression
\begin{align}
    \mathcal{M}(t) = \bra{\varphi_0 (t) }  a_0^\dagger a_0 \ket{\varphi_1(t)},
\end{align}
which is necessary to compute the QAC for our system. Here, $\ket{\varphi_1(t)}$ represents the many-body eigenstate with the lowest energy. This many-body state corresponds to a single particle-hole excitation from the instantaneous ground state and can be expressed as
\begin{align}\label{Matrix_Element_}
    \ket{\varphi_1 (t) } = g_0^\dagger g_{-1} \ket{\varphi_0(t)}.
\end{align} 

Keeping that in  mind, the matrix element $\mathcal{M}(t)$ in Eq. \eqref{Matrix_Element_} can be found by the expression
\begin{align}\label{New: AUX1}
    \mathcal{M}(t) = \bra{\varphi_0 (t) }  a_0^\dagger a_0  g_0^\dagger(t) g_{-1}(t)  \ket{\varphi_0(t) }.
\end{align}

As already discussed, the set of single-particle operators $\{g_k(t) \}$ form a complete set, allowing us to expand the tight-binding site operator $a_0$ as
\begin{equation}\label{New: AUX2}
    a_0 = \sum_k u_{0,k}(t) ~ g_k(t),
\end{equation}
where $u_{0,k}(t) = \left\{ a_0,  g_k^\dagger (t) \right\}$ are the coefficients of the expansion.

Now, we can substitute the result from Eq. \eqref{New: AUX2} into Eq. \eqref{New: AUX1}, and we obtain
\begin{align}\label{New: AUX3}
    \mathcal{M}(t) = \sum_{k,q} &u_{0,k}(t) u_{0,q}(t)  \times \nonumber \\  &\bra{\varphi_0 (t) }  g_k^\dagger(t) g_q(t)  g_0^\dagger (t) g_{-1}(t)  \ket{\varphi_0(t) }.
\end{align}

Wick's theorem \cite{Mahan2010-xj} then tells us that the terms in the summation in Eq. \eqref{New: AUX3} are nonzero only if  $k = -1$ and $q = 0$, which results in the expression
\begin{align}\label{New: AUX4}
    \mathcal{M}(t) = u_{0,-1}(t) u_{0,0}(t).
\end{align}

Now, we only need to find the coefficients $u_{0,k}(t)$ to finally find the matrix element which we are interested in. The expression in Eq. \eqref{New: AUX4} can be further analyzed using Eq. \eqref{fourier} with $n = 0$, which gives
\begin{equation}\label{New: AUX5}
    a_0 = \sum_k \frac{1}{\sqrt{N}} \tilde a_k.
\end{equation}

Keeping the calculations to find $\mathcal{M}(t)$, now we can use Eq. \eqref{New: AUX5} and we found that
\begin{align}\label{New: AUX6}
      u_{0,k}(t) &= \left\{ a_0,  g_k(t)^\dagger\right\} = \frac{1}{\sqrt{N}} \sum_q   \left\{ \tilde a_q,  g_k^\dagger (t) \right\} \nonumber \\ &=  -\frac{\sin\delta(t)}{\pi \sqrt{N}} \sum_q \frac{\Delta\varepsilon}{\varepsilon_k - \varepsilon_q - \frac{1}{\pi}\delta(t) \Delta\varepsilon} \nonumber \\ &=  -\frac{\sin\delta(t)}{\pi \sqrt{N}} \sum_q \frac{1}{k - q - \frac{1}{\pi}\delta(t)}.
\end{align}
Here, we used that $\alpha_{q,k}(t) = \left\{ \tilde a_q,  g_k^\dagger (t) \right\}$, the Eq. \eqref{t.d.coefficients}, and that $\varepsilon_q = \Delta\varepsilon \cdot q$.

To continue the calculations, we need to determine the value of the summation that appears in Eq. \eqref{New: AUX6}. This can be achieved using the results from Appendix \ref{Annex_Analytical_Diagonalization}, by equating the right-hand sides of the expressions in Eqs. \eqref{AUX_SWSUM_I} and \eqref{soma}, we can conclude that
\begin{align}\label{New: AUX7}
       \sum_q \frac{1}{k - q - \frac{1}{\pi}\delta(t)} \approx - \pi \mathrm{cot}~\delta(t).
\end{align}
Since $ |k| \ll N$, the integral contribution from Eq. \eqref{soma} is practically null.

Using the result from Eq. \eqref{New: AUX7}, we can write the Eq. \eqref{New: AUX6} as
\begin{align}\label{New: AUX8}
      u_{0,k}(t) &\approx  \frac{\cos\delta(t)}{\sqrt{N}}.
\end{align}

Finally, using Eq. \eqref{New: AUX8}, we find that the matrix element we are interested in can be written simply as
\begin{align}\label{New: AUX9}
\mathcal{M}(t)    &\approx \frac{1}{N}\cos^2\delta(t) \nonumber \\
                  &\approx \frac{\cos^2 \delta(t) }{2 \pi} \frac{T_\tau}{T_{\Delta\varepsilon}}.
\end{align}
In the last step, we used Eqs. \eqref{GAP}, \eqref{T_tau}, and \eqref{T_eps}.

\end{document}